\documentclass[3p, 11pt, authoryear]{elsarticle}

\makeatletter
\def\ps@pprintTitle{%
    \let\@oddhead\@empty
    \let\@evenhead\@empty
    \let\@evenfoot\@oddfoot
    }
\makeatother

\usepackage[table]{xcolor} 
\usepackage{tabularx}
\usepackage{colortbl}
\usepackage{amsmath}
\usepackage{amssymb}
\usepackage{amsthm}
\usepackage{mathrsfs}
\usepackage[mathscr]{eucal}
\usepackage{bm}
\usepackage{graphicx}
\usepackage[final]{hyperref}
\usepackage[ruled,vlined]{algorithm2e}
\usepackage{enumitem}
\usepackage{array}
\usepackage{multirow}
\usepackage{makecell}
\usepackage[title]{appendix}
\usepackage{empheq}
\usepackage[capitalise, nameinlink, noabbrev]{cleveref}
\usepackage{cases}
\usepackage{textcomp}
\usepackage{gensymb}
\usepackage{caption}
\usepackage{subcaption}
\hypersetup{
	colorlinks=true,       
	linkcolor=darkred,        
	citecolor=blue,        
	filecolor=magenta,     
	urlcolor=blue         
}
\usepackage{newfloat}
\usepackage{multirow}

\usepackage[colorinlistoftodos,prependcaption,textsize=normalsize]{todonotes}
\usepackage{regexpatch}

\usepackage{tcolorbox} 

\usepackage{csvsimple}

\usepackage{longtable} 

\makeatletter
\xpatchcmd{\@todo}{\setkeys{todonotes}{#1}}{\setkeys{todonotes}{inline,#1}}{}{}
\makeatother

\usepackage{etoolbox}
\usepackage{courier}
\usepackage{listings}
\definecolor{codegreen}{rgb}{0,0.6,0}
\definecolor{codegray}{rgb}{0.5,0.5,0.5}
\definecolor{codepurple}{rgb}{0.58,0,0.82}
\definecolor{backcolour}{rgb}{0.96, 0.99, 1}

\lstdefinestyle{mystyle}{
    backgroundcolor=\color{backcolour},
    basicstyle=\ttfamily\footnotesize,
    breakatwhitespace=false,         
    breaklines=true,                 
    captionpos=b,                    
    keepspaces=true,                 
    numbers=none,
    numbersep=5pt,                  
    showspaces=false,                
    showstringspaces=false,
    showtabs=false,                  
    tabsize=2,
    frame=single,
    framesep=6pt
}

\newcommand{\R}{\mathbb{R}}

\newcommand{\Rpow}[1]{\bbR^{#1}}

\newcommand{\norm}[1]{\left\lVert #1 \right\rVert}

\DeclareMathOperator*{\argmin}{arg\,min}

\newtheorem{remark}{Remark}

\newtcolorbox{algobox}[1]{
    colback=black!2,
    colframe=black!50,
    boxrule=0.5pt,
    title={#1}
}

\newcommand{\dd}{\,\textup{d}}

\newcommand{\calF}{{\mathcal{F}}}

\newcommand{\bbC}{{\mathbb{C}}}

\newcommand{\bbE}{{\mathbb{E}}}

\newcommand{\bbP}{{\mathbb{P}}}

\newcommand{\bbR}{{\mathbb{R}}}

\newcommand{\rb}[1]{{\bm{\mathrm{#1}}}}

\newcommand{\rma}{\mathrm{a}}
\newcommand{\rmb}{\mathrm{b}}

\newcommand{\rmg}{\mathrm{g}}

\newcommand{\rmm}{\mathrm{m}}

\newcommand{\rmo}{\mathrm{o}}

\newcommand{\rms}{\mathrm{s}}

\newcommand{\rmu}{\mathrm{u}}
\newcommand{\rmv}{\mathrm{v}}
\newcommand{\rmw}{\mathrm{w}}
\newcommand{\rmx}{\mathrm{x}}

\newcommand{\rmA}{{\mathrm{A}}}
\newcommand{\rmB}{{\mathrm{B}}}
\newcommand{\rmC}{{\mathrm{C}}}
\newcommand{\rmD}{{\mathrm{D}}}

\newcommand{\rmF}{{\mathrm{F}}}

\newcommand{\rmI}{{\mathrm{I}}}

\newcommand{\rmN}{{\mathrm{N}}}

\newcommand{\rmP}{{\mathrm{P}}}

\newcommand{\rmR}{{\mathrm{R}}}

\newcommand{\rmU}{{\mathrm{U}}}
\newcommand{\rmV}{{\mathrm{V}}}

\newcommand{\mscQ}{{\mathscr{Q}}}

\newcommand{\mscS}{{\mathscr{S}}}

\newcommand{\mscZ}{{\mathscr{Z}}}

\newcommand{\dualDot}[2]{\langle {#1}, {#2}\rangle}
\newcommand{\innerProd}[2]{\left( {#1}, {#2}\right)}

\renewcommand{\arraystretch}{1.5}

\crefformat{appendix}{#2#1#3}
\Crefname{lstlisting}{Listing}{Listings}

\newcommand{\ith}[2]{{#1}^{\text{#2}}}

\begin{document}

\begin{frontmatter}
\title{From Theory to Application: A Practical Introduction to Neural Operators in Scientific Computing}
\author{{{Prashant K. Jha$^{1}$}\fnref{fn1}}\\[15pt]
{\large
Published in {\it Mathematics}\,; DOI: \url{https://doi.org/10.3390/math14132421}
}
}

\fntext[fn1]{Department of Mechanical Engineering, South Dakota School of Mines and Technology, Rapid City, SD 57701, USA. \\
Email address: prashant.jha@sdsmt.edu; pjha.sci@gmail.com} 

\begin{abstract}
This review examines neural operator architectures for learning solution operators of parametric partial differential equations (PDEs), with an emphasis on conceptual clarity and practical implementation. The work analyzes key models, including DeepONet, PCANet, and the Fourier Neural Operator, highlighting their underlying representations, computational structures, and comparative performance. These architectures are demonstrated on three canonical PDE problems: the Poisson equation, a linear elasticity problem, and a hyperelasticity problem. To make the presentation self-contained, key foundational topics are introduced, including finite-dimensional representations of function spaces, singular-value decomposition, and sampling from infinite-dimensional function spaces. Beyond forward modeling, the review discusses the use of neural operators as surrogate models within a Bayesian inverse-problem framework, including prior specification, forward-map approximation, and posterior computation. The performance of the three neural-operator architectures is evaluated on in-distribution samples, out-of-distribution samples, and Bayesian inference tasks. The review also discusses challenges related to prediction accuracy and generalization, outlining emerging strategies such as residual-based error correction and multi-level training. The review concludes by positioning neural operators within broader scientific-computing workflows and by identifying directions for reliable, scalable operator learning.
\end{abstract}


\end{frontmatter}

\tableofcontents

\section{Introduction}\label{s:intro}
Neural operators have emerged as powerful tools for approximating solution operators of parametric partial differential equations (PDEs). Their key advantage lies in learning mappings between infinite-dimensional function spaces, enabling efficient surrogate modeling for complex systems. In addition, their ability to provide rapid evaluations makes them particularly attractive for applications such as optimization, control, and inverse problems.

This article provides a practical and hands-on introduction to several neural operator architectures that have become central to scientific computing. The following neural operators are considered:
\begin{enumerate}
    \item Deep Operator Network (DeepONet) \cite{wang2021learning,lu2021learning};
    \item Principal Component Analysis/Proper Orthogonal Decomposition-based Neural Operator (PCANet/PODNet) \cite{BhattacharyaHosseiniKovachkiEtAl2020,FrescaManzoni2022};
    \item Fourier Neural Operator (FNO) \cite{li2020fourier,kovachki2023neural}.
\end{enumerate}

The presentation is designed to be self-contained and implementation-focused. Neural operator architectures, variants, and applications have been surveyed from broader taxonomic and benchmarking perspectives in several recent review articles \cite{kovachki2023neural, azizzadenesheli2024neural, deHoop2022cost}. In contrast, the present article focuses on conceptual understanding, practical implementation of representative neural operator architectures, and the application of neural operator surrogates to Bayesian inverse problems, where these surrogates are used to accelerate repeated forward evaluations in MCMC simulations. To support this objective, several key topics are addressed:
\begin{itemize}
    \item Sampling random functions using Gaussian measures on function spaces;
    \item Construction of data representations tailored to different neural operator architectures;
    \item Algorithms and Python implementations for essential components, including random function sampling and Markov chain Monte Carlo (MCMC) methods for Bayesian inverse problems;
    \item Application of neural operators as surrogate models in Bayesian inverse problems.
\end{itemize}

To illustrate these ideas, two classical linear parametric PDEs and one nonlinear parametric PDE are used as model problems. The first is the Poisson equation for temperature distribution on a rectangular domain, where the input parameter field is the diffusivity. The second is a two-dimensional linear elasticity problem, where the input parameter field is Young's modulus. The third is a hyperelasticity problem that serves as a nonlinear counterpart of the second problem. For these problems, neural operators are constructed to approximate the solution operator, and their accuracy is assessed on random samples of parameter fields.

In addition, neural operators are employed as surrogate models in Bayesian inverse problems, where the underlying parameter fields are inferred from observational data. The resulting posterior estimates are compared against those obtained using the true forward model, demonstrating that neural operators can provide accurate and computationally efficient approximations in this setting. The model problems considered here exhibit rapid decay of singular values in both input and output representations, indicating a low-dimensional structure that is favorable for operator learning. This is reflected in the high predictive accuracy of the neural operators when test samples are drawn from the same distribution as the training data and when the prior for Bayesian inference is well aligned with the training distribution. However, when applying the trained neural operators to out-of-distribution samples, such as those with higher frequencies or larger magnitudes than the training data, significant errors can arise. This is shown through performance assessments on out-of-distribution samples, where the neural operators exhibit large prediction errors.

For complex settings, such as highly nonlinear PDEs or challenging inference and optimization problems, neural operators may exhibit significant errors. This limitation motivates the need for improved strategies for controlling prediction accuracy. These issues are discussed in the concluding section, where existing approaches for error control are surveyed. The discussion also highlights emerging directions, including foundation models for PDEs that aim to construct surrogate operators capable of generalizing across families of PDEs, as well as related frameworks such as multi-task and universal neural operators.

\subsection{Organization of the article}
\begin{itemize}
    \item \cref{s:prelim} introduces notation and key mathematical preliminaries, including finite-dimensional approximations of function spaces, singular value decomposition, and sampling of random functions using Gaussian measures.
    \item \cref{s:modelProblems} presents the parametric PDEs used as model problems.
    \item \cref{s:nn} discusses the architectures and implementations of DeepONet, PCANet, and FNO. 
    \item \cref{s:bayesian} presents the application of neural operators to Bayesian inverse problems, where they are used as surrogate models in MCMC simulations.
    \item \cref{s:nopPerformance} evaluates the performance of the neural operators for in-distribution prediction, out-of-distribution prediction, and Bayesian inference.
    \item \cref{s:conclusion} collects selected recent work in the field, offers a perspective on important open issues, and concludes the article. The subsection on prediction accuracy (\cref{ss:accuracy}) reviews related work and identifies future research directions, while \cref{ss:foundational} discusses emerging approaches based on foundation models for PDEs. Final remarks are provided in \cref{ss:finalThoughts}.
    \item \labelcref{s:appNNPredictions,s:appBayesianInference} present additional results on neural operator predictions and Bayesian inference, respectively.
\end{itemize}

Code and Jupyter notebooks are available at: \url{https://github.com/CEADpx/neural_operators/} (see tag \lstinline{survey26_v2}), with an accompanying Zenodo archive \cite{jha2025neural_operators}. The dataset is shared separately in the Dropbox folder:\\ 
\href{https://www.dropbox.com/scl/fo/co5v2bozvr5y8uv5kc29y/ACKiT1sBBQCTKV2wZYcAIlI?rlkey=agt87l1tf89g967gf8ofe5nik&st=3no4j03k&dl=0}{\lstinline{NeuralOperator_Survey_Shared_Data_June2026}}.

\section{Preliminaries}\label{s:prelim}

This section establishes the notation and core mathematical framework used throughout the article. It introduces finite-dimensional representations of function spaces, sampling strategies for random functions, and related concepts that are central to the formulation and implementation of neural operators.

\subsection{Notations}\label{ss:notat}
Let $\mathbb{N}, \mathbb{Z}, \R$ denote the spaces of natural numbers, integers, and real numbers, respectively, and let $\R^{+}$ denote the space of all nonnegative real numbers. $\Rpow{n}$ denotes the $n$-dimensional Euclidean space. The space of square-integrable functions $f: D \subset \Rpow{q} \to \Rpow{d}$ is denoted by $L^2(D; \Rpow{d})$. $H^s(D; \Rpow{d})$ denotes the space of functions in $L^2(D; \Rpow{d})$ whose weak derivatives up to order $s$ belong to $L^2(D; \Rpow{d})$. $L(M; U)$ denotes the space of continuous linear maps from $M$ to $U$, and $C^1(A; U)$ denotes the space of continuously differentiable maps from $A \subset M$ to $U$. Given a generic complete normed (function) Banach space $A$, $\norm{\cdot}$ denotes the norm, and, if it is a Hilbert space, $\innerProd{u}{v}$ denotes the inner product for $u,v \in A$. $A^\ast$ denotes the dual of the Banach space $A$, and $\dualDot{b}{a}$ denotes the duality pairing between $b \in A^\ast$ and $a \in A$.

Throughout the text, $m \in M$ will denote the input or parameter field in the parametric boundary value problem, $M$ the appropriate Banach function space for input fields, and $u\in U$ the solution field, where $U$ is the Banach space for solutions of the PDE. The solution operator or forward operator mapping the parameter field $m\in M$ to the PDE solution $u \in U$ is denoted by $F(m)$. The neural operator approximation of the forward operator is denoted by $F_{NOp}(m)$. The finite-dimensional approximations of functions $m\in M$ and $u \in U$ are denoted by the Roman letters $\rmm$ and $\rmu$; more generally, for any $a\in A$, the approximation is denoted by $\rma$. The forward map between finite-dimensional approximations is denoted by $\rmF$. The projections of $\rmm$ and $\rmu$ onto lower-dimensional subspaces are denoted by $\tilde{\rmm}$ and $\tilde{\rmu}$, respectively. The mapping between lower-dimensional subspaces will be denoted by placing a tilde over the corresponding symbol, e.g., $\tilde{\rmF}$.

For vectors $\rma,\rmb\in\bbR^d$, $\rma\cdot \rmb=\sum_{i=1}^d \rma_i \rmb_i$ denotes the dot product. For second-order tensors $\rmA,\rmB\in\bbR^{d\times d}$, $\mathrm{tr}(\rmA)=\sum_{i=1}^d \rmA_{ii}$ is the trace, $\rmA^T$ is the transpose, $\rmA:\rmB:=\mathrm{tr}(\rmA^T\rmB)=\sum_{i,j}\rmA_{ij}\rmB_{ij}$ is the tensor inner product, and $\mathrm{det}(\rmA)$ is the determinant of $\rmA$.

The key notations used throughout the text are collected in \cref{tab:notat} below.
\begin{center}
\begin{longtable}{|p{0.32\linewidth}|p{0.62\linewidth}|}
    \hline
    {\bf Symbol} & {\bf Description} \\
    \hline
    $m$ & Typical parameter field in the parametric PDEs. This is also the input function to the neural operator \\
    $u$ & Typical solution of the PDE given $m$. This is also the output of the neural operator \\
    $D_{A}$ & Domain of a generic function $a \in A$ \\
    $\partial D, \Gamma_{b}$ & Domain boundary and subset of $\partial D$ tagged by $b$, respectively \\
    $q_{A}$ & Dimension of the domain of function $a$ \\
    $d_{A}$ & Dimension of the pointwise values of the function $a$ \\
    $x\in D_{M}$ & A point in the domain of input/parameter functions \\
    $y\in D_{U}$ & A point in the domain of output/solution functions \\
    $M \subset \{m: D_{M} \subseteq \Rpow{q_{M}} \to \Rpow{d_{M}}\}$ & Function space of the parameters in the parametric PDEs \\
    $U \subset \{u: D_{U} \subseteq \Rpow{q_{U}} \to \Rpow{d_{U}}\}$ & Function space of solutions of the PDE \\
    $F: M \to U$ & Forward solution operator, which is also the target operator of neural operators \\
    $\phi^A = \{\phi^A_{i}\}_{i=1}^{\infty} \subset A$ & Basis functions for the function space $A$; an element $a \in A$ has a representation $a = \sum_i \rma_i \phi^A_{i}$, where $\{\rma_i\}$ are the coefficients \\ 
    $\rma \in \Rpow{p_{A}}$ & Coefficient vector for the finite-dimensional representation of the function $a$ \\
    $p_{A}$ & Dimension of the finite-dimensional representation of $a\in A$ \\
    $\rma_i$ & $\ith{i}{th}$ component of a vector $\rma$ \\
    $\rmF : \Rpow{p_{M}} \to \Rpow{p_{U}}$ & Finite-dimensional approximation of the operator $F: M \to U$ \\
    $r_{A}$ & Dimension of the reduced representation of $\rma \in \Rpow{p_{A}}$, $r_{A} < p_{A}$ \\
    $\tilde{\rma} \in \Rpow{r_{A}}$ & The reduced-dimensional representation of $\rma \in \Rpow{p_{A}}$ \\
    $\tilde{\rmP}_{A} : \Rpow{p_{A}} \to \Rpow{r_{A}}$ & Projection operator that takes $\rma \in \Rpow{p_{A}}$ to the reduced space $\tilde{\rma} = \tilde{\rmP}_{A}(\rma) \in \Rpow{r_{A}}$ \\
    $\tilde{\rmF} : \Rpow{r_{M}} \to \Rpow{r_{U}}$ & Reduced-order approximation of $\rmF$ mapping between reduced-dimensional (latent) spaces of input and output functions \\
    $N(\bar{w}, C)$ & Gaussian random field with mean $\bar{w} \in W$ and covariance operator $C: W \to W$ \\
    $\rmN(\rmw, \rmC)$ & Gaussian random field in the finite-dimensional space $\Rpow{p_{W}}$, where $\rmw \in \Rpow{p_{W}}$ and $\rmC \in \Rpow{p_{W} \times p_{W}}$ \\
    $F_{NOp}: M \times \Rpow{p_\Theta} \to U$ & Neural operator approximation of $F: M \to U$; $p_\Theta$
    is the number of trainable parameters in the neural operator \\
    $\rmF_{NOp} : \Rpow{p_{M}} \times \Rpow{p_\Theta} \to \Rpow{p_{U}}$ & Neural operator approximation of finite-dimensional representation of the map $\rmF: \Rpow{p_{M}} \to \Rpow{p_{U}}$ \\
    $N$ & Number of samples used for training and testing neural operators \\
    $m^I, u^I, \rmm^I, \rmu^I$ & $\ith{I}{th}$ input-output sample in function and finite-dimensional spaces \\
    $\rb{X}^I, \rb{Y}^I$ & $\ith{I}{th}$ sample, $1\leq I \leq N$, of the data from $\rb{X}, \rb{Y}$, respectively \\
    $\rmg \in \Rpow{d_{\rmg}}$ & Observed data, where $d_{\rmg}$ is the dimension of the data \\
    $\bar{\rb{B}} : U \to \Rpow{d_{\rmg}}$ & State-to-observable map, which maps the output function $u$ to the observed data $\rmg$ \\
    $\rb{B}: W \to \Rpow{d_{\rmg}}$ & Parameter-to-observable map, which maps the parameter field to be inferred, $w\in W$, to the predicted data $\rmg = \rb{B}(w)$ \\
    $\eta \sim \rmN({\rm{0}}, {\Gamma}_{\rmg})$ & Gaussian noise with zero mean and covariance $\Gamma_{\rmg}$. Example: $\Gamma_{\rmg}=\sigma_{\rmg}^2\rmI$, where $\sigma_{\rmg}^2$ is the variance and $\rmI$ is the identity matrix \\
    $\Phi(w)$, $\pi_{like}(\rmg \mid w)$ & Likelihood potential of $w$ and the likelihood of $\rmg$ given $w$ \\
    $\mu^{\rmo}$, $\mu^{\rmg}$ & Prior and posterior measures on $W$, respectively \\
    \hline
    \caption{Key notations.}\label{tab:notat}
\end{longtable}
\end{center}

\vspace{-20pt} 
\subsection{Series representation of functions and finite-dimensional approximation}
One of the central ideas that many neural operators leverage is the finite-dimensional representation of functions in terms of coefficients and basis functions in their respective spaces. Following the notations in the previous section and \cref{tab:notat}, suppose that $M$ and $U$ are Hilbert spaces and, therefore, have orthonormal bases $\phi^M = \{\phi^M_i\}_{i=1}^\infty \subset M$ and $\phi^U = \{\phi^U_i\}_{i=1}^\infty \subset U$. Then a function $m \in M$ can be represented as
\begin{equation}
    m(x) = \sum_{i=1}^\infty \rmm_i \phi^M_i(x)\,, \qquad \forall x \in D_{M}\,,
\end{equation}
where $\rmm_i = \innerProd{m}{\phi^M_i} \in \R$ (inner product between $m$ and $\phi^M_i$) are the coefficients or degrees of freedom associated with the $\ith{i}{th}$ mode. Note that for each $i$, $\phi^M_i(x) \in \Rpow{d_{M}}$ for $x \in D_{M}$, and $\phi^U_i(y) \in \Rpow{d_{U}}$ for $y\in D_{U}$. 

It is useful to consider the example where the domain is the unit interval $D_{M} = (0,1)$ and $m \in L^2(D_{M}; \R)$. In this case, one can write $m(x) = \sum_{i=1}^\infty \rmm_i \phi^M_i(x)$, where the basis functions take the form
\begin{equation}
    \begin{split}
    &\phi^M_1 = 1\,, \quad 
    \phi^M_2 = \sqrt{2}\cos\left(2\pi x\right)\,, \quad 
    \phi^M_3 = \sqrt{2}\sin\left(2\pi x\right)\,, \cdots, \\
    &\qquad \phi^M_{2j} = \sqrt{2}\cos\left(2 j\pi x\right)\,, \quad 
    \phi^M_{2j+1} = \sqrt{2}\sin\left(2 j\pi x\right)\,, \cdots
    \end{split}
\end{equation}
and the coefficients are given by
\begin{equation}
    \rmm_i = \innerProd{m}{\phi^M_i} = \int_0^1 m(x) \phi^M_i(x) \dd x\,, \qquad \forall i\,.
\end{equation}

Focusing on the abstract setting, let $\phi^M = \{\phi^M_i\}_{i=1}^{p_{M}}$, where $p_{M}$ is a finite integer and $\phi^M$ is a finite collection of basis functions, and let $\{\rmm_i\}_{i=1}^{p_{M}}$ be the corresponding coefficients. Then, the finite-dimensional approximation is $\sum_{i=1}^{p_{M}} \rmm_i \phi^M_i(x) \, \,\approx m(x)$ with the error given by $\norm{m - \left(\sum_{i=1}^{p_{M}} \rmm_i \phi^M_i\right)}$. Similarly, the function $u\in U$ can be approximated as $u(y) \approx \sum_{i=1}^{p_{U}} \rmu_i \phi^U_i(y)$, $y \in D_{U}$. Neural operators often leverage this finite-dimensional representation viewpoint, where the goal is to determine, or learn, the basis functions $\phi^U = \{\phi^U_i\}_{i=1}^{p_{U}}$ (or their pointwise values $\{\phi^U_i(y)\}_{i=1}^{p_{U}}$ for $y\in D_{U}$ [e.g., in DeepONet]) and the coefficients $\rmu = \{\rmu_i\}_{i=1}^{p_{U}}$ such that $\sum_{i=1}^{p_{U}} \rmu_i \phi^U_i$ provides the ``best'' approximation of $u = F(m)$, where $m\in M$ is the input function to the operator.

\begin{remark}
    PCANet explicitly constructs reduced-dimensional representations of the input and output functions via singular value decomposition and learns the map between them (see \cref{ss:dimreduction}). FNO follows a different construction, utilizing Fourier transforms to represent functions in the frequency domain and learning mappings through spectral convolutions applied to the Fourier modes.
\end{remark}

\subsubsection{Finite element approximation}\label{ss:fea}
For a general class of function spaces $M \subseteq \{m: D_{M} \subseteq \Rpow{q_{M}} \to \Rpow{d_{M}}\}$ and spatial domain $D_{M}$, the preceding construction of finite-dimensional approximations can be restrictive. A more straightforward numerical way to obtain finite-dimensional approximations is to use numerical techniques such as finite difference and finite element methods. In this work, the finite element method is used (e.g., to generate samples of input functions using Gaussian priors and solve PDE-based problems). 

To be more precise, consider a finite element discretization $D_{M_h}$ of the domain $D_{M}$ consisting of simplex elements $\{T_e\}_{e=1}^{N_e}$, so that $D_{M_h} = \cup_{e} \bar{T}_e \approx D_{M}$. Suppose $\phi^M_i : D_{M_h} \to \Rpow{d_{M}}$ denotes the basis function associated with the $i^{\text{th}}$ scalar degree of freedom and $\rmm \in \Rpow{p_{M}}$ the vector of coefficients associated with the basis functions $\{\phi^M_i\}_{i=1}^{p_{M}}$. Then, the finite-dimensional approximation of $m$ is given by $m_h(x) = \sum_{i=1}^{p_{M}} \rmm_i \phi^M_i(x)$ for $x\in D_{M_h}$. The same procedure can be used to construct the finite-dimensional approximation of $u$ using the basis functions $\{\phi^U_i\}_{i=1}^{p_{U}}$ and coefficients $\rmu \in \Rpow{p_{U}}$. 

\begin{remark}
    When $m$ is vector-valued, that is, $m(x) \in \R^{d_{M}}$ with $d_{M} > 1$, the finite element approximation must represent each component of $m$ independently while using a common spatial discretization. This leads naturally to vector-valued basis functions.

    Let $\{\psi_j : D_{M_h} \to \R\}_{j=1}^n$ denote the standard scalar Lagrange basis functions associated with the $n$ mesh vertices. Since each vertex carries $d_{M}$ degrees of freedom, the total number of degrees of freedom is $p_{M} = n \, d_{M}$.

    The vector-valued basis functions are constructed by combining the scalar spatial basis with the canonical basis of $\R^{d_{M}}$. Specifically, for each degree of freedom index $i$ with $1 \leq i \leq p_{M}$, there exist unique indices $j \in \{1,\dots,n\}$ and $k \in \{1,\dots,d_{M}\}$ such that
    \begin{equation*}
    i = (j-1)d_{M} + k,
    \end{equation*}
    where $j$ identifies the vertex in the mesh and $k$ identifies the component. The corresponding basis function associated with this degree of freedom is defined as
    \begin{equation*}
    \phi^M_i(x) = \psi_j(x)\, \mathbf{e}_k, \quad x \in D_{M_h},
    \end{equation*}
    where $\mathbf{e}_k$ is the $k^{\text{th}}$ canonical basis vector in $\R^{d_{M}}$, i.e., $\mathbf{e}_k = (0,\ldots,1,\ldots,0)^\top$ with the $1$ in the $\ith{k}{th}$ position.
\end{remark}

Let $V_{M_h} = \mathrm{span}\{\phi^M_i\}_{i=1}^{p_{M}}$, where $p_{M}$ is the number of degrees of freedom. Then, the function $m\in M$ can be approximated by a function $m_h \in V_{M_h}$ given by
\begin{equation}
    m(x) \qquad \approx \qquad m_h(x) = \sum_{i=1}^{p_{M}} \rmm_i \phi^M_i(x)\,, \qquad \forall x\in D_{M_h}\,,
\end{equation}
provided the coefficients $\rmm \in \Rpow{p_{M}}$ are chosen appropriately. For example, $\rmm$ is selected such that it minimizes the $L^2$-error $\norm{m - m_h}_{L^2}$. 

Given finite-dimensional approximations $m_h$ and $u_h$ of $m$ and $u$, respectively, the map $F(m) = u$ is also approximated by $F_h(m_h) = u_h$ in the sense that $F_h$ takes $m_h$ and returns the output $u_h$ such that the operator error
\begin{equation*}
    \norm{F(m) - F_h(m_h)}
\end{equation*}
is small for the relevant class of inputs $m$. 

Finally, note that, for a fixed mesh and the basis functions $\phi^M_i$, it is easy to see that if $\rmm \in \Rpow{p_{M}}$ is fixed, then the function $m_h$ is completely characterized. If $m_h$ is fixed, using the unique representation of $m_h$, the coefficients $\rmm$ are completely characterized. Thus, $V_{M_h}$ can be identified using $\Rpow{p_{M}}$ (and vice versa). This makes it possible to represent the finite-dimensional function space $V_{M_h}$ by the Euclidean space $\Rpow{p_{M}}$ of the coefficients. Throughout the article, functions $m$ and $u$ will be represented by the finite-dimensional approximations $\rmm$ and $\rmu$, where the corresponding functional representations, $m_h = \sum_{i=1}^{p_{M}} \rmm_i \phi^M_i$ and $u_h = \sum_{i=1}^{p_{U}} \rmu_i \phi^U_i$, are assumed implicitly. In the same spirit, using $u_h = F_h(m_h)$, the operator $F_h$ can be identified with a map $\rmF: \Rpow{p_{M}} \to \Rpow{p_{U}}$:
\begin{equation}\label{eq:finiteDimFfromOpF}
    \rmF(\rmm) = \rmu \qquad \Rightarrow \qquad F_h(m_h) = u_h = \sum_{i=1}^{p_{U}} \rmu_i \phi^U_i \quad \text{with} \quad m_h = \sum_{i=1}^{p_{M}} \rmm_i \phi^M_i\,.
\end{equation}
Here, $\rmF$ maps the coefficient vector $\rmm$ to $\rmu$ and is induced by the map $F_h$.

\subsection{Dimensional reduction and singular-value decomposition (SVD)}\label{ss:dimreduction}
While the theory is based on functions defined on a continuum domain, computer implementations introduce discretizations of the domain and, consequently, discrete approximations of functions. For example, the training data for neural operators is typically a collection of pairs $(\rmm^I, \rmu^I)$, where $\rmm^I \in \Rpow{p_{M}}$ and $\rmu^I = \rmF(\rmm^I) \in \Rpow{p_{U}}$ are discrete approximations of functions in $M$ and $U$, where $\rmF$ is the finite-dimensional mapping between $\Rpow{p_{M}}$ and $\Rpow{p_{U}}$ that approximates the operator of interest $F$. Generally speaking, the dimensions of the discretized input and target functions, $p_{M}$ and $p_{U}$, are large, and the problem of approximating the map $\rmF$ between high-dimensional spaces becomes challenging and potentially ill-posed.

The second key idea used in various neural operators, after the linear basis representation discussed earlier, is to reduce the dimensions of discretized input and output functions; see \cite{BhattacharyaHosseiniKovachkiEtAl2020}. If $\rmm \in \Rpow{p_{M}}$ and $\rmu \in \Rpow{p_{U}}$, and the goal is to determine a map $\rmm \mapsto \rmu = \rmF(\rmm)$ from the data $\{(\rmm^I, \rmu^I)\}_{I=1}^N$, then, as an alternative to learning or approximating the map $\rmF$, one could attempt to characterize the map $\tilde{\rmF}$, where
\begin{equation}
    \rmm \qquad \mapsto \qquad \rmu = \tilde{\rmP}_{U}^T\left(\tilde{\rmF} \left( \tilde{\rmP}_{M} (\rmm) \right) \right)\,.
\end{equation}
Here, $\tilde{\rmP}_{M} \in \Rpow{r_{M}\times p_{M}}$ is the projection operator that projects $\rmm \in \Rpow{p_{M}}$ into a low-dimensional subspace, $\tilde{\rmm} := \tilde{\rmP}_{M}(\rmm) \in \Rpow{r_{M}}$ (with $r_{M} \ll p_{M}$). $\tilde{\rmP}_{U} \in \Rpow{r_{U} \times p_{U}}$ has the same role as $\tilde{\rmP}_{M}$ but for target functions $\rmu\in \Rpow{p_{U}}$. The transpose of $\tilde{\rmP}_{U}$, $\tilde{\rmP}_{U}^T$, projects an element in $\Rpow{r_{U}}$ to $\Rpow{p_{U}}$. Note that $\tilde{\rmF}: \Rpow{r_{M}} \to \Rpow{r_{U}}$, which needs to be learned, is the mapping between two low-dimensional spaces, and hence identifying $\tilde{\rmF}$ is less daunting than identifying $\rmF$. In summary, using $\tilde{\rmP}_{M}$ and $\tilde{\rmP}_{U}$, the dimensions of the operator inference problem are significantly reduced, and, by controlling $r_{M}$ and $r_{U}$, one can balance the trade-off between accuracy and computational cost. 

\subsubsection{Projectors via SVD}\label{sss:svd}
The projectors $\tilde{\rmP}_{M}$ and $\tilde{\rmP}_{U}$ used for dimensional reduction can be obtained via singular-value decomposition (SVD). Focusing on the input space $\Rpow{p_{M}}$, let $\rmR$ denote a $p_{M} \times N$ matrix such that, for $1\leq I\leq N$, $\rmm^I$ makes up the $\ith{I}{th}$ column of the matrix $\rmR$. 
Here, $N$ may be smaller or larger than $p_{M}$, and since the samples may be linearly dependent, the rank of $\rmR$ need not be equal to $\min\{p_{M},N\}$. Let
\begin{equation}
    r = \mathrm{rank}(\rmR) \leq \min \{p_{M}, N\}\,.
\end{equation}
Consider the compact SVD
\begin{equation}
    \rmR = \rmU \rmD \rmV^T\,,
\end{equation}
where $\rmU\in\Rpow{p_{M}\times r}$ and $\rmV\in\Rpow{N\times r}$ have orthonormal columns, and $\rmD\in\Rpow{r\times r}$ is diagonal with positive singular values. Denote the columns of $\rmU$ by $\{\rmw^I\}_{I=1}^{r}$, i.e.,
\begin{equation}
    \rmU =
    \begin{bmatrix}
        \mid & \mid & & \mid \\
        \rmw^1 & \rmw^2 & \cdots & \rmw^{r} \\
        \mid & \mid & & \mid
    \end{bmatrix}.
\end{equation}
The vectors $\{\rmw^I\}_{I=1}^{r}$ form an orthonormal basis for the column space of $\rmR$, which is an $r$-dimensional subspace of $\Rpow{p_{M}}$.

Let $r_{M}>0$ be the prescribed reduced dimension, with $r_{M}\leq r$. Given $r_{M}$, define $\rmU_{r_{M}}\in\Rpow{p_{M}\times r_{M}}$ by retaining the first $r_{M}$ columns of $\rmU$:
\begin{equation}
	\rmU_{r_{M}} = \begin{bmatrix}
		\mid & \mid & & \mid \\
		\rmw^1 & \rmw^2 & \cdots & \rmw^{r_{M}} \\
		\mid & \mid & & \mid
	\end{bmatrix}.
\end{equation}
Noting the properties of $\rmU_{r_{M}}$ (e.g., see [\citealp{jha2024residual}](Section 3.2.1)), $\rmU^T_{r_{M}}$ is taken as the projector, i.e.,
\begin{equation}
    \tilde{\rmP}_{M} := \rmU^T_{r_{M}}.
\end{equation}

\begin{remark} 
    The rank of $\rmR$ depends on the linear independence of the sampled coefficient vectors $\{\rmm^I\}_{I=1}^N$. If these vectors are linearly dependent, then $\mathrm{rank}(\rmR)<\min\{p_{M},N\}$. In contrast, if the samples are drawn independently from a distribution on $\Rpow{p_{M}}$ that is not supported on a proper linear subspace, then $\mathrm{rank}(\rmR)=\min\{p_{M},N\}$. Numerically, the effective rank is determined by the decay of the singular values and the tolerance used to distinguish nonzero singular values. 
\end{remark}

For $\rmu$, the projector $\tilde{\rmP}_{U}$ is obtained following the same procedure as above using a matrix $\rmR$ of size $p_{U} \times N$ with $\rmu^I$ making its $I^{\text{th}}$ column. The left singular-vector matrix from the SVD of $\rmR$, denoted by $\rmU$, is truncated by retaining the first $r_{U}$ columns of $\rmU$. If the truncated matrix is $\rmU_{r_{U}}$, then $\tilde{\rmP}_{U} := \rmU^T_{r_{U}}$. 

\begin{figure}[h]
    \centering
    \vspace{-10pt} 
    \includegraphics[width=0.55\linewidth]{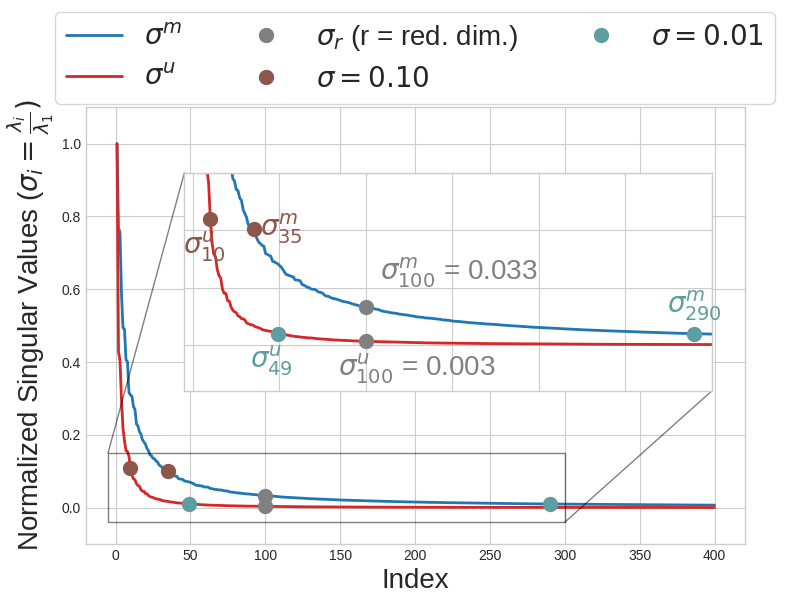}
    \vspace{-5pt} 
    \caption{Normalized singular values of the centered and normalized input and output data $\{\rmm^I = \exp(\rmw^I) \mid \rmw^I \sim \rmN({\rm{0}}, \rmC)\}$ and $\{\rmu^I = \rmF(\rmm^I)\}$, where $\rmN({\rm{0}}, \rmC)$ is the $p_{M}$-dimensional Gaussian distribution obtained from the finite element discretization of of $N(0,C)$ with $C=L_\Delta^{-2}$; see \cref{sss:poissonSetup} for details. Here, $\rmm^I$ denotes a discretized input and $\rmF$ denotes the discretized solution operator associated with the PDE. $\sigma^a$, $a\in\{m,u\}$, represents the normalized singular values. Small dots indicate the modes at which the normalized singular value is $0.1$ or $0.01$, as well as at specified reduced dimension $r = 100$ ($\ith{100}{th}$ mode).}
    \label{fig:svdSec2}
\end{figure}

In \cref{fig:svdSec2}, the normalized singular values of representative centered and normalized input and output data are shown. The grey, brown, and cadet blue dots on each curve represent the singular value at mode $100$, and the mode with a normalized singular value of $0.1$ and $0.01$. Based on the plot, projecting $u$ onto SVD subspaces of dimensions $r_{U}=10$, $49$, and $100$ gives approximate projection errors of $10\%$, $1\%$, and $0.3\%$, respectively, relative to the largest singular value. Similarly, for $m$, projection into reduced dimension $r_{M} = 35$, $100$, and $290$ will result in average projection errors of approximately $10\%$, $3.3\%$, and $1\%$, respectively. 

\vspace{-25pt} 
\subsection{Sampling functions as infinite-dimensional random variables}\label{ss:sampling}

The final topic in this section is the sampling of random parameters, which are functions. Consider a probability space $(\Omega, \calF, \bbP)$, where $\Omega$ is a sample space and $\calF$ is a $\sigma$-algebra on which the probability measure $\bbP$ is defined with $\bbP(\Omega) = 1$. The goal is to draw $W$-valued random fields, where $W$ is assumed to be a separable Hilbert space. Equivalently, the goal is to define a random field $\mscZ: \Omega \to W$ such that, for a given $z \in \Omega$, $w = \mscZ(z) \in W$. Suppose such a $\mscZ$ is available; then, the probability that $\mscZ$ takes values in a subset $A\subseteq W$ defines the pushforward measure $\mu_{\mscZ}$, given by
\begin{equation}
    \mu_{\mscZ}(A) = \text{probability of } \mscZ \in A = \bbP\left(\{z\in \Omega: \mscZ(z) \in A\}\right) = \bbP\left( \mscZ^{-1}(A) \right)\,,
\end{equation}
where $\mscZ^{-1}(A) \in \calF$ is assumed to be measurable in the probability space $(\Omega, \calF, \bbP)$. Thus, $\mu_{\mscZ}$ is a measure on $W$ induced by the random field $\mscZ$. Now, suppose that a random field $\mscZ$ is such that the measure $\mu_\mscZ$ is Gaussian in the sense of \cite{Dashti2017,mandel2023introduction}; specifically, see \cite[Definition 6 and Lemma 23]{Dashti2017}. In this case, $\mu_\mscZ$ is written as $\mu_\mscZ = N(\bar{w}, C)$. Here, $\bar{w}\in W$ is the mean function, and $C: W \to W$ is called the covariance operator. At the outset, $C$ is assumed to be a trace-class operator ([\citealp[Lemma 23]{Dashti2017}] and [\citealp[Theorem 7]{mandel2023introduction}]).

Our next goal is to consider specific examples of $\mscZ$ such that $\mu_\mscZ$ is Gaussian, as mentioned above, and understand how random function samples are generated. In this direction, it is useful to highlight the role of $C$ in generating samples $w = \mscZ(z) \in W$; this helps explain why commonly used forms of $C$ are effective. Consider another random field $\mscS: \Omega \to W$ such that $\mu_\mscS = N(0, 1)$, where $0 \in W$ is the mean function and $1: W\to W$ is the identity covariance operator. Given a function $\bar{w} \in W$ and a covariance operator $C: W \to W$, let $\mscZ$ be such that $\mu_\mscZ = N(\bar{w}, C)$. Then,
a sample $w = \mscZ(z)$ can be obtained by transforming the sample $s = \mscS(z)$ as follows:
\begin{equation}\label{eq:sampleZ}
    \mscZ(z) = w := \bar{w} + C^{1/2} s\,.
\end{equation}
Thus, $C^{1/2}$, the square root of the covariance operator, plays a key role in generating random functions. It is therefore useful to choose $C$ so that $C^{1/2}$ is computationally tractable while ensuring that the $W$-valued samples are well defined, have the desired regularity, and exhibit controlled correlation between pointwise values.

\subsubsection{Gaussian measures based on Laplacian-like operators}\label{sss:gaussian}
Following \cite{bui2013computational}, let $C = L_\Delta^{-2}$, where $L_\Delta: W_{L_\Delta} \subseteq W \to W$ is a Laplacian-like operator given by
\begin{equation}\label{eq:LDelta}
    L_\Delta := \begin{cases}
        -\mathsf{a}_c \nabla \cdot \mathsf{b}_c \nabla + \mathsf{c}_c\,, \qquad\qquad  &  \text{in } D_{W}\,,\\
        \mathsf{a}_c n \cdot \mathsf{b}_c \nabla\,, \qquad \qquad & \text{on }\partial D_{W}\,.
    \end{cases}
\end{equation}
Here, $D_{W}$ is the domain of functions $w$, and $\mathsf{a}_c, \mathsf{b}_c, \mathsf{c}_c$ are parameters in the operator (they could vary over the domain or be taken as constants). In this work, $\mathsf{a}_c$ and $\mathsf{c}_c$ will be constant, and in some situations, $\mathsf{b}_c$ will be considered to be a spatially varying scalar-valued $L^2(D_{W})$ function. In the literature, it is also common to take $\mathsf{b}_c$ as $\Rpow{q_{W} \times d_{W}}\times \Rpow{q_{W}\times d_{W}}$-valued function allowing one to encode anisotropy and inhomogeneous behavior in the prior. In the above, $W_{L_\Delta}\subseteq W$ is the domain of the operator $L_\Delta$ such that $L_\Delta(w)$ is well-defined for all $w \in W_{L_\Delta}$. The natural choice is $W_{L_\Delta} = \{w\in W: \norm{w}_{H^2} < \infty \}$. If $L_\Delta$ is defined through the bilinear form
\begin{equation}
    \innerProd{v}{L_\Delta w}
    =
    \int_{D_{W}}
    \left[
        \mathsf{a}_c \mathsf{b}_c\nabla w \cdot \nabla v
        +
        \mathsf{c}_c w v
    \right] \dd x\,,
\end{equation}
then, the operator domain can be taken as
\begin{equation}
    \quad W_{L_\Delta} = W \cap H^1(D_{W}; \Rpow{d_{W}})\,,
\end{equation}
and take $v, w \in W_{L_\Delta}$. 
In the bilinear form above, note that the boundary term in \eqref{eq:LDelta} cancels out the boundary term after integration by parts.

With this specific form of $C$, the sampling of functions $w\in W$ can be summarized as follows:

\begin{algobox}{Sampling a Gaussian random field with covariance operator $C=L_\Delta^{-2}$}
\begin{enumerate}
    \item Fix $\bar{w}\in W$, $C = L_\Delta^{-2}$, and let $\mscS: \Omega \to W$ be a random field such that $\mu_\mscS = N(0,1)$.
    \item Draw a sample $s = \mscS(z)$, $z\in \Omega$.
    \item Generate a new sample $w\sim N(\bar{w}, C)$ as follows:
        \begin{align}
            & w = \bar{w} + C^{1/2}(s) \notag\\
            \Rightarrow \; & w = \bar{w} + L_{\Delta}^{-1}(s) \notag\\
            \Rightarrow \; & w = \bar{w} + v\,, \text{ where }v \text{ satisfies } s = L_\Delta (v) \notag\\
            \Rightarrow \; & w = \bar{w} + v\,, \text{ where }v \text{ satisfies } \innerProd{g}{s} = \innerProd{g}{L_\Delta (v)}\,, \quad \forall g \in W_{L_\Delta}\,, \notag\\
            \Rightarrow \; & w = \bar{w} + v\,, \text{ where }v \text{ satisfies } \notag\\
            & \qquad \qquad \int_{D_{W}} sg \dd x = \int_{D_{W}} \left[\mathsf{a}_c \mathsf{b}_c \nabla v \cdot \nabla g + \mathsf{c}_c v g \right] \dd x\,, \quad \forall g \in W_{L_\Delta}\,,
        \end{align}
        where the last two equations express the weak-form definition of $L_\Delta$ that can be solved using the finite element method. 
\end{enumerate}
\end{algobox}

In terms of the numerical implementation, consider a finite element mesh $D_{W_h}$ and a finite element function space $V_h \subset W$ with basis functions $\phi^W = \{\phi^W_i\}_{i=1}^{p_{W}}$. Let the mean function $\bar{w} \in W$ be projected onto $V_h$, with projection given by $\bar{w}_h = \sum_i \bar{\rmw}_{h_i} \phi^W_i$. The procedure to sample $w_h \sim N(\bar{w}_h, C)$ is as follows:

\vspace{5pt} 
\begin{algobox}{Sampling a Gaussian random field with $C=L_\Delta^{-2}$ in a finite element setting}
\begin{enumerate}
    \item To draw a sample $s = \mscS(z) \in V_h$, consider the following steps:
    \begin{enumerate}
        \item For each $i$, $1\leq i \leq p_{W}$, draw a number $\rms_i \sim \rmN(0, 1)$, where $\rmN(0,1)$ is the standard normal distribution on $\R$; and
        \item Take $s = \sum_{i} \rms_i \phi^W_i$, i.e., $s$ is a finite element function with coefficients $\rms$.
    \end{enumerate}
    \item Solve for $v_h\in V_h$ such that
    \begin{equation}
        \int_{D_{W_h}} sg\,\dd x
        =
        \int_{D_{W_h}}
        \left[
            \mathsf a_c \mathsf b_c \nabla v_h\cdot\nabla g
            +
            \mathsf c_c v_h g
        \right]\,\dd x,
        \qquad
        \forall g\in V_h .
    \end{equation}
    This yields coefficients $\rmv_h\in\R^{p_W}$ and the finite element function $v_h=\sum_i \rmv_{h_i}\phi_i^W$.
    \item A new sample $w_h\sim N(\bar{w}_h, C)$ is given by
    \begin{equation}
        w_h = \sum_i \rmw_{h_i} \phi^W_i = \bar{w}_h + v_h, \qquad \text{where} \qquad \rmw_{h_i} = \bar{\rmw}_{h_i} + \rmv_{h_i}.
    \end{equation}
\end{enumerate}
\end{algobox}
\vspace{5pt} 

\vspace{10pt} 
\begin{remark}
Spectral methods, including Karhunen--Lo\`eve expansions, can also be used to sample Gaussian random fields with covariance operators defined through Laplacian-like operators. These methods are often efficient on simple domains with compatible boundary conditions, especially when the eigenpairs of the covariance operator or the associated differential operator are available. The finite element construction is used here because it is straightforward and directly compatible with the PDE discretizations used later for data generation and Bayesian inference. It also extends naturally to general geometries and boundary conditions, unstructured and higher-order meshes, spatially varying coefficients such as $\mathsf{b}_c$, and vector-valued finite element spaces. 
\end{remark}
\vspace{10pt} 

\begin{figure}[h]
\centering
\subfloat[Diffusivity $\mathsf{b}_c$ in $L_\Delta$\label{fig:subSampling1}]{
  \begin{minipage}{.22\textwidth}
  \centering
  \includegraphics[width=.98\linewidth]{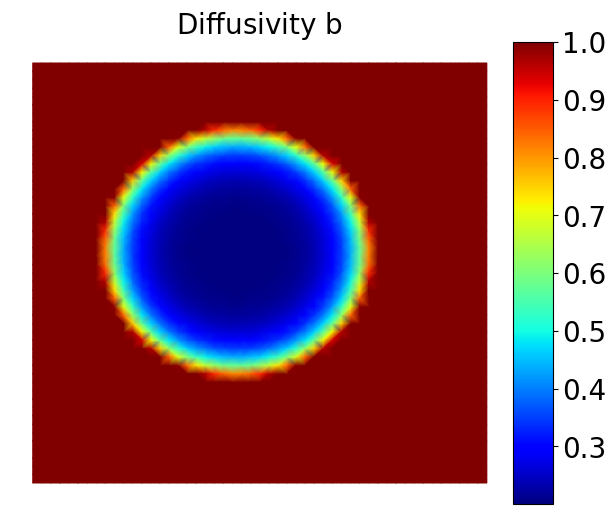}\\[50pt]
  \end{minipage}
}%
\subfloat[Random samples\label{fig:subSampling2}]{
  \begin{minipage}{.78\textwidth}
  \centering
  \includegraphics[width=.98\linewidth]{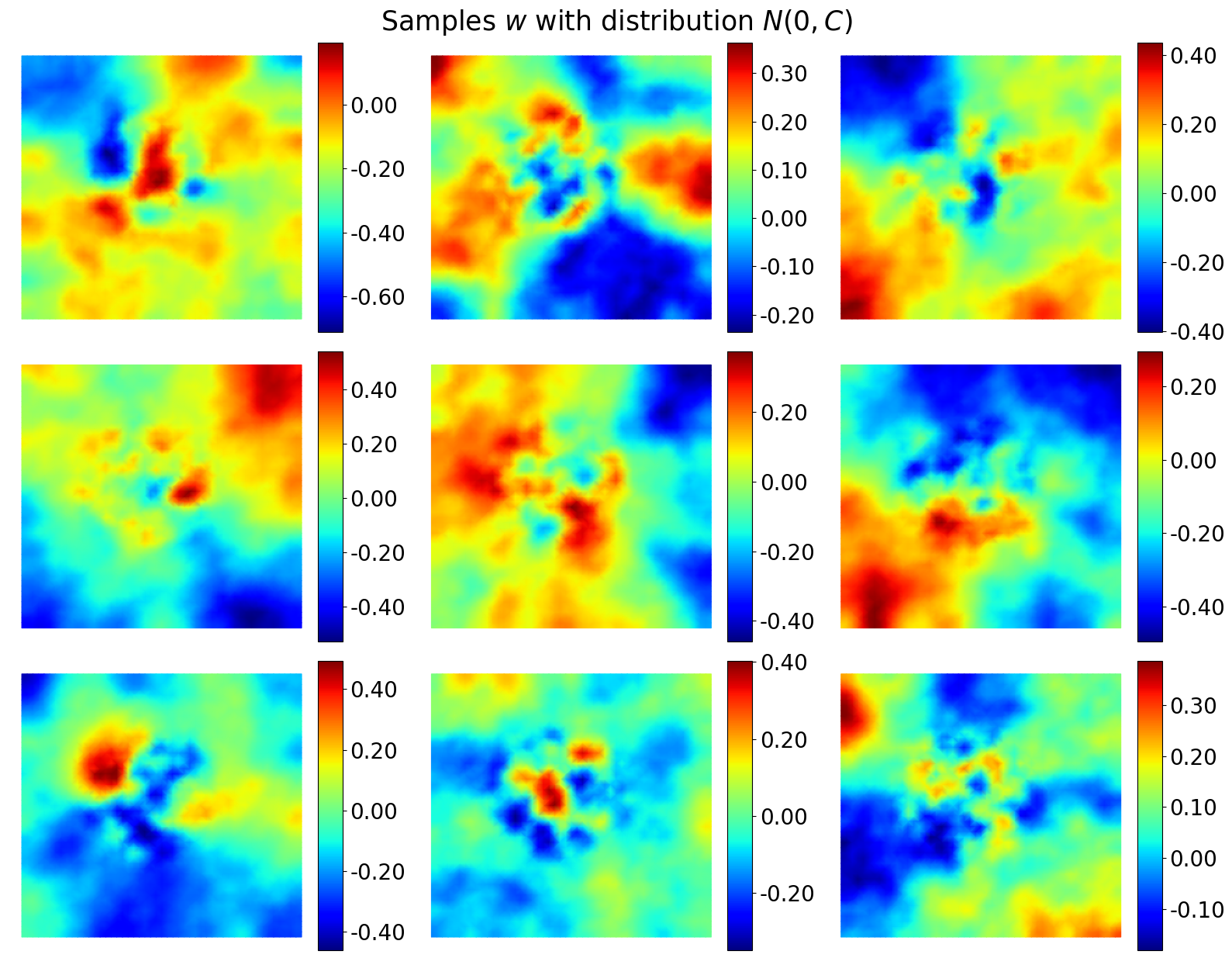}
  \end{minipage}
}
\caption{Random samples $w$ generated using a Gaussian measure based on the Laplacian-like operator $L_\Delta$ defined in \eqref{eq:LDelta}.}
\label{fig:sampling}
\end{figure}

In \cref{lst:sampling}, the above algorithm is implemented using the Python library \lstinline{FEniCSx}\footnote{Implementation uses \lstinline{FEniCSx} (version 0.10.0), which is the next generation of \lstinline{FEniCS}; see \cite{baratta2023dolfinx}.}. In \cref{fig:sampling}, results from a Gaussian sampler with parameters $\mathsf{a}_c = 0.01$, $\mathsf{c}_c = 0.2$, an inhomogeneous diffusivity function $\mathsf{b}_c$, and zero mean $\bar{w} = 0\in W$ are shown. This example of a prior is considered in the the notebook \lstinline{PriorSampler.ipynb}\footnote{\url{https://github.com/CEADpx/neural_operators/blob/survey26_v2/survey_work/common/prior_sampler/PriorSampler.ipynb}}.

\vspace{10pt}
\vspace{20pt} 
\begin{lstlisting}[language=Python, caption={Finite element implementation of Gaussian random field sampling with covariance $C=L_\Delta^{-2}$.}, label={lst:sampling}]
...
import numpy as np
import ufl
from dolfinx import default_scalar_type, fem
from dolfinx.fem.petsc import assemble_matrix, assemble_vector, create_vector
...

class PriorSampler:
    
    def __init__(self, V, a, c, seed=0):
        ...
        # function space
        self.V = V

        # vertex to dof vector and dof vector to vertex maps
        self.V_vec2vv, self.V_vv2vec = build_vector_vertex_maps(self.V)
        ...
        self.a_form = self.a*self.b_fn\
                        *ufl.inner(ufl.grad(self.u_trial), \
                                  ufl.grad(self.u_test))*ufl.dx \
                    + self.c*self.u_trial*self.u_test*ufl.dx
        self.L_form = self.s_fn*self.u_test*ufl.dx
        ...        
        
    def __call__(self, m = None):
        # forcing term
        self.s_fn.x.array[:] = np.random.normal(0.0, 1.0, self.s_dim)
        ...
        # assemble rhs only
        self._assemble_rhs()
        
        # solve
        self.u_fn.x.array[:] = 0.0
        self._solve(self.u_fn)

        # add mean
        self.u_fn.x.petsc_vec.axpy(1.0, self.mean_fn.x.petsc_vec)

        # vertex_dof ordered
        self.u = self.u_fn.x.array[self.V_vec2vv].copy()

        log_prior = self._log_prior_from_source()
        ...
        if m is not None:
            m = self.u.copy()
            return m, log_prior
        else:
            return self.u.copy(), log_prior
\end{lstlisting}

\section{Model problems}\label{s:modelProblems}
Neural operators will be discussed in the context of solving three PDE problems. These are presented in the following subsections. 

\subsection{Poisson problem}\label{ss:poisson}
Consider a two-dimensional domain given by a rectangle $D_{U} = (0, L_1) \times (0, L_2) \subset \bbR^2$ and suppose $u: D_{U} \to \bbR$ denotes the temperature field. The governing equation based on the balance of energy reads:
\begin{equation}\label{eq:poisson}
    \begin{aligned}
        -\nabla \cdot \left( m(x) \nabla u(x) \right) &= f(x)\,, \qquad &&\forall x \in D_{U}\,, \\
        u(x) &= 0\,, \qquad &&\forall x \in \Gamma_{u_d}\,, \\
        m(x) \nabla u(x)\cdot n(x) &= q(x)\,, \qquad && \forall x \in \Gamma_{u_q} = \partial D_{U} - \Gamma_{u_d}\,,
    \end{aligned}
\end{equation}
where $\Gamma_{u_d}:= \left\{ x \in \partial D_{U}: x_1 < L_1 \right\}$ (i.e., all sides except the right side of the rectangular domain) and $\Gamma_{u_q}:= \partial D_{U} - \Gamma_{u_d}$. Here, $f(x)$ is the specified heat source (J/m$^3$/s), $m(x)$ is the diffusivity (J/K/m/s), $n(x)$ is the unit outward normal, and $q(x)$ is the specified heat flux (J/m$^2$/s). To be consistent with the notations in \cref{ss:notat}, here $D_{M} = D_{W} = D_{U}$ (domains of functions $m$, $w$ ($w$ to be introduced shortly), and $u$), $q_{W} = q_{M} = q_{U} = 2$ (dimensions of the domains of functions), and $d_{M} = d_{W} = d_{U} = 1$ (dimensions of the pointwise values of functions). Working in the variational setting, the appropriate function spaces for diffusivity and temperature are:
\begin{equation}
    M:=\left\{ m: D_{M} \to \bbR^{+}: \norm{m}_{L^2} < \infty \right\}\,,\; U:= \left\{u \in H^1(D_{U}; \bbR): u = 0\; \text{ on }\Gamma_{u_d} \right\}\,.
\end{equation}
The weak form of the problem is: given $m \in M$, find $u\in U$ such that
\begin{equation}\label{eq:poissonWeak}
    \int_{D_{U}} m \nabla u \cdot \nabla v \dd x
    =
    \int_{D_{U}} f v \dd x
    +
    \int_{\Gamma_{u_q}} q v \dd s\,,
    \qquad \forall v \in U\,.
\end{equation}

The diffusivity field $m$ is assumed to be the unknown and uncertain parameter field. To ensure that $m$ is a positive function (diffusivity cannot be zero or negative), consider a random field $\mscZ: \Omega \to W:= L^2(D_{W}; \bbR)$ (with $D_{W} = D_{M} = D_{U}$) and suppose the associated measure $\mu_\mscZ$ on $W$ is Gaussian $N(0, C)$ ($0$ mean and $C$ covariance operator defined in \cref{sss:gaussian} with parameters $\mathsf{a}_c, \mathsf{b}_c, \mathsf{c}_c$). Now, given a sample $w = \mscZ(z) \in W$, sample $m = \mscQ(z)$ is computed as follows:
\begin{equation}
    \mscQ(z) = m = \alpha_m \exp(w) + \beta_m\,, \qquad \text{where }\quad  z\in \Omega\,, w = \mscZ(z)\,, \text{ and } \mu_\mscZ = N(0, C)\,.
\end{equation}
Here, $\alpha_m$ and $\beta_m$ are constants. Thus, $m$ is a positive function and comes from a log-normal distribution. The generation of samples $w$ is discussed in \cref{ss:sampling}, and given $w$, computing $m$ using the formula above is straightforward; see \lstinline{transform_gaussian_pointwise()} in \cref{lst:poisson}.

Finally, given $m\in M$, let $F(m) = u\in U$ be the solution of boundary value problem (BVP) \eqref{eq:poisson}. This $F: M \to U$ is the solution/forward operator and target of neural operator learning.

\vspace{20pt} 
\subsubsection{Setup details and data generation}\label{sss:poissonSetup}
\begin{figure}[h!]
    \centering
    \includegraphics[width=0.9\linewidth]{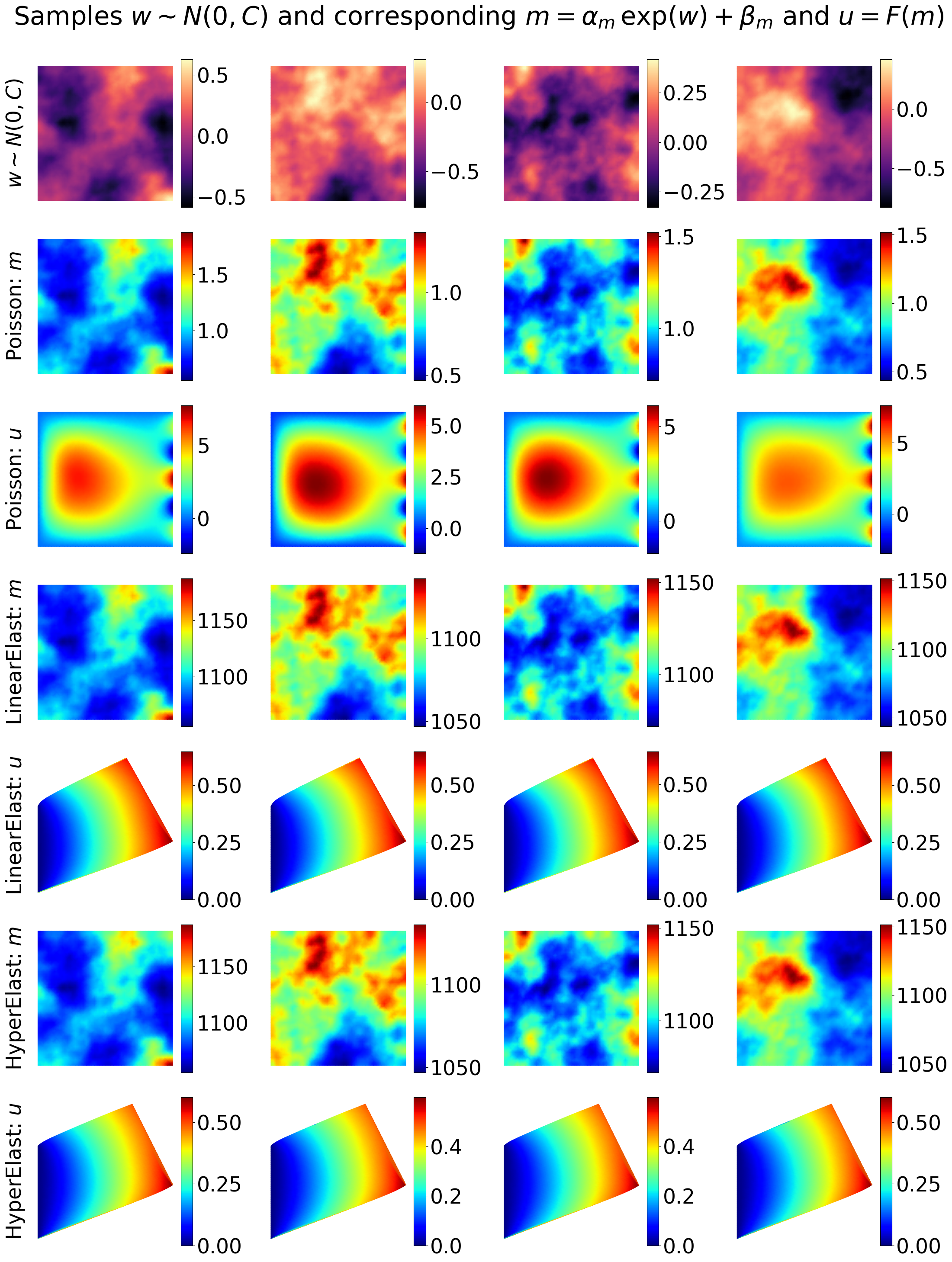}
    \caption{Some representative data samples for the three problems. For the linear elasticity and hyperelasticity problems, the displacement (PDE solution) $u$ is visualized by plotting the deformed configuration of the domain with coordinates $z = x + u(x)$ for $x\in D_{U}$. The color map indicates the displacement magnitude $|u(x)|$.}\label{fig:samplesAllModels}
\end{figure}

Let $L_1 = L_2 = 1$, and consider the following $f$ and $q$:
\begin{equation}
    f(x) = 1000 (1-x_2)x_2(1 - x_1)^2 \qquad \text{and} \qquad q(x) = 50\sin(5\pi x_2)\,.
\end{equation}
The covariance operator in $\mu_{\mscZ} = N(0, C)$ is taken as $C = L_{\Delta}^{-2}$, where $L_\Delta$ is defined in \eqref{eq:LDelta}. The parameters in $L_{\Delta}$ and the transformation of $w$ into $m$ are
\begin{equation}
\mathsf{a}_c = 0.005\,, \quad \mathsf{b}_c = 1\,, \quad \mathsf{c}_c = 0.2\,,\quad \alpha_m = 1\,, \quad \beta_m = 0\,.
\end{equation}
Data for neural operator training is generated using finite element discretization with triangle elements and linear interpolation for both input and output functions. The number of elements is 5000, the number of nodes is $N_{nodes} = 2601$, and $p_{M} = p_{U} = N_{nodes} = 2601$. \cref{lst:poisson} details the setup and solution of the Poisson model problem \eqref{eq:poisson}, building on the abstract class \lstinline{PDEModel} shown in \cref{lst:pdemodel}. The sampling of uncertain parameter $m = \alpha_m \exp(w) + \beta_m \sim \mu_\mscQ$, where $w\sim \mu_\mscZ = N(0, C)$, is straightforward using the method and implementation of sampling $w\sim \mu_\mscZ$ discussed in \cref{sss:gaussian} and \cref{lst:sampling}. In \cref{fig:samplesAllModels}, samples of $w$ and corresponding $(m, u)$ pairs are shown. The notebook \lstinline{Poisson.ipynb}\footnote{\url{https://github.com/CEADpx/neural_operators/blob/survey26_v2/survey_work/problems/poisson/Poisson.ipynb}} implements methods to generate and post-process data for neural operator training.

\vspace{10pt}
\vspace{30pt} 
\begin{lstlisting}[language=Python, caption={Abstract PDE class.}, label={lst:pdemodel}]
import numpy as np
from dolfinx import fem
from fenicsUtilities import build_vector_vertex_maps

class PDEModel:

    def __init__(self, Vm, Vu, prior_sampler, seed = 0):
        ...
        # prior and transform parameters
        self.prior_sampler = prior_sampler
        
        # FE setup
        self.Vm, self.Vu, self.mesh = Vm, Vu, Vm.mesh
        ...

    # following functions must be defined in the inherited class
    # boundaryU(x, on_boundary) 
    # is_point_on_dirichlet_boundary(x) 
    # assemble(self, assemble_lhs=True, assemble_rhs=True)
    # compute_mean(self, m)
    # solveFwd(self, u=None, m=None, transform_m=False)
    # samplePrior(self, m=None, transform_m=False)
\end{lstlisting}

\begin{lstlisting}[language=Python, caption={Key components of the Poisson model implementation.}, label={lst:poisson}]
...
import ufl
from dolfinx import default_scalar_type, fem
from dolfinx.fem.petsc import (
    apply_lifting,
    assemble_matrix,
    ...
)
from pdeModel import PDEModel
...

class PoissonModel(PDEModel):
    def __init__(self, Vm, Vu, prior_sampler, logn_scale=1., logn_translate=0., seed=0):
        super().__init__(Vm, Vu, prior_sampler, seed)
        ...
        # External heat source and boundary flux
        self.f_expr = 1000 * (1 - x[1]) * x[1] * (1 - x[0]) * (1 - x[0])
        self.q_expr = 50 * ufl.sin(5 * ufl.pi * x[1])
        ...
        # variational form
        self.a_form = self.m_fn \
                    * ufl.inner(ufl.grad(self.u_trial),  ufl.grad(self.u_test)) * dx
        self.L_form = self.f_expr * self.u_test * dx + self.q_expr * self.u_test * ds
        ...
        # Dirichlet boundary condition
        ...
        dofs = fem.locate_dofs_geometrical(self.Vu, self._dirichlet_boundary)
        self.bc = [fem.dirichletbc(default_scalar_type(0.0), dofs, self.Vu)]
        ...
    ...

    def transform_gaussian_pointwise(self, w, m_local=None):
        if m_local is None:
            self.m_transformed = self.logn_scale * np.exp(w) + self.logn_translate
            return self.m_transformed.copy()
        return self.logn_scale * np.exp(w) + self.logn_translate

    def compute_mean(self, m):
        return self.transform_gaussian_pointwise(self.prior_sampler.mean, m)
    
    def solveFwd(self, u=None, m=None, transform_m=False):
        ...
        # solve
        self.u_fn.x.petsc_vec.set(0.0)
        self._setup_solver()
        self._ksp.solve(self.rhs, self.u_fn.x.petsc_vec)
        ...
        return self.function_to_vertex(self.u_fn, u, is_m=False)

    def samplePrior(self, m=None, transform_m=False):
        if transform_m:
            w, _ = self.prior_sampler()
            self.m_transformed = self.transform_gaussian_pointwise(w, \
                                        self.m_transformed)
        else:
            self.m_transformed = self.prior_sampler()[0]

        if m is None:
            return self.m_transformed.copy()
        m = self.m_transformed.copy()
        return m
\end{lstlisting}

\subsection{Linear elasticity problem}\label{ss:elastModel}
The second problem concerns deformation of a two-dimensional rectangular domain $D_{U} = (0, L_1) \times (0, L_2) \subset\bbR^2$. The deformation is governed by the linear elasticity equations. Suppose $m(x)$ denotes the Young's modulus field at a point $x\in D_{U}$, $u = (u_1, u_2) : D_{U} \to \bbR^2$ denotes the displacement field, $e(x) = (\nabla u + \nabla u^T)/2$ denotes the linearized strain, $\sigma(x)$ denotes the Cauchy stress, and $b(x)$ denotes the body force per unit volume. The equation for $u$ is based on the balance of linear momentum and reads:
\begin{equation}\label{eq:elastModel}
    \begin{aligned}
        -\nabla \cdot \sigma(x) &= b(x)\,, \qquad & &\forall x\in D_{U}\,,\\
        \sigma(x) &= \lambda(x) \mathrm{tr}(e) \rb{I} + 2\mu(x) e\,, \qquad & & \forall x \in D_{U}\,,\\
        u(x) &= 0\,,\qquad & & \forall x\in \Gamma_{u_d}\,,\\
        \sigma(x) n(x) &= t(x)\,, \qquad & & \forall x\in \Gamma_{u_q} = \partial D_{U} - \Gamma_{u_d}\,,
    \end{aligned}
\end{equation}
where $\mathrm{tr}(e) = e_{11} + e_{22}$ is the trace of the strain tensor, $\rb{I}$ is the identity second order tensor in $\Rpow{2}$, and $\lambda$ and $\mu$ are Lam\'e parameters related to the Young's modulus $m$ and the Poisson ratio $\nu$ as follows:
\begin{equation}\label{eq:lame}
    \lambda(x) = \frac{m(x) \nu}{(1+\nu)(1-2\nu)} \quad \text{and} \quad \mu(x) = \frac{m(x)}{2(1+\nu)}\,.
\end{equation}
In \eqref{eq:elastModel}, $\Gamma_{u_d} := \left\{x\in \partial D_{U}: x_1 = 0 \right\}$ and $\Gamma_{u_q} := \partial D_{U} - \Gamma_{u_d}$, $n$ is the unit outward normal, and $t$ is the specified traction. The scalar field $m\in M$ is considered to be uncertain, and the forward map $F: M \to U$ is defined such that, given $m\in M$, the function $u = F(m) \in U$ solves the BVP \eqref{eq:elastModel}. To be consistent with the notations, here $D_{W} = D_{M} = D_{U}$, $q_{W} = q_{M} = q_{U} = 2$, $d_{W} = d_{M} = 1$, and $d_{U} = 2$. Working in the variational setting, appropriate function spaces for the parameter field and solution are as follows:
\begin{equation}
    M:=\left\{ m: D_{M} \to \bbR^{+}: \norm{m}_{L^2} < \infty \right\}\,, \; U:= \left\{u \in H^1(D_{U}; \bbR^2): u = 0\; \text{ on } \Gamma_{u_d} \right\}\,.
\end{equation}
The weak form of the problem is: given $m\in M$, find $u\in U$ such that
\begin{equation}\label{eq:elastWeak}
    \int_{D_{U}} \sigma : \frac{(\nabla v + \nabla v^T)}{2} \dd x
    =
    \int_{D_{U}} b\cdot v \dd x
    +
    \int_{\Gamma_{u_q}} t\cdot v \dd s\,,
    \qquad \forall v\in U\,,
\end{equation}
where $\rmA : \rmB = \sum_{i,j=1}^{2} \rmA_{ij} \rmB_{ij}$ denotes the tensor inner product between second-order tensors $\rmA$ and $\rmB$. To ensure positivity, a log-normal distribution is used for $m$:
\begin{equation}
    \mscQ(z) = m = \alpha_m \exp(w) + \beta_m\,, \qquad \text{where }\quad  z\in \Omega\,, w = \mscZ(z)\,, \text{ and } \mu_\mscZ = N(0, C)\,.
\end{equation}

\vspace{10pt} 
\subsubsection{Setup details and data generation}\label{sss:elasticitySetup}
The domain is the unit square, $L_1=L_2=1$. The body force is zero, and a constant traction is applied on $\Gamma_{u_q}$:
\begin{equation}
    b(x) = 0\hat{e}_1 + 0\hat{e}_2
    \quad \text{and} \quad
    t(x) = 20\hat{e}_1 + 50\hat{e}_2\,.
\end{equation}
The Poisson ratio is fixed at $\nu = 0.45$. The covariance operator is taken as $C = L^{-2}_{\Delta}$, with the parameters
\begin{equation}
\mathsf{a}_c = 0.005\,, \quad \mathsf{b}_c = 1\,, \quad \mathsf{c}_c = 0.2\,,\quad \alpha_m = 100\,, \quad \beta_m = 1000\,.
\end{equation}
As in the Poisson problem, a finite element approximation with triangle elements and linear interpolation is used for both the input and output functions. The number of elements is 5000, the number of nodes is $N_{nodes} = 2601$, $p_{U} = 2N_{nodes} = 5202$, and $p_{M} = N_{nodes} = 2601$. \cref{lst:elastsolve} outlines crucial steps in solving the problem. The sampling of $m$ is similar to problem 1. \cref{fig:samplesAllModels} shows representative samples of $w$ and corresponding $(m, u)$ pairs. The notebook \lstinline{LinearElasticity.ipynb}\footnote{\url{https://github.com/CEADpx/neural_operators/blob/survey26_v2/survey_work/problems/linear_elasticity/LinearElasticity.ipynb}} implements methods to generate and post-process data for neural operator training.

\vspace{10pt}
\vspace{15pt} 
\begin{lstlisting}[language=Python, caption={Class for the linear elasticity problem. Here, only the initialization part is shown, as all other functions are similar to the implementation of the Poisson problem in \cref{lst:poisson}.}, label={lst:elastsolve}]
from pdeModel import PDEModel
...
class LinearElasticityModel(PDEModel):
    
    def __init__(self, Vm, Vu, prior_sampler, logn_scale=1., logn_translate=0., seed=0):
        ...
        self.nu = 0.45
        self.lam_fact = self.nu / ((1 + self.nu) * (1 - 2 * self.nu))
        self.mu_fact = 1.0 / (2 * (1 + self.nu))

        # External body force and boundary traction
        self.b = ufl.as_vector((0.0, 0.0))
        self._traction_x = fem.Constant(self.mesh, default_scalar_type(20.0))
        self._traction_y = fem.Constant(self.mesh, default_scalar_type(50.0))
        self.t = ufl.as_vector((self._traction_x, self._traction_y))

        # variational form
        I = ufl.Identity(spatial_dim)
        sigma = self.lam_fact * ufl.tr(ufl.grad(self.u_trial)) * I \
                + 2 * self.mu_fact * ufl.sym(ufl.grad(self.u_trial))
        self.a_form = self.m_fn * ufl.inner(sigma, ufl.sym(ufl.grad(self.u_test))) * dx
        self.L_form = ufl.inner(self.b, self.u_test) * dx \
                      + ufl.inner(self.t, self.u_test) * ds
        ...
    ...    
    def solveFwd(self, u = None, m = None, transform_m = False):
        # similar to the solveFwd() for the Poisson problem (Listing 3)

    def samplePrior(self, m = None, transform_m = False):
        # similar to the samplePrior() for the Poisson problem (Listing 3)
\end{lstlisting}

\subsection{Hyperelasticity problem}\label{ss:hyperElastModel}

The third problem concerns hyperelastic deformation on the same rectangular domain considered in \cref{ss:elastModel}. The uncertain parameter remains the Young's modulus field $m(x)$, and the displacement field is $u:D_{U}\to\bbR^2$. The deformation gradient, right Cauchy--Green tensor, first invariant, and Jacobian are
\begin{equation}
    \rb{F}=\rb{I}+\nabla u\,,\qquad
    \rb{C}=\rb{F}^{\top}\rb{F}\,,\qquad
    I_1=\mathrm{tr}(\rb{C})\,,\qquad
    J=\mathrm{det}(\rb{F})\,.
\end{equation}
A compressible Neo-Hookean strain-energy density \cite{korobeynikov2025two} is used:
\begin{equation}\label{eq:strainEnergyHyperelasticity}
    W(\rb{F})=\frac{\mu}{2}\left(I_1-3-2\ln J\right)+\frac{\lambda}{2}(\ln J)^2\,,
\end{equation}
where $\lambda$ and $\mu$ are the Lam\'e parameters related to the Young's modulus $m$ and Poisson ratio $\nu$ as in \eqref{eq:lame}. 
Using the chain rule and the identities (Eqs.~(1.226), (1.228), and (1.232) in [\citealp{jog2015continuum}])
\begin{equation}
    \frac{\partial I_1}{\partial \rb{F}} = 2\rb{F}\,,
    \qquad
    \frac{\partial J}{\partial \rb{F}} = J\rb{F}^{-T}\,,
    \qquad
    \frac{\partial \ln J}{\partial \rb{F}} = \rb{F}^{-T}\,,
\end{equation}
the first Piola--Kirchhoff stress for the strain-energy density in \eqref{eq:strainEnergyHyperelasticity} is
\begin{equation}
    \rb{P}
    =
    \frac{\partial W}{\partial \rb{F}}
    =
    \mu \rb{F}
    +
    \left(\lambda \ln J - \mu\right)\rb{F}^{-T}\,.
\end{equation}

The strong form of the hyperelasticity problem on the displacement field $u$ is written in the reference configuration as
\begin{equation}\label{eq:hyperElastModel}
    \begin{aligned}
        -\nabla \cdot \rb{P}(x) &= b(x)\,, \qquad & &\forall x\in D_{U}\,,\\
        u(x) &= 0\,, \qquad & &\forall x\in \Gamma_{u_d}\,,\\
        \rb{P}(x)n(x) &= t(x)\,, \qquad & &\forall x\in \Gamma_{u_q} = \partial D_{U} - \Gamma_{u_d}\,,
    \end{aligned}
\end{equation}
As in the case of linear elasticity, consider the function spaces 
\begin{equation}
    M:=\left\{ m: D_{M} \to \bbR^{+}: \norm{m}_{L^2} < \infty \right\}\,, \; U:= \left\{u \in H^1(D_{U}; \bbR^2): u = 0\; \text{ on } \Gamma_{u_d} \right\}\,.
\end{equation}
Then, the weak form of the problem is: given $m\in M$, find $u\in U$ such that
\begin{equation}\label{eq:hyperElastWeak}
    \int_{D_{U}} \rb{P}:\nabla v \dd x
    =
    \int_{D_{U}} b\cdot v \dd x
    +
    \int_{\Gamma_{u_q}} t\cdot v \dd s\,, \qquad \forall v\in U\,.
\end{equation}
As before, the forward map $F:M\to U$ is defined by $u=F(m)$, where $m\in M$ is the uncertain Young's modulus field.

The transformation from $w$ to $m$ and the prior model are the same as in \cref{ss:elastModel}.

\vspace{10pt} 
\subsubsection{Setup details and data generation}\label{sss:hyperelasticitySetup}
The setup, including the domain, external loading, Poisson ratio, and sampling details, is the same as in \cref{sss:elasticitySetup}. The main computational change is that each forward solve requires a Newton iteration because the hyperelasticity problem is nonlinear. To improve robustness, the traction load is applied incrementally over $20$ load steps during the nonlinear solve; see \lstinline{solveFwd} in \cref{lst:hyperelastsolve} for the key implementation steps. \cref{fig:samplesAllModels} shows representative samples of $w$ and corresponding $(m,u)$ pairs. The notebook \lstinline{Hyperelasticity.ipynb}\footnote{\url{https://github.com/CEADpx/neural_operators/blob/survey26_v2/survey_work/problems/hyperelasticity/Hyperelasticity.ipynb}} implements methods to generate and post-process data for neural operator training.

\vspace{10pt}
\vspace{10pt} 
\begin{lstlisting}[language=Python, caption={Class for the hyperelasticity problem. Only the initialization and nonlinear-solver components are shown because the sampling and data-handling functions are similar to the linear elasticity implementation.}, label={lst:hyperelastsolve}]
...
from pdeModel import PDEModel

class HyperelasticityModel(PDEModel):

    def __init__(self, Vm, Vu, prior_sampler, logn_scale=1., logn_translate=0., seed=0):
        # most setup details (nu, b, t, ...) follow LinearElasticityModel
        ...
        # variational form
        spatial_dim = self.mesh.geometry.dim
        I = ufl.variable(ufl.Identity(spatial_dim))
        F = ufl.variable(I + ufl.grad(self.u_fn))
        C = ufl.variable(F.T * F)
        Ic = ufl.variable(ufl.tr(C))
        J = ufl.variable(ufl.det(F))

        mu = self.m_fn * self.mu_fact
        lam = self.m_fn * self.lam_fact
        # Stable compressible Neo-Hookean
        W = (mu / 2.0) * (Ic - 3.0 - 2.0 * ufl.ln(J)) + (lam / 2.0) * (ufl.ln(J)) ** 2
        P = ufl.diff(W, F)

        self._residual_form = ufl.inner(self.b, self.u_test) * dx \
                                + ufl.inner(self.t, self.u_test) * ds \
                                - ufl.inner(ufl.grad(self.u_test), P) * dx

        self._nonlinear_problem = None
        self._newton_solver = None
        ...

    def _setup_solver(self):
        if self._nonlinear_problem is None:
            self._nonlinear_problem = NewtonSolverNonlinearProblem(
                self._residual_form, self.u_fn, bcs = self.bc)
            self._newton_solver = \
                NewtonSolver(self.mesh.comm, self._nonlinear_problem)
            ...

    def _run_newton(self):
        self._setup_solver()
        n_iter, converged = self._newton_solver.solve(self.u_fn)
        if not converged:
            raise RuntimeError("Hyperelasticity Newton solver did not converge.")
        return n_iter

    def solveFwd(self, u = None, m = None, transform_m = False):
        ...
        # load stepping for robust convergence
        target_tx = float(self._traction_x.value)
        target_ty = float(self._traction_y.value)
        n_steps = self.n_load_steps if self.reset_u else 1

        for step in range(1, n_steps + 1):
            if n_steps > 1:
                load_frac = step / n_steps
                self._traction_x.value = target_tx * load_frac
                self._traction_y.value = target_ty * load_frac
            self._run_newton()
        ...
        return self.function_to_vertex(self.u_fn, u, is_m = False)

    def samplePrior(self, m = None, transform_m = False):
        # similar to the samplePrior() for the Poisson problem (Listing 3)
\end{lstlisting}

\section{Neural operators as surrogates for the forward problem}\label{s:nn}

Over the years, several neural operator architectures have emerged that leverage neural networks to construct efficient approximations of $F$. This section presents representative approaches, emphasizing their underlying principles and implementation details to develop a practical understanding of their methods.

The following subsections introduce DeepONet, PCANet, and FNO, along with their implementation details.

\subsection{Deep Operator Network (DeepONet)}\label{ss:deeponet}
\begin{figure}[h!]
    \centering
    \includegraphics[width=0.95\linewidth]{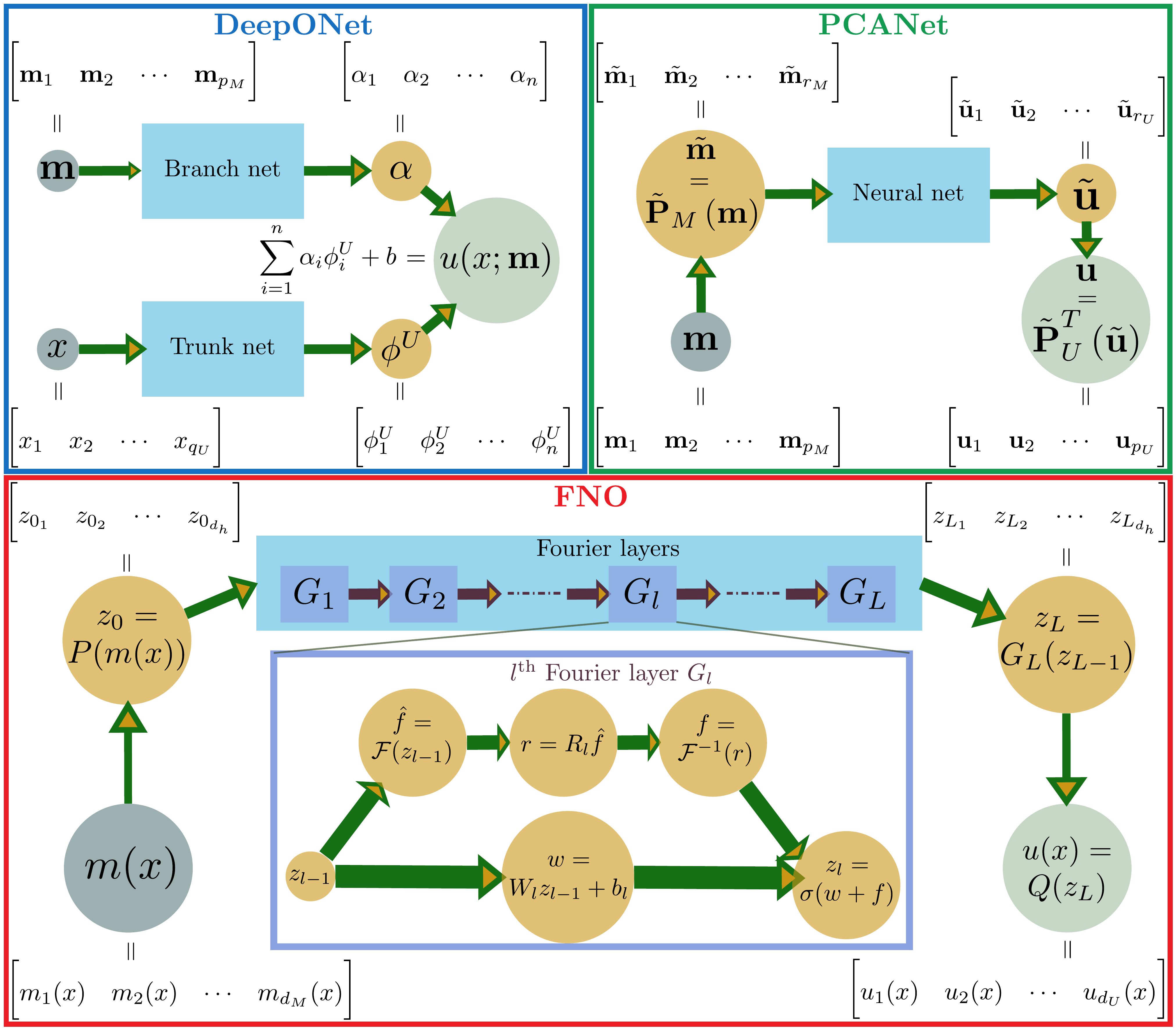}
    \caption{Schematics of three neural operators, DeepONet, PCANet, and FNO. Grey and light green circles represent the input and output of neural operators. The blue box includes a parameterized neural network-based map. In this work, the blue boxes for DeepONet and PCANet employ fully connected multi-layer perceptron neural networks. In the case of FNO, trainable parameters (namely, $R_l, W_l, b_l$) appear within each Fourier layer.}
    \label{fig:schematicsNOp}
\end{figure}

DeepONet was first introduced in \cite{lu2019deeponet}, and over the years, various extensions of the general framework and applications have been realized. Consider some $m$, and suppose $\{\phi^U_i\}_{i=1}^{p_{U}} \subset U$ is a finite collection of basis functions, then, at $x \in D_{U}$,
\begin{equation}\label{eq:FbasisApprox}
    F(m)(x) = u(x)  \approx \sum \;\underbrace{\innerProd{F(m)}{\phi^U_i}}_{=: \alpha_i(m)} \;\phi^U_i(x)\,,
\end{equation}
where, coefficients $\alpha_i = \alpha_i(m)$ depend on $m$. DeepONet's underlying idea is to learn the above finite-dimensional representation of the output of operator $F$. That is, identify the linear bases or more precisely learn the values of basis functions $\phi^U_i(x)$ at coordinates $x \in D_{U}$, and the coefficients associated with the bases dependent on the input $m$, $\{\alpha_i(m)\}$ so that $\sum_i \alpha_i(m) \phi^U_i(x)$ is approximately equal to $F(m)(x) = u(x)$. Towards this, DeepONet considers two neural networks, so-called branch and trunk networks; see \cref{fig:schematicsNOp}. The branch network takes discretization of the input function, denoted by $\rmm\in \Rpow{p_{M}}$, and the neural network produces $N_{br}$ number of coefficients, which are used as $\{\alpha_i\}_{i=1}^{N_{br}}$. On the other hand, the trunk network takes as input the spatial coordinate, $x$, and outputs $d_{U}N_{tr}$ numbers, which play the role of $\{\phi^U_i(x)\}_{i=1}^{N_{tr}}$ in \eqref{eq:FbasisApprox}. If $m$ and $u$ are scalar fields then $N_{tr} = N_{br}$. Finally, the loss function is defined in terms of the norm of the difference between the ground truth and approximation by the joint output from the branch and trunk networks, $\sum_{i=1}^{N_{br}}\alpha_i(m)\phi^U_i(x)$. 

In summary, the operator approximation in DeepONet takes the form, for $m \in M$ (input actually is the finite-dimensional discretization $\rmm \in \Rpow{p_{M}}$ of $m\in M$) and $x \in D_{U}$,
\begin{equation}\label{eq:deeponet}
    F_{NOp}(m)(x) = \sum_{i=1}^{N_{br}} \alpha_i(m; \theta_{br}) \phi^U_i(x; \theta_{tr}) + b\,,
\end{equation}
where 
\begin{itemize}
    \item[(i)] $\alpha_i = \alpha_i(m; \theta_{br}) \in \R$, $1\leq i \leq N_{br}$, are outputs of the branch network with $\theta_{br}$ neural network trainable parameters,
    \item[(ii)] $\phi^U_i(x; \theta_{tr}) \in \Rpow{d_{U}}$, $1\leq i\leq N_{br}$, are outputs of the trunk network (total $N_{tr} = d_{U} N_{br}$ numbers) with $\theta_{tr}$ neural network trainable parameters, and
    \item[(iii)] $b \in \Rpow{d_{U}}$ is a bias. 
\end{itemize}
A key feature of DeepONet is the separation between learning input-dependent coefficients and spatial representations via the branch and trunk networks. Another crucial property of DeepONet is that an approximation of $u = F(m)$ at any arbitrary point $x \in D_{U}$ can be computed. 

\subsubsection{Implementation of DeepONet}
To simplify the presentation, the input and output functions, $m\in M$ and $u \in U$, respectively, are assumed to be scalar-valued, and they are appropriately discretized, e.g., using a finite element approximation. Extending the cases when one or both of these functions are vector-valued is trivial; see \cref{rem:vectorvalued}.

Let $\rmm \in \Rpow{p_{M}}$ be the input discretized function, and let $\rmu =\rmF(\rmm) \in \Rpow{p_{U}}$ be the corresponding output discretized function, where $\rmF$ is the discretization of the operator of interest $F$. The following constitutes data for DeepONet:
\begin{enumerate}
    \item Consider the $N_M\times p_{M}$-matrix, $\rb{X}_{br}$, where $N_M$ is the number of input function samples and $\ith{I}{th}$ row of $\rb{X}_{br}$ is the sample $\rmm^I \in \Rpow{p_{M}}$. Each row of $\rb{X}_{br}$ will be the input to the branch network.
    
    \item Select $N_x$ number of locations, $\{x^I\}_{I=1}^{N_x}$, where $x^I \in D_{U} \subseteq \Rpow{q_{U}}$ and $q_{U}$ is the dimension of the domain. Each location $x^I$ will serve as input to the trunk neural network, so the DeepONet output is the prediction of the target function's value at $x^I$. In the present implementation, the typical input coordinate $x^I$ corresponds to the $\ith{I}{th}$ discretization grid or node of a mesh so that the value of the output function at $x^I$ is represented by the element $\rmu_I$ in the vector $\rmu$ corresponding to that grid/node (this is true only for scalar-valued functions). The matrix $\rb{X}_{tr}$ of size $N_x \times q_{U}$ is the data for the trunk network, and each row of $\rb{X}_{tr}$ is the input to the trunk network.
    
    \item $N_M\times N_x$-matrix $\rb{Y}$, such that an element $\rb{Y}_{IJ}$ is the pointwise value of the output data corresponding to $\rmm^I$ at location $x^J$, i.e.,
    \begin{equation}
        \rb{Y}_{IJ} = (\rmu^I)_J \approx F(m^I)(x^J)\,.
    \end{equation}
    I.e., $\rb{Y}_{IJ}$ is the pointwise value of the output data corresponding to the branch network input data function $\rmm^I$ ($I^\text{th}$ row of $\rb{X}_{br}$) at trunk network input location $x^J$ ($J^\text{th}$ row of $\rb{X}_{tr}$).
\end{enumerate}
Next, the combined training parameters $\theta = \{\theta_{br}, \theta_{tr}, b\} \in \Rpow{p_\Theta}$ are optimized to minimize the error:
\begin{equation}
\begin{aligned}
    \theta^\ast &= \argmin_{\theta = \{\theta_{br}, \theta_{tr}, b\} \in \Rpow{p_\Theta}}\; \frac{1}{N_M N_x} \sum_{I=1}^{N_M} \sum_{J=1}^{N_x} \left\vert \rb{Y}_{IJ} - \underbrace{\left(\sum_{k=1}^{N_{br}} \alpha_k(\rmm^I; \theta_{br}) \phi^U_k(x^J; \theta_{tr}) + b \right)}_{\text{DeepONet output}} \right\vert^2\,.
\end{aligned}
\end{equation}
Here, $p_\Theta$ is the total number of trainable parameters in the DeepONet architecture. 

\begin{remark}\label{rem:vectorvalued}
    The DeepONet framework described so far can be easily extended to the case where the operator's target function is vector-valued. Suppose $u(x) = (u_1(x), u_2(x), ..., u_{d_{U}}(x)) \in \Rpow{d_{U}}$, $u_j$ being the $\ith{j}{th}$ component function. One way to extend the DeepONet framework discussed above to vector-valued target functions is by considering $N_{br}$ branch outputs $\{\alpha_k\}_{k=1}^{N_{br}}$ and $N_{tr} = d_{U} N_{br}$ trunk outputs $\{\phi^U_k\}_{k=1}^{N_{br}}$ and using \eqref{eq:deeponet}. Here, $\phi^U_k$ are $\Rpow{d_{U}}$-valued functions, so the number of scalar outputs from the trunk network is $N_{tr}=d_{U}N_{br}$.
    
    The alternative construction considered in this work uses a trunk network that produces $N_{tr}$ outputs $\{\psi_k\}_{k=1}^{N_{tr}}$, where $\psi_k(x)\in\R$. The branch network produces $N_{br}=d_{U}N_{tr}$ outputs $\{\alpha_k\}_{k=1}^{N_{br}}$. Then, for the $x^J$ row of $\rb{X}_{tr}$ and the $\rmm^I$ row of $\rb{X}_{br}$, the prediction $u_{pred}=(u_{pred,1},u_{pred,2},\ldots,u_{pred,d_{U}})$ of $u$ is given by
    \begin{equation}\label{eq:deeponetVectorValued}
    \begin{aligned}
        u_{pred, 1} &= \sum_{k=1}^{N_{tr}} \alpha_k(\rmm^I; \theta_{br})\psi_{k}(x^J; \theta_{tr}) + b_1\,\\
        u_{pred, 2} &= \sum_{k=1}^{N_{tr}} \alpha_{k+N_{tr}}(\rmm^I; \theta_{br})\psi_{k}(x^J; \theta_{tr}) + b_2\,\\
        &\vdots \\
        u_{pred, d_{U}} &= \sum_{k=1}^{N_{tr}} \alpha_{k+(d_{U} - 1)N_{tr}}(\rmm^I; \theta_{br})\psi_{k}(x^J; \theta_{tr}) + b_{d_{U}}\,,
    \end{aligned}
    \end{equation}
    i.e., the first $N_{tr}$ outputs of the branch are used to predict $u_1$ component, the next $N_{tr}$ branch outputs are used to predict $u_2$ component, and so on. In the above, the same trunk outputs are used for all components of the target functions.
\end{remark}

Multi-layer perception (MLP) is used as branch and trunk networks. The implementation used in this work is based on the \lstinline{DeepONet} GitHub repository\footnote{\url{https://github.com/GideonIlung/DeepONet}} with some minor modifications. \cref{lst:mlp} shows the implementation of MLP and \cref{lst:DeepONetImpl} implements the DeepONet framework. 

\vspace{10pt}
\begin{lstlisting}[language=Python, caption={Multi-layer Perceptron (MLP) implementation.}, label={lst:mlp}]

import torch
import torch.nn as nn

class MLP(nn.Module):

    def __init__(self, input_size, hidden_size, num_classes, depth, act):
        super(MLP, self).__init__()
        self.layers = nn.ModuleList()
        self.act = act 

        # input layer
        self.layers.append(nn.Linear(input_size, hidden_size))
        
        # hidden layers
        for _ in range(depth - 2):
            self.layers.append(nn.Linear(hidden_size, hidden_size))
        
        # output layer
        self.layers.append(nn.Linear(hidden_size, num_classes))

    def forward(self, x, final_act=False):
        for i in range(len(self.layers) - 1):
            x = self.act(self.layers[i](x))
        
        # last layer
        x = self.layers[-1](x) 
        if final_act == False:
            return x
        else:
            return torch.relu(x)            
\end{lstlisting}

\begin{lstlisting}[language=Python, caption={Core implementation of the DeepONet architecture.}, label={lst:DeepONetImpl}]
...
import torch
import torch.nn as nn
from torch.utils.data import DataLoader
...
from torch_mlp import MLP
from dataMethods import DataHandler
...

class DeepONet(nn.Module):
        
    def __init__(self, ...):
        super(DeepONet, self).__init__()
        ...
        # branch network
        self.branch_net = MLP(input_size=num_inp_fn_points, hidden_size=num_neurons, \
                              num_classes=num_br_outputs, depth=num_layers, act=act)
        self.branch_net.float()

        # trunk network
        self.trunk_net = MLP(input_size=out_coordinate_dimension, \
                             hidden_size=num_neurons, num_classes=num_tr_outputs, \
                             depth=num_layers, act=act)
        self.trunk_net.float()
        
        # bias added to the product of branch and trunk networks
        self.bias = [nn.Parameter(torch.ones((1,)),requires_grad=True) \
                            for i in range(num_Y_components)]
        ...
    ...    
    def forward(self, X, X_trunk):
        ...
        br_out = self.branch_net.forward(X)
        tr_out = self.trunk_net.forward(X_trunk,final_act=True)
        ntr = self.num_tr_outputs

        if self.num_Y_components == 1:
            output = br_out @ tr_out.t() + self.bias[0]
        else:
            # if d_u > 1, split the branch output and compute the product
            output = []
            for i in range(self.num_Y_components):
                output.append(br_out[:,i*ntr:(i+1)*ntr] @ tr_out.t() + self.bias[i])
            
            # stack and reshape 
            output = torch.stack(output, dim=-1)
            output = output.reshape(-1, X_trunk.shape[0] * self.num_Y_components)

        return output

    def train(self, train_data, test_data, batch_size, epochs, lr, ...):
        ...
        # train and test dataloaders to sample batches of data
        dataset = DataHandler(train_data['X_train'], \
                    train_data['X_trunk'], train_data['Y_train'])
        dataloader = DataLoader(dataset, batch_size=batch_size, shuffle=True) 

        test_dataset = DataHandler(test_data['X_train'], \
                    test_data['X_trunk'], test_data['Y_train'])
        test_dataloader = DataLoader(test_dataset, batch_size=batch_size,shuffle=True)

        # loss and optimizer setup
        criterion = nn.MSELoss()
        optimizer = optim.Adam(self.parameters(), lr=lr, weight_decay=1e-4)
        scheduler = optim.lr_scheduler.StepLR(optimizer, step_size=100, gamma=0.5)
        ...
        # number of trainable parameters in the model
        self.trainable_params = sum(p.numel() for p in \
                        self.parameters() if p.requires_grad)
        ...        
        for epoch in range(1, epochs+1):
            ...
            # training loop
            for X_train, _, Y_train in dataloader:
                ...
                # clear gradients
                optimizer.zero_grad()

                # forward pass through model
                Y_train_pred = self.forward(X_train, X_trunk)

                # compute and save loss
                loss = criterion(Y_train_pred, Y_train)
                train_losses.append(loss.item())

                # backward pass
                loss.backward()

                # update parameters
                optimizer.step()
                
            # update learning rate
            scheduler.step()

            # testing loop
            with torch.no_grad():
                for X_test, _, Y_test in test_dataloader:
                    # forward pass through model
                    Y_test_pred = self.forward(X_test, X_trunk)

                    # compute and save test loss
                    test_loss = criterion(Y_test_pred, Y_test)
                    test_losses.append(test_loss.item())
            # log losses, print progress, save model checkpoints, etc.
            ...
    def predict(self, X, X_trunk):
        with torch.no_grad():
            return self.forward(X, X_trunk)
\end{lstlisting}

\subsubsection{Architecture and preliminary results}\label{sss:DeepONet}
A three-layer fully connected network is considered for both the branch and trunk networks. Neural network and optimization-related parameters are tabulated in \cref{tab:parameters}. For the linear elasticity and hyperelasticity problems, $N_{br} = 200$ and $N_{tr} = 100$, and the prediction of vector-valued outputs is based on \cref{rem:vectorvalued}.

\begin{figure}[h!]
    \centering
    \includegraphics[width=0.95\linewidth]{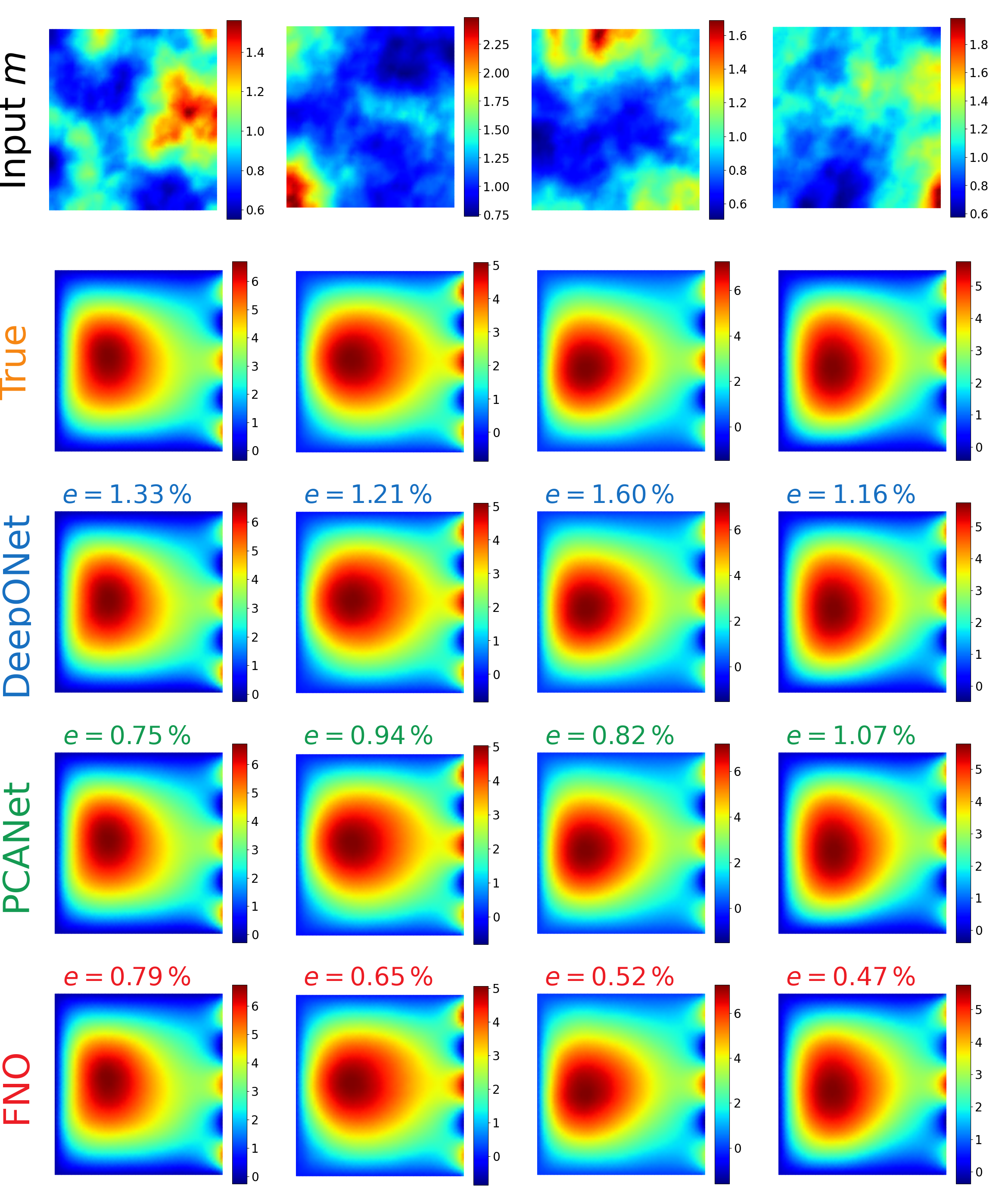}
    \caption{Comparison of DeepONet, PCANet, and FNO predictions with the finite element solution for four samples of diffusivity in the Poisson problem. Here, $e$ denotes the relative percentage $\ell^2$ error between the true finite element solution and the surrogate solution.}
    \label{fig:poissonCompareNOps}
\end{figure}

For the Poisson problem, \cref{fig:poissonCompareNOps} shows the typical prediction error using DeepONet. In \cref{fig:elasticityCompareNOps} and \cref{fig:hyperelasticityCompareNOps}, the results for the linear elasticity and hyperelasticity problems are shown. These figures also compare the accuracy of PCANet and FNO, which will be discussed in the following two subsections. The notebooks \lstinline{DeepONet-Poisson}\footnote{\url{https://github.com/CEADpx/neural_operators/blob/survey26_v2/survey_work/problems/poisson/DeepONet/training_and_testing.ipynb}}, \lstinline{DeepONet-Linear Elasticity}\footnote{\url{https://github.com/CEADpx/neural_operators/blob/survey26_v2/survey_work/problems/linear_elasticity/DeepONet/training_and_testing.ipynb}}, and \lstinline{DeepONet-Hyperelasticity}\footnote{\url{https://github.com/CEADpx/neural_operators/blob/survey26_v2/survey_work/problems/hyperelasticity/DeepONet/training_and_testing.ipynb}} describe the steps used to construct, train, and test the DeepONet models for the three problems.

\vspace{10pt} 
\begin{table}[!h]
    \centering
    \begin{tabular}{|p{0.12\linewidth}|c|p{0.65\linewidth}|}
            \hline
         {\bf Parameter} & {\bf Value} & {\bf Description} \\
         \hline
         Layers $L$ & $3$ & Number of layers in DeepONet (branch and trunk each), PCANet, and FNO \\
         Width & $64$ & Number of neurons per hidden layer (DeepONet)\\
         Width & $300$ & Number of neurons per hidden layer (PCANet) \\
         $d_h$ & 22 & Dimension of FNO layer outputs \\
         $N_{train}$ & $3200$ & Number of training data pairs $(\rmm^I, \rmu^I)$ \\
         $N_{test}$ & $800$ & Number of testing data pairs $(\rmm^I, \rmu^I)$ \\
         $N_m$ & $3200$ & $N_m$ is same as $N_{train}$ (notation used for DeepONet) \\
         $N_x$ & $2601$ & Number of coordinates for evaluation of $u$ for DeepONet \\
         $p_{M}$, $p_{U}$ & $2601$, $2601 d_{U}$ & Dimensions of discretized functions ($d_{U} = 2$ for elasticity) \\
         $r_{M}$, $r_{U}$ & $200$, $100$ & Reduced dimensions for the three problems for PCANet \\
         $N_{br}$, $N_{tr}$ & $100 d_U$, $100$ & Number of branch and trunk outputs in DeepONet \\
         Batch & $20$ & Neural networks are trained on ``batch size'' samples \\
         Epochs & $1000$ & Number of optimization steps \\
         Epochs & $400$ & Number of optimization steps (FNO) \\
         LR & $0.001$ & Learning rate: controls parameter update during training \\
         Activation & ReLU & Activation function \\
         $k_{max}$ & 8 & Number of Fourier modes to keep in FNO \\
         $n_1$, $n_2$ & 51, 51 & Number of grid points in $D_{U} = (0,1)^2$ for FNO\\
         \hline 
    \end{tabular}
    \caption{Summary of parameters used in neural operator training and testing.}
    \label{tab:parameters}
\end{table}
\vspace{10pt} 

\subsection{Principal Component Analysis-based Neural Operator (PCANet)}\label{ss:pcanet}
The second neural operator of interest is the PCANet introduced in \cite{BhattacharyaHosseiniKovachkiEtAl2020}, which utilizes Principal Component Analysis (PCA) to reduce the dimensions of input and output discretized functions and poses an operator learning problem in the reduced-dimensional spaces, thereby making learning more efficient. Consider $\{(\rmm^I, \rmu^I = \rmF(\rmm^I))\}_{I=1}^N$ set of paired data, where $\rmm^I \in \Rpow{p_{M}}$ and $\rmu^I \in \Rpow{p_{U}}$. In general, learning the approximation of the target map $\rmF: \Rpow{p_{M}} \to \Rpow{p_{U}}$ becomes challenging if $p_{M}$ and $p_{U}$ are large. PCANet alleviates this challenge by introducing a low-dimensional approximation of $\rmF$. Specifically, in PCANet, the dimensions of input and output functions are reduced using SVD, and the neural network between reduced-dimensional inputs and outputs is introduced to approximate the mapping. The effectiveness of this approach depends on whether the input and output data admit accurate low-dimensional representations, which is typically reflected in the decay of the singular values of the corresponding data matrices. When the singular values decay slowly, a much larger reduced dimension may be required, and the efficiency and accuracy advantages of PCANet can be reduced.

To make this precise, let $r_{M}$ and $r_{U}$ denote the reduced dimensions for input and output functions, respectively, and $\tilde{\rmP}_{M} \in \Rpow{r_{M}\times p_{M}}$ and $\tilde{\rmP}_{U} \in \Rpow{r_{U}\times p_{U}}$ are the projectors based on SVD for dimension reduction. Now consider a parameterized neural network map $\tilde{\rmF}_{\theta} : \Rpow{r_{M}} \to \Rpow{r_{U}}$ to construct the approximation $\rmF_{NOp}$ of $\rmF$ as follows, for a given $\rmm\in \Rpow{p_{M}}$, 
\begin{equation}\label{eq:PCANet}
    \Rpow{p_{U}} \ni \;\rmu = \rmF(\rmm) \qquad \approx \qquad \rmF_{NOp}(\rmm) = \underbrace{\tilde{\rmP}_{U}^T \left( \underbrace{\tilde{\rmF}_{\theta} \left(\underbrace{\tilde{\rmP}_{M}(\rmm)}_{=: \tilde{\rmm} \text{ (project $\rmm$)}})\right)}_{=: \tilde{\rmu} \text{ (low-dim map)}} \right)}_{\text{lift $\tilde{\rmu}$}} \; \in \Rpow{p_{U}} \,.
\end{equation} 
Conceptually, the parameters $\theta$ can be determined via the optimization problem:
\begin{equation}\label{eq:pcanetOpt}
    \theta^\ast = \argmin_{\theta \in \Rpow{p_{\Theta}}} \frac{1}{N} \sum_{I=1}^N \norm{\rmF(\rmm^I) - \tilde{\rmP}_{U}^T\left(\tilde{\rmF}_\theta \left( \tilde{\rmP}_{M} (\rmm^I)\right) \right)}^2\,,
\end{equation}
where $\norm{\cdot}$ denotes the $\ell^2$-norm and $p_\Theta$ is the number of trainable parameters. Here, the projection operators are fixed, and the only learnable component is the reduced map $\tilde{\rmF}_\theta$.

Thus, unlike DeepONet, which learns basis representations explicitly, PCANet relies on fixed data-driven bases obtained via SVD and focuses on learning the mapping in the reduced space. In practice, the optimization problem \eqref{eq:pcanetOpt} is reformulated in the reduced space, i.e., the parameters $\theta$ in $\tilde{\rmF}_\theta$ are trained using projected data. \cref{fig:schematicsNOp} presents the schematic of PCANet and the associated projection steps.

\subsubsection{Implementation of PCANet}
Let $\rb{X}$ and $\rb{Y}$ are two $N\times p_{M}$ and $N\times p_{U}$ matrices, respectively, such that rows of $\rb{X}$ and $\rb{Y}$ are input and output function samples, $\rmm^I \in \Rpow{p_{M}}$ and $\rmu^I\in \Rpow{p_{U}}$, where $1 \leq I \leq N$. The data for the neural network-based map $\tilde{\rmF}_\theta$ in \eqref{eq:PCANet} is constructed using the following steps, which include the preprocessing to construct SVD-based projectors: 
\begin{enumerate}
    \item \textit{Center and scale} the data matrices $\rb{X}$ and $\rb{Y}$ by subtracting the sample mean and dividing elementwise by the sample standard deviation. For example, if $\bar{\rb{X}}$ and $\sigma_X$ are $1\times p_{M}$ row vectors containing the mean and standard deviation of the columns of $\rb{X}$, then the normalized input data $\hat{\rb{X}}$ is obtained by
    \begin{equation*}
        \hat{\rb{X}} = \frac{\rb{X} - \bar{\rb{X}}}{\sigma_X + \mathrm{tol}}\,,
    \end{equation*}
    where each row of $\rb{X}$ is subtracted by the row vector $\bar{\rb{X}}$, and the division by $\sigma_X+\mathrm{tol}$ is performed elementwise. The small number $\mathrm{tol}$ is introduced to ensure numerical stability. The normalized output data $\hat{\rb{Y}}$ is obtained in the same way.

    \item \textit{SVD projectors} for input and output data are determined following the procedure in \cref{sss:svd}. For example, take $\rmR = \hat{\rb{X}}^T$ and compute the projector $\tilde{\rmP}_{M}\in \Rpow{r_{M}\times p_{M}}$, where $r_{M}$ is the specified reduced dimension. Similarly, the projector $\tilde{\rmP}_{U}$ is obtained using $\rmR = \hat{\rb{Y}}^T$.
    
    \item \textit{Projected data for the neural network} are computed by projecting the rows of $\hat{\rb{X}}$ into the reduced space. To be specific, let the $I^{\text{th}}$ row of $\hat{\rb{X}}$ be $(\hat{\rmm}^I)^T \in \Rpow{1\times p_{M}}$, where $\hat{\rmm}^I$ is the centered and scaled $I^{\text{th}}$ input sample. Its projection is $\tilde{\rmm}^I = \tilde{\rmP}_{M}\left(\hat{\rmm}^I\right) \in \Rpow{r_{M}}$. Using this projection, a new matrix $\tilde{\rb{X}}$ of size $N\times r_{M}$ is formed, where $(\tilde{\rmm}^I)^T \in \Rpow{1\times r_{M}}$ is the $I^{\text{th}}$ row. Similarly, $\tilde{\rb{Y}}$ of size $N\times r_{U}$ is obtained by applying $\tilde{\rmP}_{U}$ to each centered and scaled output sample.
    
    \item For the reduced-dimensional map $\tilde{\rmF}_\theta:\Rpow{r_{M}}\to \Rpow{r_{U}}$, defined by $\tilde{\rmu}=\tilde{\rmF}_\theta(\tilde{\rmm})$ (see \eqref{eq:PCANet}), the matrices $\tilde{\rb{X}}$ and $\tilde{\rb{Y}}$ constitute the input and output data, respectively.
\end{enumerate}

Given the projected data matrices $\tilde{\rb{X}}$ and $\tilde{\rb{Y}}$ and a neural network-based map $\tilde{\rmF}_\theta$, the optimization problem for determining $\theta$ is
\begin{equation}
    \theta^\ast = \argmin_{\theta \in \Rpow{p_{\Theta}}} \frac{1}{N} \sum_{I=1}^N \left\vert\left\vert \tilde{\rb{Y}}_{I} - \tilde{\rmF}_\theta(\tilde{\rb{X}}_{I})\right\vert\right\vert^2\,,
\end{equation}
where $(\cdot)_{I}$ denotes the $I^\text{th}$ row of a matrix. Once $\theta^\ast$ is determined, the map $\tilde{\rmF}_{\theta^\ast}$ can be used as in \eqref{eq:PCANet} to obtain the neural operator surrogate $\rmF_{NOp}$ of the target operator $\rmF$. The core steps in implementing PCANet, i.e., $\tilde{\rmF}_\theta$, are shown in \cref{lst:PCANetImpl}.

\vspace{10pt}
\begin{lstlisting}[language=Python, caption={PCANet implementation. Its interface is similar to that of DeepONet, except that PCANet does not take spatial coordinates as input.}, label={lst:PCANetImpl}]
...
from torch_mlp import MLP
...

class PCANet(nn.Module):
    
    def __init__(self, ...):
        super(PCANet, self).__init__()
        ...
        self.net = MLP(input_size=num_inp_red_dim, hidden_size=num_neurons, \
                            num_classes=num_out_red_dim, depth=num_layers, act=act)
        self.net.float()
        ...

    def forward(self, X):
        X = self.convert_np_to_tensor(X)
        return self.net.forward(X)

    def train(self, train_data, test_data, batch_size, epochs, lr, ...):
        ...
        # train and test dataloaders to sample batches of data
        dataset = DataHandler(train_data['X_train'], None, train_data['Y_train'])
        dataloader = DataLoader(dataset, batch_size=batch_size, shuffle=True) 

        test_dataset = DataHandler(test_data['X_train'], None, test_data['Y_train'])
        test_dataloader = DataLoader(test_dataset, batch_size=batch_size,shuffle=True)

        # rest is similar to DeepONet in Listing 7 without X_trunk
        ...
        
    def predict(self, X):
        with torch.no_grad():
            return self.forward(X)
\end{lstlisting}

\begin{figure}[h!]
\centering
\subfloat[\centering Poisson equation\label{fig:svdSub1}]{
  \begin{minipage}{.48\textwidth}
  \centering
  \includegraphics[width=\linewidth]{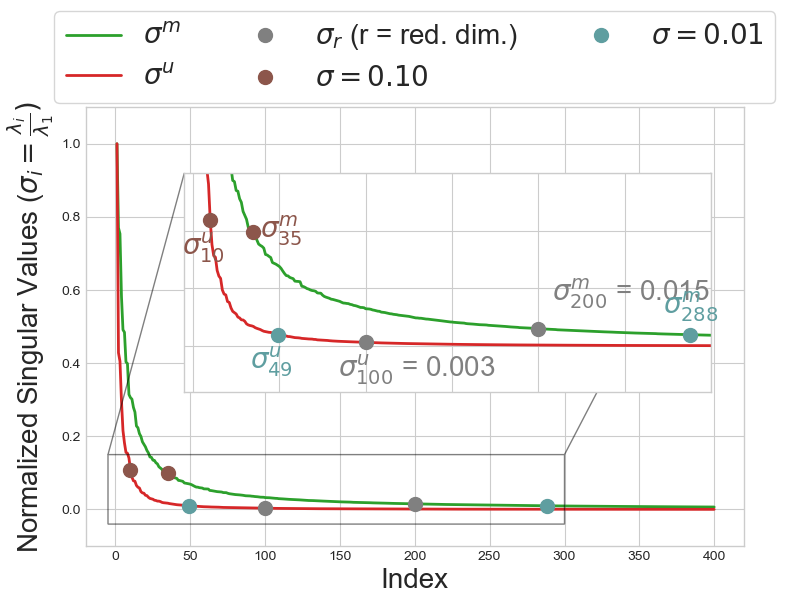}
  \end{minipage}
}%
\subfloat[\centering Linear elasticity\label{fig:svdSub2}]{
  \begin{minipage}{.48\textwidth}
  \centering
  \includegraphics[width=\linewidth]{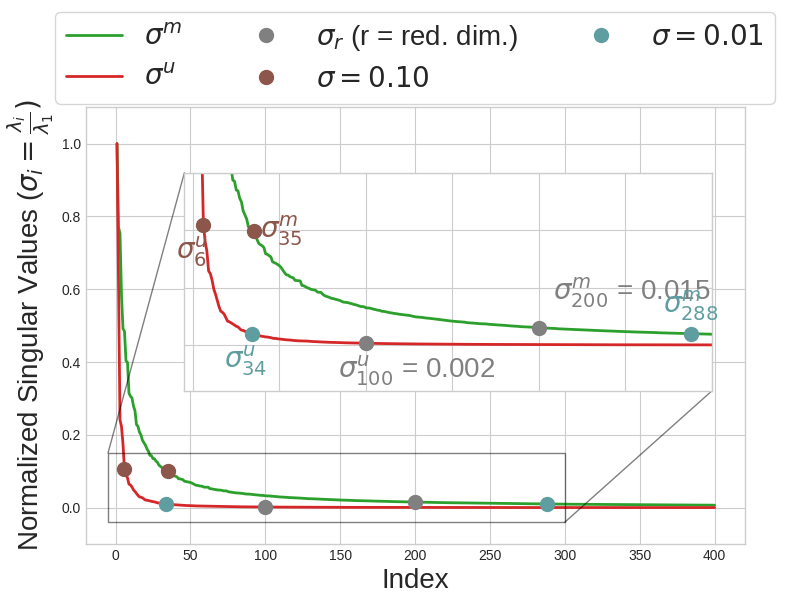}
  \end{minipage}
}
\caption{Normalized singular values of the centered and scaled input and output data matrices $\hat{\rb{X}}$ and $\hat{\rb{Y}}$. The green and red curves represent the normalized singular values of the input and output data, respectively. For both curves, annotations are shown near the $10\%$ (brown dots) and $1\%$ (blue dots) levels, i.e., $0.1$ and $0.01$ fractions of the largest singular value. The grey dots correspond to the fixed reduced dimensions used in PCANet. The rapid decay of singular values shows that the Poisson and linear elasticity problems are inherently low-dimensional. For all problems, the input samples are drawn from the same distribution using the same random-number-generator seed; therefore, the singular-value curves for the input data are the same in the left and right panels. The hyperelasticity problem exhibits a similar spectral decay to the linear elasticity problem.}
\label{fig:svd}
\end{figure}

\subsubsection{Architecture and preliminary results}\label{sss:PCANet}
A fully connected neural network with three layers is used to test PCANet. Other parameters, including the reduced dimensions, are listed in \cref{tab:parameters}. First, the singular values of the input and output data are analyzed in \cref{fig:svd} to identify reasonably small reduced dimensions without significantly compromising accuracy. For all three problems, both input and output data show a rapid decay of singular values. Since the input data exhibit a relatively slower decay, the reduced dimension for the input data is selected to be higher than that for the output data; see \cref{tab:parameters}.

The predictions from trained PCANet are compared with the finite element solution and the other two neural operators in \cref{fig:poissonCompareNOps} for the Poisson problem. The results for the linear elasticity and hyperelasticity problems are shown in \cref{fig:elasticityCompareNOps} and \cref{fig:hyperelasticityCompareNOps}, respectively. The notebooks \lstinline{PCANet-Poisson}\footnote{\url{https://github.com/CEADpx/neural_operators/blob/survey26_v2/survey_work/problems/poisson/PCANet/training_and_testing.ipynb}}, \lstinline{PCANet-Linear Elasticity}\footnote{\url{https://github.com/CEADpx/neural_operators/blob/survey26_v2/survey_work/problems/linear_elasticity/PCANet/training_and_testing.ipynb}}, and \lstinline{PCANet-Hyperelasticity}\footnote{\url{https://github.com/CEADpx/neural_operators/blob/survey26_v2/survey_work/problems/hyperelasticity/PCANet/training_and_testing.ipynb}} describe the construction, training, and testing of PCANet models for the three problems.

\subsection{Fourier Neural Operator (FNO)}\label{ss:fno}
Fourier neural operator considers the composition of layers, with a typical layer involving an affine transformation and an integral kernel operator followed by nonlinear activation. These affine and integral kernel operations are parameterized. While there are multiple choices of integral kernel operator \cite{kovachki2023neural}, this work uses the Fourier transform. Consider the case of $D_{M} = D_{U} = D \subset \Rpow{q}$ (therefore, $q_{M} = q_{U} = q$) and the neural operator approximation $F_{NOp}$ such that
\begin{equation}\label{eq:fnoGeneral}
    u(x) = F(m)(x) \quad \approx \quad F_{NOp}(m)(x) := Q \left( \underbrace{G_L \left(G_{L-1} \cdots G_1\left(\underbrace{P \left(m(x)\right)}_{= z_0 \in \Rpow{d_h}}\right)\right)}_{=z_L(x) \in \Rpow{d_h}}\right) \,,
\end{equation}
or, concisely,
\begin{equation}
    F_{\mathrm{NOp}} = Q\circ G_L \circ G_{L-1} \circ \cdots \circ G_1 \circ P\,,
\end{equation}
where $\circ$ denotes the composition of maps. The map $F_{NOp}$ involves the following operations:
\begin{itemize}
    \item Lifting $m(x) \in \Rpow{d_{M}}$ to $\Rpow{d_h}$, where $d_h$ is the dimension of outputs from hidden layers via the trainable matrix $P$ of size $d_h \times d_{M}$.
    \item Projecting the final hidden-layer output $z_L(x) \in \Rpow{d_h}$ onto $\Rpow{d_{U}}$ (space of $u(x)$) via the $d_{U} \times d_h$ trainable matrix $Q$. Thus, $z_L$ is a $d_h$-dimensional feature field defined over the domain $D$.
    \item Application of operators $G_l$, $1\leq l \leq L$, where $G_l$ is defined via:
    \begin{equation}\label{eq:fnoLayerGeneral}
        \Rpow{d_h} \ni z_l = G_l(z_{l-1}) = \sigma_l \left(\underbrace{W_l z_{l-1} + b_l}_{\text{linear/local operation}} + \underbrace{K_l(z_{l-1})}_{\text{nonlocal operation}}\right)\,.
    \end{equation}
    Here, $z_l = z_l(x)$ denotes a $d_h$-dimensional feature field defined over the domain $D$. Note that $z_{l-1}$ is the output of the preceding layer, $\sigma_l$ activation function, $W_l$ weight matrix, $b_l$ a bias vector, and $K_l(\cdot)$ an integral kernel operator. 
\end{itemize}

Up until now, the map $F_{NOp}$ in \cref{eq:fnoGeneral} with the above definitions of $P$, $Q$, $G_l$, $l=1,...,L$, is abstract due to the generality of $K_l$. The linear operation in \eqref{eq:fnoLayerGeneral} captures the local effects on the reconstructed function, while the $K_l$ is designed to capture the non-local effects (interaction with other degrees of freedom) via the integral kernel operation. There are various choices for $K_l$, as discussed in \cite{kovachki2023neural}. For example:
\begin{itemize}
    \item[(1)] Low-rank Neural Operator (LNO) in which $K_l$ takes the form 
    $$K_l(z) = \sum_{j=1}^r \innerProd{\psi^{(i)}}{z} \varphi^{(i)}(x),$$
    where $\varphi^{(i)}$ and $\psi^{(i)}$ are some parameterized functions;
    \item[(2)] Graph Neural Operator (GNO) in which 
    $$K_l = \frac{1}{|B(x, r)|} \sum_{y_i \in B(x,r)} k(x, y_i) z(y_i),$$
    where $k(\cdot, \cdot)$ is a kernel function; and 
    \item[(3)] Fourier Neural Operator (FNO) maps $z$ to Fourier space, followed by mapping the weighted Fourier modes back to the real space. Since FNO is the main focus of this article, it is discussed in more detail next. Readers are referred to \cite{li2020fourier, kovachki2023neural} for further discussion of FNO and associated ideas.
\end{itemize}

In FNO, $K_l(z)$, for $z : D \to \Rpow{d_h}$, takes the form:
\begin{equation}
    K_l(z) = \calF^{-1}\left( R_l \calF(z)\right)\,,
\end{equation}
where
\begin{itemize}
    \item $\calF(z)$ is the Fourier transform applied to each component of the feature field $z(x)$ over the spatial domain. Only the first $k_{max}$ modes are retained, so the output of $\calF(z)$ is in $\bbC^{d_h \times k_{max} \times k_{max}}$.
    \item $R_l$ applies the weighting to different Fourier modes. $R_l$ is a complex-valued $d_h \times d_h \times k_{max} \times k_{max}$ trainable tensor and $R_l \calF(z) \in \bbC^{d_h \times k_{max} \times k_{max}}$. 
    \item $\calF^{-1}(\cdot) \in \Rpow{d_h}$ is the inverse Fourier transform applied component-wise to reconstruct the feature field in physical space. 
\end{itemize}

\cref{fig:schematicsNOp} displays the FNO framework based on Fourier transforms. In practice, in the discrete setting, one evaluates the FNO output at all grid points to have
\begin{equation}\label{eq:fno}
    \Rpow{p_{U}} \ni \rmu = \rmF(\rmm) \quad \approx \quad \rmF_{NOp}(\rmm) := Q \left( G_L\left(G_{L-1} \cdots G_1\left(P(\rmm)\right) \right) \right)\,,\quad \text{for } \rmm \in \Rpow{p_{M}}\,.
\end{equation}
The optimization problem to determine $\theta$ is given by
\begin{equation}
    \theta^\ast = \argmin_{\theta \in \Rpow{p_\Theta}} \frac{1}{N_m} \sum_{I=1}^{N_m} \left\|\rmu^I - \rmF_{NOp}(\rmm^I)\right\|^2\,,
\end{equation}
where $p_\Theta$ is the number of trainable parameters in FNO.

\subsubsection{Implementation of FNO}
The implementation of FNO requires function values at grid locations, and thus, preprocessing is required to obtain these values at grid points. Suppose $D_{M} = D_{U} = D = (0,1)^2$ and consider the grid division of $\mathrm{closure}(D)$ consisting of $n_{1}$ and $n_{2}$ number of points in $x^1$ and $x^2$ directions, respectively. Linear interpolation is used to map finite element solutions onto a structured grid required by FNO. The following describes the data:
\begin{enumerate}
    \item Let $\rb{X}$ be an $N\times n_{1} \times n_{2} \times 3$ matrix, where the outer dimension corresponds to the number of samples. The element of $\rb{X}$ at $I^{\text{th}}$ outer index is a $n_{1} \times n_{2} \times 3$ matrix containing the interpolated values of a function $m^I$ (or discretization of this function given by $\rmm^I$ on a mesh) at all grid $(x^1, x^2)$ points and the coordinates $x^1$ and $x^2$ of all grid points; thus, the inner dimension is three. The spatial coordinates are included to allow the network to learn non-translation-invariant features. Next, the function values are centered and scaled using the mean and standard deviation computed from the samples $1\leq I\leq N$. 
    \item Let $\rb{Y}$ be $N\times n_1 \times n_2\times d_{U}$ matrix such that the element of $\rb{Y}$ at $I^{\text{th}}$ outer index is a $n_{1} \times n_{2} \times d_{U}$ matrix containing the interpolated values of $d_{U}$-valued function $u^I$ (or discretization of this function given by $\rmu^I$ on a mesh) at all grid points. Note that the elements of $\rb{X}$ and $\rb{Y}$ at the $I^{\text{th}}$ outer index correspond to $\rmm^I$ and $\rmu^I = \rmF(\rmm^I)$ on a grid, thereby establishing the correspondence between input and output. Function values at grid points are also centered and scaled using the sample mean and standard deviation. 
\end{enumerate}

Given a sample from $\rb{X}$, denoted $\rmx = (m^I(x^1, x^2), x^1, x^2) \in \Rpow{3}$, it is first lifted into $\Rpow{d_h}$, where $d_h > 3$ is a dimension of the input and output spaces of FNO layers, to get $z_0 = P(\rmx) = W_P\, \rmx + b_P$, where $W_P \in \Rpow{d_h\times 3}$ and $b_P \in \Rpow{d_h}$. This representation is applied pointwise at each grid location, and the full input consists of all such values arranged over the grid. Next, $z_0$ goes through $L$ layers such that given $z_{l-1}$ an output of $(l-1)^{\text{th}}$ layer, $z_{l} = G_{l}(z_{l-1})$. The \cref{lst:FNOLayer} presents the implementation of an FNO layer $G_l$ based on the \lstinline{Operator-Learning} repository\footnote{\url{https://github.com/Zhengyu-Huang/Operator-Learning}} \cite{deHoop2022cost}. The $l^{\text{th}}$ layer involves a linear transformation $z_l^1 = W_l z_{l-1} + b_l$ and Fourier-based transformation 
\[
z_l^2 = \calF^{-1} \big( R_l \, \calF(z_{l-1}) \big),
\]
where $R_l$ is a complex matrix of size $d_h \times d_h \times k_{max} \times k_{max}$, $k_{max}$ denoting the number of Fourier modes that are retained after the Fourier transform. Here, $W_l$ and $b_l$ are the weight and bias parameters, respectively. The output of the final layer $z_L = G_L(z_{L-1}) \in \Rpow{d_h}$ represents a $d_h$-dimensional feature field evaluated at all grid points and is projected to $u_{NOp}(x^1, x^2) = Q(z_L) \in \Rpow{d_{U}}$; the projection operator $Q$ introduces $W_Q \in \Rpow{d_{U}\times d_h}$ and $b_Q \in \Rpow{d_{U}}$ parameters. In \cref{lst:FNO}, lift, FNO layer applications, and projection are combined to create an FNO model; see the \lstinline{forward} method. The trainable parameters are as follows:
\begin{equation}\label{eq:fnoParams}
\begin{aligned}
\theta := &\left\{  (W_P, b_P) \in \Rpow{d_h \times 3}\times \Rpow{d_h}\,,\quad (W_Q, b_Q) \in \Rpow{d_{U}\times d_h}\times \Rpow{d_{U}}\,, \right. \\
    &\quad \left. \left\{ W_l \in \Rpow{d_h\times d_h}\,, b_l\in \Rpow{d_h}\,, R_l \in \bbC^{d_h\times d_h\times k_{max}\times k_{max}}\,, \quad 1\leq l \leq L\right\}\right\}\,.
\end{aligned}
\end{equation}
It should be noted that extending the above to a vector-valued function (for a linear elasticity problem) as input and output is relatively straightforward. 

Before writing the loss function, note that while an input to FNO is a triplet $(m(x^1, x^2), x^1, x^2)$, during training and testing, the input is processed as a batch of samples, each defined over the full spatial grid. The FNO is applied to the input matrix $\rb{X}$ of size $N\times n_1\times n_2\times 3$ altogether, i.e., $N$ samples of $m^I$ and grid locations. FNO produces $N\times n_1 \times n_2 \times d_{U}$ outputs, corresponding to $N$ number of $d_{U}$-valued functions $u^I$ at all grid locations. Noting this, the optimization problem to train parameters $\theta$ reads:
\begin{equation}
    \theta^\ast = \argmin_{\theta \in \Rpow{p_\Theta}} \frac{1}{N\, n_1\, n_2} \sum_{I=1}^N \sum_{j=1}^{n_1} \sum_{k=1}^{n_2} \left\| u^I(x^1_{jk}, x^2_{jk}) - F_{NOp}(m^I)(x^1_{jk}, x^2_{jk}) \right\|^2\,.
\end{equation}

\vspace{5pt}
\begin{lstlisting}[language=Python, caption={Implementation of an FNO layer based on the Operator-Learning repository \cite{deHoop2022cost}.}, label={lst:FNOLayer}]
import torch
import torch.nn as nn
import torch.nn.functional as nnF

class FNO2DLayer(nn.Module):
    def __init__(self, in_channels, out_channels, modes1, modes2, apply_act, act):
        super(FNO2DLayer, self).__init__()
        ...
        # parameters in nonlocal transformation
        self.scale = (1 / (in_channels * out_channels))
        self.weights1 = nn.Parameter(self.scale * torch.rand(in_channels, \
                out_channels, self.modes1, self.modes2, dtype = torch.cfloat))
        self.weights2 = nn.Parameter(self.scale * torch.rand(in_channels, \
                out_channels, self.modes1, self.modes2, dtype = torch.cfloat))

        # parameters in linear transformation
        self.w = nn.Conv2d(self.out_channels, self.out_channels, 1)

    # complex multiplication
    def compl_mul2d(self, a, b):
        # (batch, in_channel, x,y ), (in_channel, out_channel, x,y) 
        # -> (batch, out_channel, x,y)
        op = torch.einsum("bixy,ioxy->boxy", a, b)
        return op

    def fourier_transform(self, x):
        batchsize = x.shape[0]
        
        # compute Fourier coefficients
        x_ft = torch.fft.rfft2(x)

        # multiply retained Fourier modes
        out_ft = torch.zeros(batchsize, self.out_channels,  x.size(-2), \
                x.size(-1)//2 + 1, device=x.device, dtype=torch.cfloat)
        out_ft[:, :, :self.modes1, :self.modes2] = \
            self.compl_mul2d(x_ft[:, :, :self.modes1, :self.modes2], self.weights1)
        out_ft[:, :, -self.modes1:, :self.modes2] = \
            self.compl_mul2d(x_ft[:, :, -self.modes1:, :self.modes2], self.weights2)
        
        # return to physical space
        x = torch.fft.irfft2(out_ft, s=(x.size(-2), x.size(-1)))
        return x
    
    def linear_transform(self, x):
        return self.w(x)
    
    def forward(self, x):
        x = self.fourier_transform(x) + self.linear_transform(x)
        if self.apply_act:
            return self.act(x)
        else:
            return x
\end{lstlisting}

\begin{lstlisting}[language=Python, caption={Core implementation of the FNO architecture.}, label={lst:FNO}]
...
from torch_fno2dlayer import FNO2DLayer

class FNO2D(nn.Module):
    def __init__(self, num_layers, width, fourier_modes1, fourier_modes2, \
                 num_Y_components, save_file=None):
        super(FNO2D, self).__init__()
        ...
        # create hidden layers (FNO layers)
        self.fno_layers = nn.ModuleList()
        for _ in range(num_layers):
            self.fno_layers.append(FNO2DLayer(self.width, self.width, \
                                            self.fourier_modes1, self.fourier_modes2))
        
        # no activation in the last hidden layer
        self.fno_layers[-1].apply_act = False 

        # define input-to-hidden projector: input has 3 components: m(x,y), x, y
        self.input_projector = nn.Linear(3, self.width)

        # define hidden-to-output projector
        self.output_projector = nn.Linear(self.width, self.num_Y_components)
        ...
    ...
    def forward(self, X):
        x = self.convert_np_to_tensor(X)

        # input-to-hidden projector
        x = self.input_projector(x)
        
        # rearrange x so that it has the shape (batch, width, x, y)
        x = x.permute(0, 3, 1, 2)
        
        # pass through hidden layers
        for i in range(self.num_layers):
            x = self.fno_layers[i](x)
        
        # rearrange x so that it has the shape (batch, x, y, width)
        x = x.permute(0, 2, 3, 1)

        # hidden-to-output projector
        x = self.output_projector(x)

        return x
    
    def train(self, train_data, test_data, batch_size, epochs, lr, ...):
        # similar to the train() of PCANet in Listing 8 and DeepONet in Listing 7
        
    def predict(self, X):
        # same as the predict() of PCANet in Listing 8
\end{lstlisting}

\subsubsection{Architecture and preliminary results}\label{sss:FNO}
In the implementation, three FNO layers are considered with hidden-layer output dimension $d_h = 22$. Other relevant parameters are listed in \cref{tab:parameters}. \cref{fig:poissonCompareNOps}, \cref{fig:elasticityCompareNOps}, and \cref{fig:hyperelasticityCompareNOps} display FNO predictions for test input samples for the three model problems, together with the corresponding finite element solutions. FNO training and testing for the three problems are implemented in the following notebooks: \lstinline{FNO-Poisson}\footnote{\url{https://github.com/CEADpx/neural_operators/blob/survey26_v2/survey_work/problems/poisson/FNO/training_and_testing.ipynb}}, \lstinline{FNO-Linear Elasticity}\footnote{\url{https://github.com/CEADpx/neural_operators/blob/survey26_v2/survey_work/problems/linear_elasticity/FNO/training_and_testing.ipynb}}, and \lstinline{FNO-Hyperelasticity}\footnote{\url{https://github.com/CEADpx/neural_operators/blob/survey26_v2/survey_work/problems/hyperelasticity/FNO/training_and_testing.ipynb}}.

\section{Neural operators for Bayesian inference}\label{s:bayesian}
Bayesian inference is a powerful framework for solving inverse problems in which the goal is to infer unknown parameters or fields from observed data. In PDE-constrained inverse problems, posterior sampling procedures such as Markov chain Monte Carlo (MCMC) require repeated evaluations of the forward model, and the computational cost is often dominated by the associated likelihood evaluations. Neural operators can therefore serve as efficient surrogate forward models, accelerating these evaluations while maintaining acceptable approximation accuracy. The goal of this section is to demonstrate the integration of neural operator surrogates into the Bayesian inference framework and to evaluate the resulting inference performance.

We first formulate the Bayesian inverse problem in an abstract infinite-dimensional setting, following \cite{Dashti2017, bui2013computational, stuart2010inverse}. We then present the inference results for the Poisson problem in the main text, while the corresponding results for the linear elasticity and hyperelasticity models are provided in \labelcref{s:appBayesianInference}. The neural operator architectures and training procedures are described in the previous section, and the trained models are used here as surrogate forward models within the Bayesian inference problems.

\subsection{Abstract Bayesian inference problem in infinite dimensions}\label{ss:abstractBayesian}
Consider a parameter field $m \in M$ to be inferred from observational data $\rmg \in \Rpow{d_{\rmg}}$ and the corresponding solution of the PDE, $u = F(m) \in U$. Let $\bar{\rb{B}}: U \to \Rpow{d_{\rmg}}$ denote the state-to-observable map such that $\bar{\rb{B}}(u)$ gives the predicted observations corresponding to the state $u$. The Bayesian inverse problem is formulated through the relation
\begin{equation}\label{eq:dataAndModel}
    \rmg = \bar{\rb{B}}(u) + \eta = \bar{\rb{B}}(F(m)) + \eta\,,
\end{equation}
where $\eta \sim \rmN({\rm{0}}, \Gamma_{\rmg})$ represents observational and modeling noise, and $\Gamma_{\rmg} \in \Rpow{d_{\rmg} \times d_{\rmg}}$ is the covariance matrix. For simplicity, the covariance matrix is taken to be $\Gamma_{\rmg} = \sigma_{\rmg}^2 \rmI$, where $\sigma_{\rmg}^2$ is the noise variance and $\rmI$ is the identity matrix.

In many applications, the admissible parameter space $M_{ad}$ is a subset of $M$ incorporating physical constraints. For example, diffusivity and Young's modulus must remain strictly positive. A transformation $m = m(w)$ with $w \in W$ is used to enforce the positivity constraint:
\begin{equation}\label{eq:transformW}
    m(x) = \alpha_m \exp(w(x)) + \beta_m\,, \quad x \in D_{W}\,,
\end{equation}
where $\alpha_m$ and $\beta_m$ are constants. The inference problem is thus posed in terms of $w \in W$, where $W$ is a Hilbert space. 

Define the parameter-to-observable map $\rb{B}: W \to \Rpow{d_{\rmg}}$ by
\begin{equation}
\rb{B}(w) = \bar{\rb{B}}(F(m(w)))\,.
\end{equation}
Using $\rb{B}$, the goal is to infer the probability distribution of $w$ from
\begin{equation}\label{eq:dataAndModelFinal}
    \rmg = \rb{B}(w) + \eta\,.
\end{equation}
Together with the noise model, the above equation defines the conditional probability distribution of the data $\rmg$ given $w$. Since the noise is assumed to be Gaussian, $\eta \sim \rmN({\rm{0}}, \Gamma_{\rmg})$, the likelihood, i.e., the distribution of $\rmg$ conditioned on $w$, is also Gaussian: $\rmg \mid w \sim \rmN(\rb{B}(w), \Gamma_{\rmg})$. Explicitly,
\begin{equation}
    \pi_{like}(\rmg \mid w) =
    \frac{1}{\sqrt{\det(\Gamma_{\rmg})} \left(2\pi\right)^{\frac{d_{\rmg}}{2}}}
    \exp \left[ -\frac{1}{2} (\rmg - \rb{B}(w))^T \Gamma_{\rmg}^{-1} (\rmg - \rb{B}(w)) \right]\,.
\end{equation}

It is useful to define the likelihood potential function
\begin{equation}\label{eq:likePotential}
    \Phi(w) := -\log \pi_{like}(\rmg \mid w)\,,
\end{equation}
where $\rmg$ is fixed, so that $\Phi$ is a function of $w$.

To solve \eqref{eq:dataAndModelFinal}, that is, to identify the distribution of $w$ conditioned on the data $\rmg$ and the noise model, one must specify a prior measure $\mu^{\rmo}$ on $W$ that encodes prior knowledge. For example, this prior knowledge may include assumptions on the mean and covariance. Next, suppose the prior is Gaussian, $\mu^{\rmo} = N(0,C)$, where the mean is zero and $C$ is the covariance operator. With this Gaussian prior on $w$, the transformation \eqref{eq:transformW} induces a log-normal prior on the parameter field $m \in M_{ad}$. The choice of $C$, $\alpha_m$, and $\beta_m$ reflects prior knowledge of $m$.

Bayes' rule relates the prior measure $\mu^{\rmo}$, the likelihood $\pi_{like}$, and the posterior measure $\mu^{\rmg}$, which is the distribution of $w$ conditioned on the data $\rmg$, as follows:
\begin{equation}\label{eq:Bayes}
    \frac{\dd \mu^{\rmg}}{\dd \mu^{\rmo}}(w) = \frac{1}{Z} \pi_{like}(\rmg \mid w) = \frac{1}{Z} \exp\left(-\Phi(w)\right)\,,
\end{equation}
where $Z$ is the normalizing constant given by
\begin{equation}
    Z = \int_W \exp\left(-\Phi(w)\right) \dd \mu^{\rmo}(w)\,.
\end{equation}
Bayes' rule \eqref{eq:Bayes} indicates how the posterior and prior measures are related. Given \eqref{eq:Bayes}, it is straightforward to see that
\begin{equation}
    \begin{split}
        \bbE_{\mu^{\rmg}}\left[ f \right] &= \int_W f(w)\, \dd \mu^{\rmg}(w) \\
        &= \int_W f(w) \frac{\exp\left(-\Phi(w)\right)}{Z} \dd \mu^{\rmo}(w)\\
        &= \bbE_{\mu^{\rmo}}\left[ f(\cdot)\, \frac{\exp\left(-\Phi(\cdot)\right)}{Z} \right]\,.
    \end{split}
\end{equation}
The above identity shows that posterior expectations can be expressed as prior expectations weighted by the likelihood factor. In practice, direct evaluation of such expectations is generally not feasible in high or infinite dimensions, which motivates the use of sampling methods such as Markov chain Monte Carlo (MCMC). This is discussed next.

\subsubsection{Markov chain Monte Carlo (MCMC) method to sample from the posterior measure}
To sample from the posterior measure $\mu^{\rmg}$ (i.e., to generate samples $w\in W$ distributed according to $\mu^{\rmg}$), several algorithms are available (see [\citealp[Section 4]{cotter2013mcmc}] and [\citealp{Dashti2017}]). This work uses the preconditioned Crank--Nicolson (pCN) method due to its simplicity. The pCN algorithm, based on \cite{cotter2013mcmc}, is as follows:

\vspace{5pt} 
\begin{algobox}{Preconditioned Crank--Nicolson (pCN) MCMC algorithm}
\begin{enumerate}
    \item Compute an initial sample $w^{(0)} \sim \mu^{\rmo} = N(0, C)$.
    \item For $k = 0, 1, \dots, k_{mcmc}-1$:
    \begin{enumerate}
        \item Propose $v^{(k)} = w^{(k)} \sqrt{1 - \beta_{pCN}^2} + \beta_{pCN} \xi$, where $\xi \sim \mu^{\rmo}$.
        \item Compute $a = \Phi(w^{(k)}) - \Phi(v^{(k)})$, and draw $b \sim \mathrm{Uniform}[0,1]$.
        \item If $\exp(a) > b$ (equivalently $a > \log(b)$), accept $v^{(k)}$ and set $w^{(k+1)} = v^{(k)}$. Otherwise, set $w^{(k+1)} = w^{(k)}$.
    \end{enumerate}
    \item Discard the initial $k_{burn}$ samples as burn-in, and use $\{w^{(k)}\}_{k=k_{burn}+1}^{k_{mcmc}}$ as samples from the posterior. The posterior mean is approximated by
    \begin{equation}\label{eq:posteriorMean}
        \bar{w} = \frac{1}{k_{mcmc} - k_{burn}} \sum_{k=k_{burn} + 1}^{k_{mcmc}} w^{(k)}\,.
    \end{equation}
\end{enumerate}
\end{algobox}
\vspace{5pt} 

In the above algorithm, $\beta_{pCN}$ is the proposal hyperparameter. To verify that the acceptance rule behaves as intended, consider two cases. First, if the proposed sample $v^{(k)}$ has higher cost than the current sample $w^{(k)}$ (i.e., $\Phi(v^{(k)}) > \Phi(w^{(k)})$), then $a<0$ and $\exp(a)<1$, so the proposed sample may be accepted or rejected depending on the draw $b$. Second, if the proposed sample has lower cost than the current sample, then $a>0$ and $\exp(a)>1$, so the proposed sample is accepted regardless of $b$.

\Cref{lst:mcmc} shows the MCMC implementation based on the pCN method. Functions \lstinline{proposal} and \lstinline{sample} contain the core pCN implementation. The listing also shows how surrogate models are used to compute the forward solution in the \lstinline{solveFwd} function.

\begin{lstlisting}[language=Python, caption={Core implementation of the pCN-based MCMC sampler.}, label={lst:mcmc}]
...
import numpy as np
from scipy.interpolate import griddata
...
from state import State
from tracer import Tracer
...

class MCMC:
    def __init__(self, model, prior, data, sigma_noise, pcn_beta = 0.2, \
                 surrogate_to_use = None, surrogate_models = None, seed = 0):
        # model class that provides solveFwd()
        self.model = model
        ...        
        # prior class that provides () and logPrior()
        self.prior = prior

        # data (dict) that provides x_obs, u_obs, m_true, u_true, etc.
        self.data = data

        # noise (std-dev) in the observations
        self.sigma_noise = sigma_noise

        # preconditioned Crank-Nicolson parameter
        self.pcn_beta = pcn_beta
        ...
        # current and proposed input parameter and state variables
        self.current = State(self.m_dim, self.u_dim, self.u_obs_dim)
        self.proposed = State(self.m_dim, self.u_dim, self.u_obs_dim)
        self.init_sample = State(self.m_dim, self.u_dim, self.u_obs_dim)
        ...
        # surrogate models
        self.surrogate_to_use = surrogate_to_use
        self.surrogate_models = surrogate_models

        # tracer
        self.tracer = Tracer(self) 
        self.log_file = None

    def solveFwd(self, current):
        if self.surrogate_to_use is not None:
            current.u = self.surrogate_models[self.surrogate_to_use].solveFwd(current.m)
        else:
            current.u = self.model.solveFwd(u = current.u, m = current.m, transform_m = True)
        
        return current.u
    
    def state_to_obs(self, u):
        # interpolate PDE solution on observation grid
        if self.u_comps == 1:
            return griddata(self.u_nodes, u, self.x_obs, method='linear')

        num_nodes = self.u_nodes.shape[0]
        u_xy = u.reshape(num_nodes, self.u_comps)
        obs_components = [griddata(self.u_nodes, u_xy[:, i], self.x_obs, \
                            method='linear') for i in range(self.u_comps)]

        return np.column_stack(obs_components).reshape(-1)
    
    def logLikelihood(self, current):
        current.u = self.solveFwd(current)
        current.u_obs = self.state_to_obs(current.u)
        current.err_obs = current.u_obs - self.u_obs
        current.cost = 0.5 * np.linalg.norm(current.err_obs)**2 / self.sigma_noise**2
        current.log_likelihood = -current.cost

        return current.log_likelihood
    
    def logPosterior(self, current):
        current.log_prior = self.prior.logPrior(current.m)
        current.log_likelihood = self.logLikelihood(current)
        current.log_posterior = current.log_prior + current.log_likelihood
        return current.log_posterior
    
    def proposal(self, current, proposed):
        # preconditioned Crank-Nicolson
        proposed.m, proposed.log_prior = self.prior(proposed.m)
        return self.pcn_beta * proposed.m \
            + np.sqrt(1 - self.pcn_beta**2) * current.m
    
    def sample(self, current):
        # compute the proposed state
        self.proposed.m = self.proposal(current, self.proposed)
        self.proposed.log_posterior = self.logPosterior(self.proposed)
        
        # accept or reject based on the likelihood ratio
        alpha = current.cost - self.proposed.cost
        if alpha > np.log(np.random.uniform()):
            current.set(self.proposed)
            return 1

        return 0
    
    def run(self, init_m, n_samples, n_burnin, pcn_beta, sigma_noise, ...):
        ...
        # run the MCMC
        init_done = False
        for i in range(n_samples + n_burnin):
            # sample
            accept = self.sample(self.current)

            # postprocess/print
            self.process_and_print(i)

            if i < n_burnin: 
                continue
            ...
            self.save(i, self.current, accept)

        # save the final state
        self.tracer.append(i, self.current, accept, force_save=True)
        
        # print message
        ...
\end{lstlisting}

\subsection{Inference of diffusivity in the Poisson problem}\label{ss:poissonBayesian}

This section considers Bayesian inference of the diffusivity field $m$ in the Poisson problem using neural operators as surrogate forward models. For a given $m$, let $u=F(m)$ denote the finite element solution of the Poisson problem \eqref{eq:poisson}. The neural operator approximation is denoted by $F_{NOp}(m)$, where $NOp \in \{\text{DeepONet}, \text{PCANet}, \text{FNO}\}$. The inference problem is posed in terms of $w\in W$, with $m=\alpha_m\exp(w)+\beta_m$ used to enforce positivity of the diffusivity. The prior measure and synthetic data generation are described first, followed by the inference results.

\subsubsection{Setup of the forward problem, prior measure, and synthetic data}\label{sss:setupInferencePoisson}

The forward problem, prior measure, covariance operator, and transformation parameters are the same as those described in \cref{sss:poissonSetup}. Thus, $w\sim\mu_{\mscZ}=N(0,C)$ and $m$ is obtained from $w$ through \eqref{eq:transformW}.

\begin{figure}[h]
    \centering
    \includegraphics[width=0.95\linewidth]{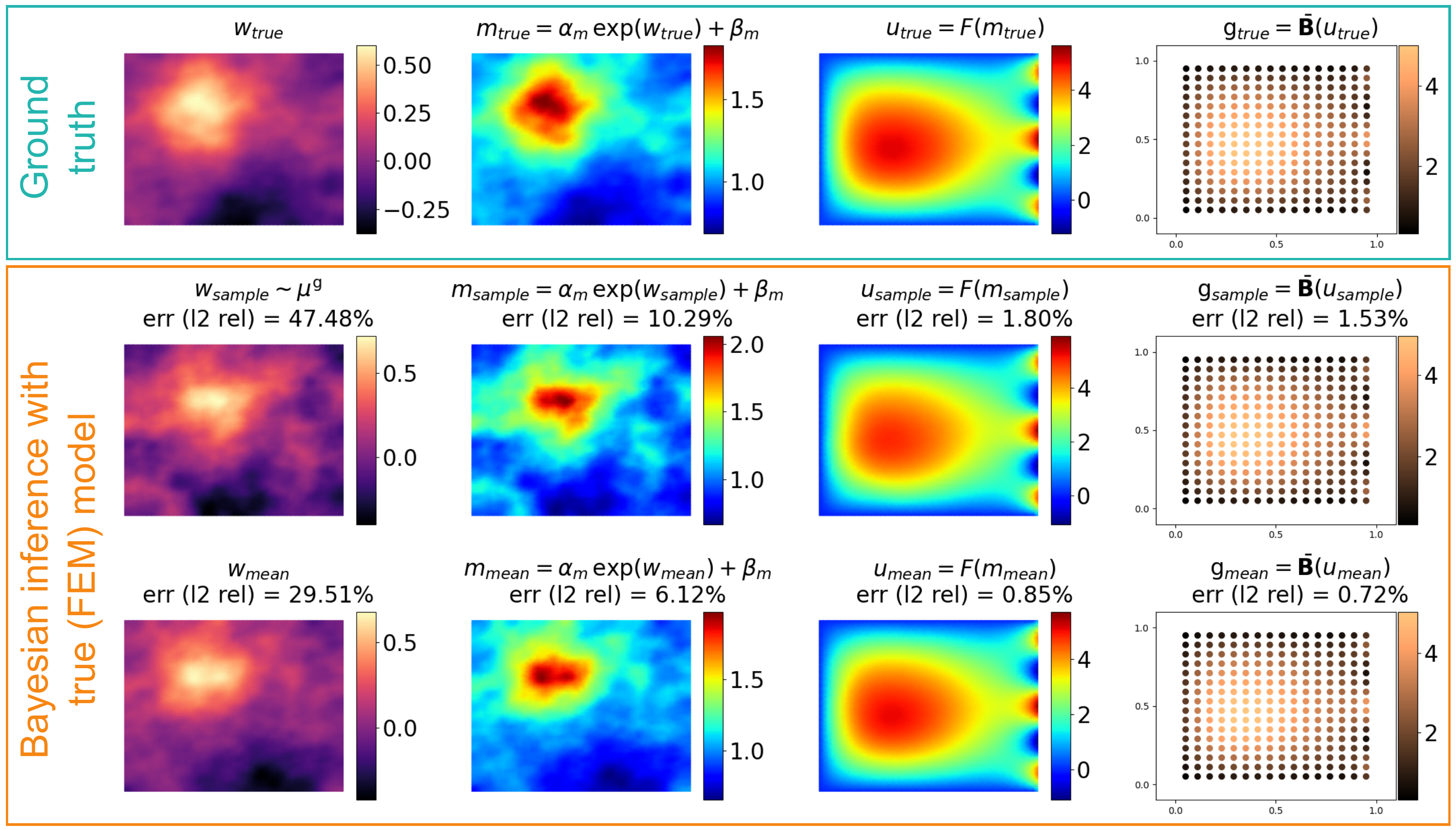}
    \caption{Bayesian inference of diffusivity in the Poisson problem using the \textit{true} forward model (numerical approximation of the PDE). The top panel displays the synthetic field $w_{true}$, the corresponding diffusivity $m_{true}=\alpha_m\exp(w_{true})+\beta_m$, the PDE solution $u_{true}=F(m_{true})$, and the observations $\rmg_{true}=\bar{\rb{B}}(u_{true})$. The observations are obtained by interpolating $u_{true}$ on a $16\times16$ grid over $D_{U}=(0,1)^2$, so that $\rmg_{true}\in\Rpow{d_{\rmg}}$ with $d_{\rmg}=16^2$. The lower panel displays a posterior sample $w_{sample}$ drawn from the posterior measure $\mu^{\rmg}$ (one realization from the MCMC chain) and the posterior mean $w_{mean}$ defined in \eqref{eq:posteriorMean}. The remaining columns display the corresponding diffusivity field $m$, the forward solution $u=F(m)$ used in the MCMC simulation, and the predicted observations $\rmg=\bar{\rb{B}}(u)$ associated with the corresponding $w$.}
    \label{fig:poissonBayesianInferenceTrue}
\end{figure}

Synthetic data are generated by selecting a fixed field $w_{true}$; see $w_{true}$ in top panel of \cref{fig:poissonBayesianInferenceTrue}. The corresponding true diffusivity field is obtained as $m_{true}=\alpha_m\exp(w_{true})+\beta_m$, and the true solution is computed as $u_{true}=F(m_{true})$. The observational data $\rmg_{true}\in\Rpow{d_{\rmg}}$, with $d_{\rmg}=16^2$, are obtained by interpolating $u_{true}$ on a $16\times16$ grid over $D_{U}=(0,1)^2$. The top panel of \cref{fig:poissonBayesianInferenceTrue} also displays $m_{true}$, $u_{true}$, and $\rmg_{true}$. The implementation can be found in the notebook \lstinline{Generate_GroundTruth.ipynb}\footnote{\url{https://github.com/CEADpx/neural_operators/blob/survey26_v2/survey_work/applications/bayesian_inverse_problem_poisson/Generate_GroundTruth.ipynb}}.

\subsubsection{Inference results}

The MCMC parameters are $k_{mcmc}$ = 10,500, $k_{burn}=500$, $\beta_{pCN}=0.2$, and $\sigma_{\rmg}=1.329$, where the covariance matrix in the noise model is $\Gamma_{\rmg}=\sigma_{\rmg}^2\rmI$. Here, $\sigma_{\rmg}$ is chosen as $5\%$ of the mean of the observational data, i.e.,
\begin{equation}
    \sigma_{\rmg} = 0.05 \times \frac{1}{d_{\rmg}}\sum_{i=1}^{d_{\rmg}}\rmg_{{true}_i}\,.
\end{equation}

The same MCMC simulation is performed using four different forward models. In the first case, the state field $u$ is computed using the finite element approximation of the Poisson problem, referred to here as the \textit{true} forward model. In the remaining three cases, DeepONet, PCANet, and FNO are used as surrogate forward models. The trained neural operators from \cref{s:nn} are used, and their prediction accuracy for random samples from the prior is shown in \cref{fig:poissonCompareNOps}. The notebook \lstinline{BayesianInversion.ipynb}\footnote{\url{https://github.com/CEADpx/neural_operators/blob/survey26_v2/survey_work/applications/bayesian_inverse_problem_poisson/BayesianInversion.ipynb}} sets up the problem, loads the trained neural operators, and runs the MCMC simulations.

\begin{figure}[!h]
    \centering
    \includegraphics[width=0.95\linewidth]{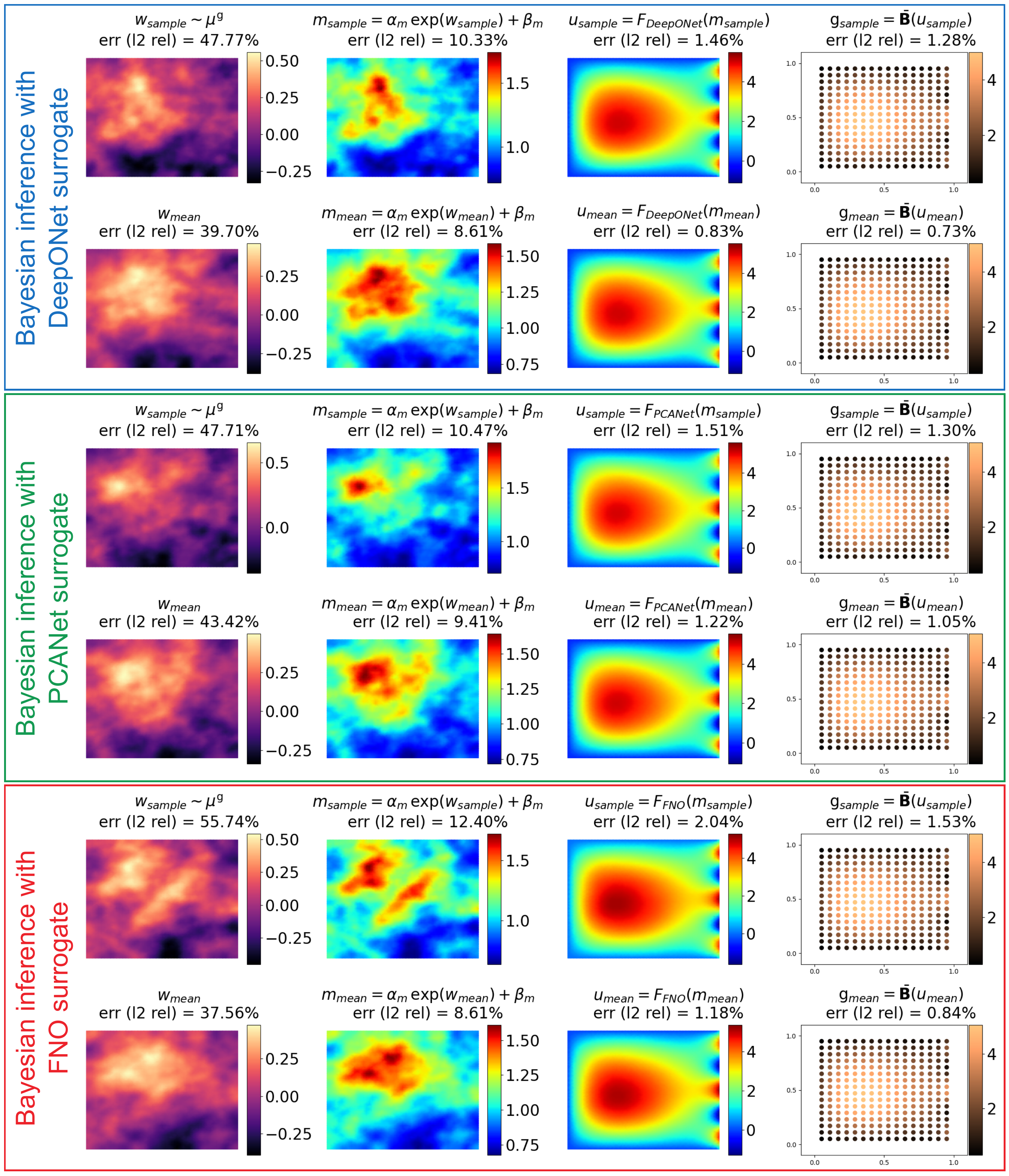}
    \caption{Comparison of Bayesian inference results for the Poisson problem using DeepONet, PCANet, and FNO surrogates. Each panel displays a posterior sample $w_{sample}$ drawn from the posterior measure $\mu^{\rmg}$ and the posterior mean $w_{mean}$. The corresponding diffusivity field $m$, forward solution $u$, and predicted observations $\rmg = \bar{\rb{B}}(u)$ are also shown, where $F_{NOp}$, $NOp \in \{DeepONet, PCANet, FNO\}$, is the neural operator approximation of the forward operator used in the MCMC simulation and $\bar{\rb{B}}$ is the state-to-observable map. These results should be compared with the inference results obtained using the \textit{true} forward model in \cref{fig:poissonBayesianInferenceTrue}.}
    \label{fig:poissonBayesianInferenceNOps}
\end{figure}

\begin{figure}[h]
\centering
\subfloat[\centering Acceptance rate\label{fig:poissonMCMCStatSub1}]{
  \begin{minipage}{.47\textwidth}
  \centering
  \includegraphics[width=\linewidth]{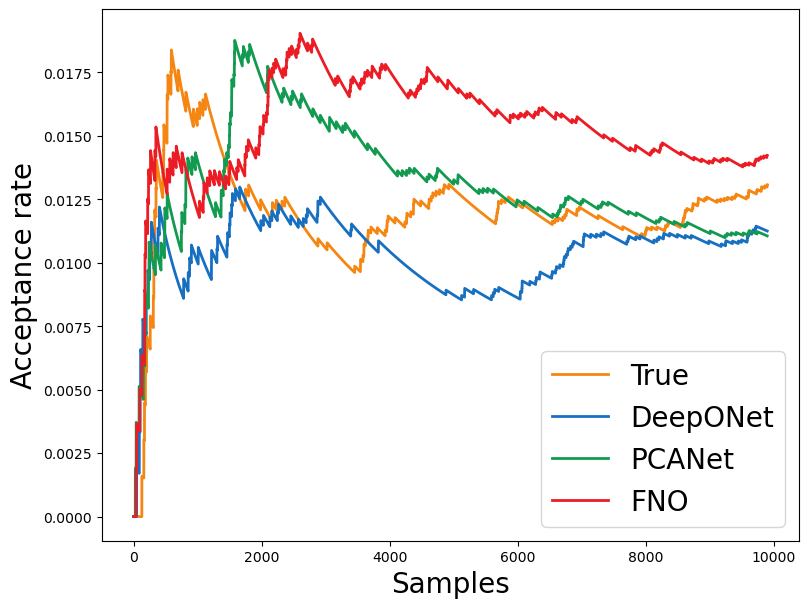}
  \end{minipage}
}%
\subfloat[\centering Cost\label{fig:poissonMCMCStatSub2}]{
  \begin{minipage}{.47\textwidth}
  \centering
  \includegraphics[width=\linewidth]{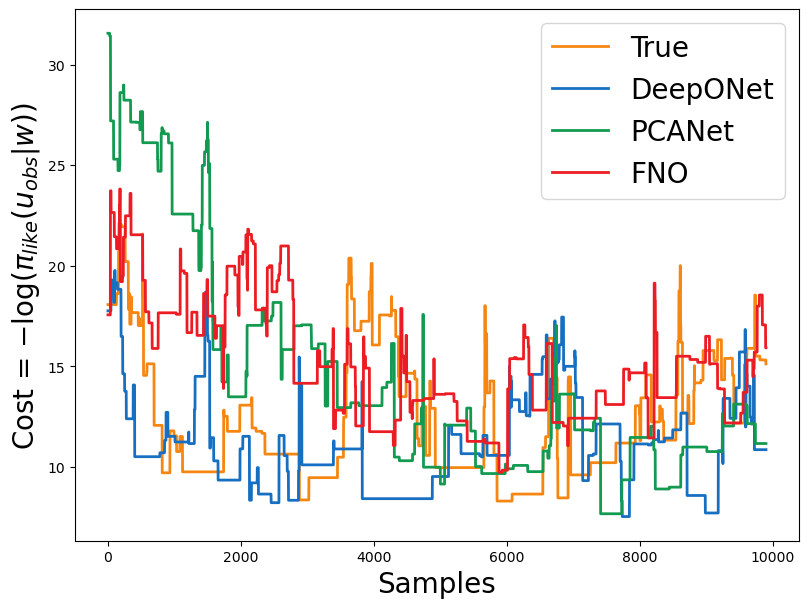}
  \end{minipage}
}
\caption{Acceptance rate and cost during MCMC simulation for the Poisson problem.}
\label{fig:poissonMCMCStat}
\end{figure}

The inference results obtained using the \textit{true} forward model, including the posterior mean and a representative posterior sample, are shown in \cref{fig:poissonBayesianInferenceTrue}. The corresponding results obtained using DeepONet, PCANet, and FNO surrogates are shown in \cref{fig:poissonBayesianInferenceNOps}. The acceptance rate and cost histories for the four MCMC simulations are shown in \cref{fig:poissonMCMCStat}.

Compared with the \textit{true} forward model, all three neural operator surrogates produce posterior means of $w$ with slightly higher error. However, the error in the posterior mean of the transformed diffusivity field $m$ remains approximately one percent. This indicates that neural operators provide effective surrogate forward models for the current Bayesian inference problem. This performance is aided by the fact that the ground-truth parameter field is close to the prior distribution used for training and inference. More substantial deviations from the training distribution would likely lead to larger surrogate-induced inference errors.

\section{Neural operator performance comparison}\label{s:nopPerformance}
In this section, the performance of the trained neural operators is compared in terms of predictive accuracy for in-distribution and out-of-distribution input samples, as well as inference accuracy for Bayesian inverse problems. The architectures for the three neural operators are listed in \cref{tab:parameters}; they are selected to have comparable numbers of trainable parameters, as indicated by $p_\Theta$ in \cref{tab:nopPerformance}. The Bayesian inference problems discussed in the previous section and the in- and out-of-distribution prediction tests considered here all use the same trained neural operators described in \cref{s:nn}.

\subsection{In-distribution performance}
In this case, test samples are drawn from the same prior distribution as the training data for the neural operators in \cref{s:nn}. Recall that for all three problems, a log-normal prior is used to generate the training data. Specifically, a sample $w\sim N(0,C)$ is first drawn and then transformed to $m$ through $m = \alpha_m \exp(w) + \beta_m$. The parameters used to generate the training data are listed below for reference:

\vspace{5pt} 
\begin{algobox}{Prior parameters for training data}
\[
\begin{aligned}
\textbf{Common parameters:}\quad
& \mathsf{a}_c = 0.005,\quad \mathsf{b}_c = 1,\quad \mathsf{c}_c = 0.2,\\
\textbf{Poisson problem:}\quad
& \alpha_m = 1,\quad \beta_m = 0,\\
\textbf{Linear and hyperelasticity problems:}\quad
& \alpha_m = 100,\quad \beta_m = 1000.
\end{aligned}
\]
\end{algobox}
\vspace{5pt} 

The rows corresponding to \lstinline{In-distribution error} in \cref{tab:nopPerformance} are computed by taking the mean relative $\ell^2$-error  over $N_{test}=800$ test samples. For the Poisson problem, all three neural operators achieve mean errors below $1.5\%$, with FNO achieving the smallest error. For the two elasticity problems, DeepONet and FNO yield errors close to $0.03\%$, while PCANet yields errors that are approximately one order of magnitude smaller.

\subsection{Out-of-distribution (OOD) performance}\label{ss:oodPerformance}
For OOD tests, two priors are considered, where the second prior is more distant from the training distribution than the first. In both cases, the covariance operator parameters are kept the same as those used to generate the training data, while the transformation parameters are changed. The transformation parameters for the two OOD priors are listed below.

\vspace{10pt} 
\begin{algobox}{Case 1 OOD prior parameters}
\[
\begin{aligned}
\textbf{Poisson problem:}\quad
& \alpha_m = 1,\quad \beta_m = 1,\\
\textbf{Linear and hyperelasticity problems:}\quad
& \alpha_m = 100,\quad \beta_m = 1500.
\end{aligned}
\]
\end{algobox}
\vspace{5pt} 

\vspace{5pt} 
\begin{algobox}{Case 2 OOD prior parameters}
\[
\begin{aligned}
\textbf{Poisson problem:}\quad
& \alpha_m = 2,\quad \beta_m = 1,\\
\textbf{Linear and hyperelasticity problems:}\quad
& \alpha_m = 1000,\quad \beta_m = 1000.
\end{aligned}
\]
\end{algobox}
\vspace{10pt} 

For both OOD cases, $100$ samples are drawn, and the mean relative percentage $l^2$ errors are reported in \cref{tab:nopPerformance}. The rows corresponding to \lstinline{Out-of-distribution error (Case 1)} show that the errors increase substantially relative to the in-distribution case. For the Poisson problem, the errors range from about $13.5\%$ to $34.1\%$, while for the elasticity problems, the errors are generally between about $13\%$ and $25\%$. In Case 2, the errors increase further, as expected, since this prior is farther from the training distribution. In this case, all three neural operators produce errors above $100\%$ for the Poisson problem, above $65\%$ for linear elasticity, and above $50\%$ for hyperelasticity. 

\begin{figure}[h!]
    \centering
    \includegraphics[width=0.95\linewidth]{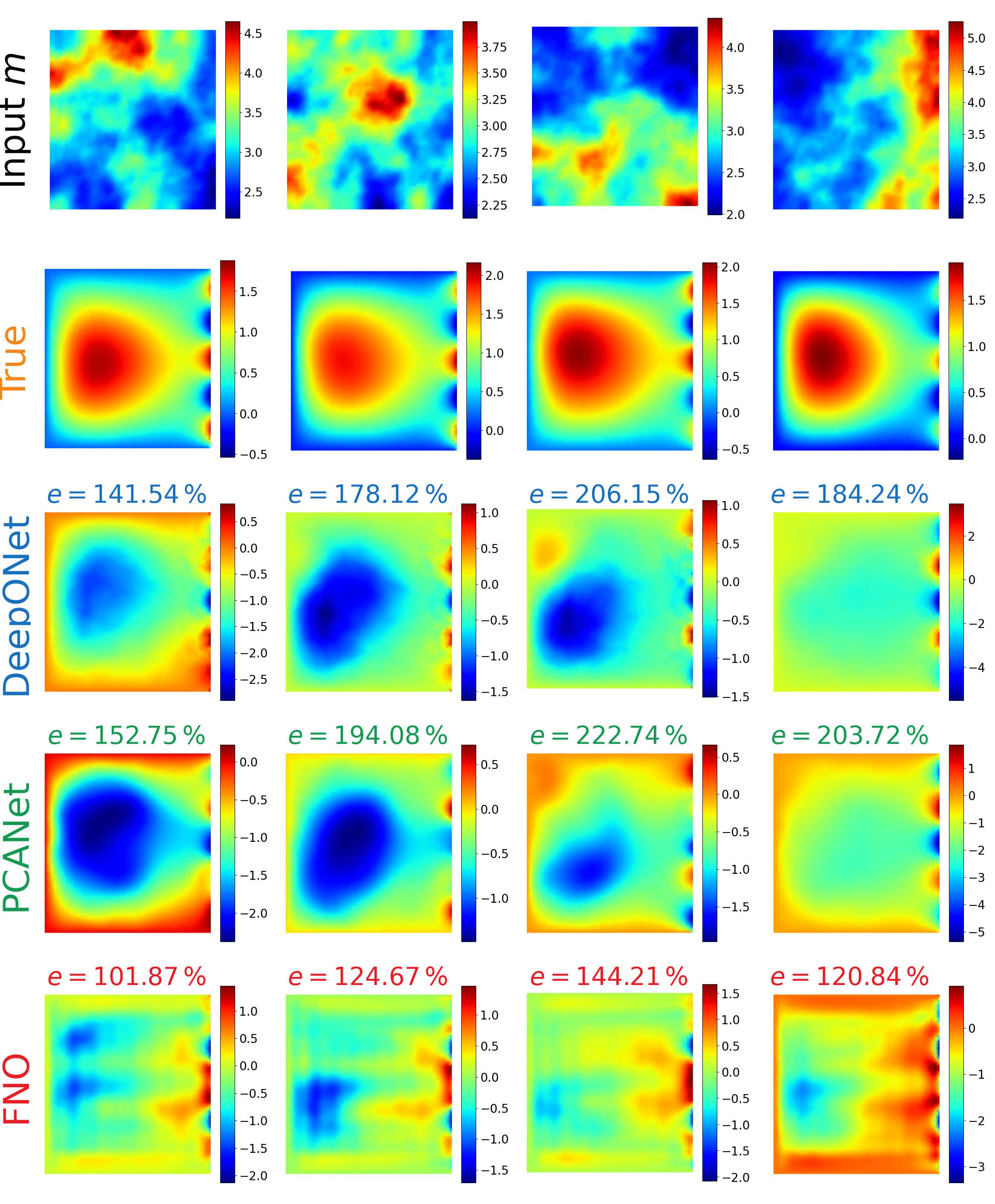}
    \caption{Representative out-of-distribution input samples and corresponding forward and surrogate solutions for the Poisson problem. Here, $e$ denotes the relative percentage $\ell^2$ error. The OOD samples correspond to Case 2 in \cref{ss:oodPerformance}.}
    \label{fig:poissonOutDistNopSamples}
\end{figure}

In \cref{fig:poissonOutDistNopSamples}, representative OOD input samples and the corresponding forward and surrogate solutions for the Poisson problem are shown. The OOD samples correspond to Case 2. Similar plots for the elasticity problems under Case 2 OOD sampling are provided in \cref{fig:elasticityOutDistNopSamples,fig:hyperelasticityOutDistNopSamples} of \labelcref{ss:appNopPerformance}. These figures show that the surrogate solutions deviate significantly from the true finite element solutions, which is consistent with the large errors reported in \cref{tab:nopPerformance}.

\vspace{20pt} 
\subsection{Bayesian inference error} 
The Bayesian inference problems and results have been described in the previous section. Here, the main inference errors are included alongside the prediction errors to provide an overall performance comparison. Suppose $\rmw_{true}$ is the true discretized parameter field and $\rmw_{post}$ is the posterior mean obtained from MCMC sampling using either a neural operator surrogate or the \textit{true} finite element model. The relative $\ell^2$ percentage error is defined as
\begin{equation*}
    \text{Inference error} = \frac{\norm{\rmw_{post} - \rmw_{true}}}{\norm{\rmw_{true}}} \times 100\%\,.
\end{equation*}
The rows corresponding to \lstinline{Bayesian inference error} in \cref{tab:nopPerformance} show the inference error for the \textit{true} model and the three neural operator surrogates. The \textit{true} forward model gives the smallest inference error, as expected. The three surrogates give inference errors reasonably close to the errors obtained using the \textit{true} model, although PCANet performs relatively poorly for the Poisson problem. One reason the surrogates perform reasonably well is that the ground-truth parameter field $\rmw_{true}$ is close to the training distribution and the prior used in Bayesian inference is the same as that used to generate the training data. If a significantly different $\rmw_{true}$ or a different prior were considered, the inference error from the surrogates would likely increase relative to that from the \textit{true} model.

\vspace{30pt} 

\begin{table}[!h]
    \centering
\renewcommand{\arraystretch}{1.2}

\begin{tabular}{| @{\hspace{0.4em}} >{\raggedright\arraybackslash}p{5.8cm} @{\hspace{1em}}|@{\hspace{1em}}>{\centering\arraybackslash}p{2.4cm} @{\hspace{1em}}|@{\hspace{1em}}>{\centering\arraybackslash}p{2.4cm} @{\hspace{1em}}|@{\hspace{1em}}>{\centering\arraybackslash}p{2.4cm} @{\hspace{0.4em}} |}
\hline
\textbf{Metric} & Poisson & Linear Elasticity & Hyperelasticity \\
\hline
\multicolumn{4}{|c|}{\textbf{True}} \\
\hline
Bayesian inference error & 29.5\% & 21.1\% & 19.7\% \\
\hline
\hline
\multicolumn{4}{|c|}{\textbf{DeepONet}} \\
\hline
Trainable Params $p_{\Theta}$ & 188{,}040 & 194{,}540 & 194{,}540 \\
\cline{1-4}
In-distribution error & 1.319\% & 0.034\% & 0.032\% \\
\cline{1-4}
Out-of-distribution error (Case 1) & 31.814\% & 18.064\% & 14.007\% \\
\cline{1-4}
Out-of-distribution error (Case 2) & 186.6\% & 66.778\% & 52.654\% \\
\cline{1-4}
Bayesian inference error & 39.7\% & 21.6\% & 21.9\% \\
\hline
\hline
\multicolumn{4}{|c|}{\textbf{PCANet}} \\
\hline
Trainable Params $p_{\Theta}$ & 180{,}700 & 180{,}700 & 180{,}700 \\
\cline{1-4}
In-distribution error & 0.884\% & 0.008\% & 0.007\% \\
\cline{1-4}
Out-of-distribution error (Case 1) & 34.111\% & 17.551\% & 14.118\% \\
\cline{1-4}
Out-of-distribution error (Case 2) & 202.861\% & 65.929\% & 52.973\% \\
\cline{1-4}
Bayesian inference error & 43.4\% & 23.8\% & 21.5\% \\
\hline
\hline
\multicolumn{4}{|c|}{\textbf{FNO}} \\
\hline
Trainable Params $p_{\Theta}$ & 187{,}485 & 187{,}508 & 187{,}508 \\
\cline{1-4}
In-distribution error & 0.575\% & 0.028\% & 0.027\% \\
\cline{1-4}
Out-of-distribution error (Case 1) & 13.512\% & 24.9\% & 13.238\% \\
\cline{1-4}
Out-of-distribution error (Case 2) & 136.091\% & 96.265\% & 69.212\% \\
\cline{1-4}
Bayesian inference error & 37.6\% & 22.2\% & 21.4\% \\
\hline
\end{tabular}
\caption{Summary of neural operator performance for in-distribution input samples, out-of-distribution input samples, and Bayesian inference. The row below \lstinline{True} lists the Bayesian inference errors obtained using the \textit{true} forward model and provides benchmark values for assessing the neural operator surrogate results in the following rows.}\label{tab:nopPerformance}
\end{table}

\newpage 
\section{Perspectives and conclusion}\label{s:conclusion}
While the focus of this article has been on the foundational aspects of operator learning and their practical implementation, the subsections below briefly review recent developments in neural operators, with particular emphasis on the growing family of architectures, emerging foundation models for PDEs, and the critical issue of controlling prediction accuracy.

\subsection{The growing field of neural operators}

The field of neural operators continues to evolve through a range of frameworks that adapt the basic operator-learning idea to different discretizations, physical constraints, geometries, and application settings. One direction is to incorporate additional physical or differential information during training. Physics-Informed Neural Operators (PINO) include the residual of the governing PDE in the loss function, in addition to losses associated with model outputs and data mismatch \cite{li2024physics, goswami2023physics, wang2021learning}. Derivative-Informed Neural Operators (DINO) \cite{o2024derivative} incorporate derivative information into the training process to improve learning efficiency and accuracy, particularly for problems where gradients play a dominant role.

Another line of work concerns the representation of the nonlocal operator and the underlying domain. Graph Neural Operators (GNO) \cite{kovachki2023neural} are based on graph discretizations of the spatial domain and are therefore well suited for problems defined on irregular geometries. Their structure is similar to that of the FNO, with a different form of the nonlocal kernel operator. The Wavelet Neural Operator (WNO) \cite{tripura2022wavelet} follows a related idea, in which the nonlocal operation is defined using wavelet transforms. The Laplace Neural Operator \cite{cao2024laplace} extends operator learning to transient problems and non-periodic domains. Other works use attention-based or transformer-type parameterizations of the operator map, including GNOT \cite{hao2023gnot}, the operator transformer of \cite{li2023oformer}, and factorized transformer models such as FactFormer \cite{li2023factformer}. These approaches broaden the class of admissible input representations beyond the regular-grid setting for which classical FNO is most natural.

Multi-resolution and multi-scale architectures form another important direction. U-NO \cite{rahman2023uno} introduces a U-shaped encoder-decoder structure for deeper operator networks, while Factorized FNO \cite{tran2023ffno} aims to improve parameter efficiency and scalability for high-resolution PDEs. Multiwavelet constructions \cite{gupta2021multiwavelet} provide a principled way to separate coarse and fine scales and have been effective for multi-scale and coupled systems.

Geometry and domain complexity have also motivated several extensions. Local neural operators \cite{li2024localneuraloperator} and diffeomorphism neural operators \cite{zhao2025dno} aim to preserve discretization flexibility while improving transfer across domains and boundary configurations. In \cite{vemparala2024deep}, a deep learning-driven domain decomposition (DLD$^3$) method is proposed, where an FNO is trained to learn the solution map on smaller subdomains of a larger problem, thereby allowing the framework to be applied to different geometries and boundary conditions. Recently, the Kolmogorov--Arnold Neural Network (KAN) has motivated alternative parameterizations of nonlinear maps \cite{liu2024kan}.

There has also been considerable effort to extend neural operator frameworks to Bayesian settings, where uncertainty is modeled through probabilistic formulations of the learned operators; see \cite{psaros2023uncertainty, jospin2022hands, garg2022variational, singh2024framework}. These approaches aim to provide probabilistic predictions that reflect uncertainty in both the data and model parameters, alongside point estimates.

Digital twins are another application area where neural operators may be useful, especially when repeated state predictions are needed under varying model or control parameters. Digital twin frameworks are expanding due to advances in computational capacity, multi-physics modeling, sensing, and control technologies; see \cite{juarez2021digital, wagg2020digital}. Neural operators can serve as fast surrogate models for predicting system states within such frameworks \cite{kobayashi2024deep}.

Neural operators have also demonstrated potential in Bayesian inverse problems, optimization, and control. In Bayesian inverse problems, they can accelerate posterior sampling by replacing expensive PDE solves with surrogate evaluations \cite{cao2023residual, cao2024derivative, gao2024adaptive}. In optimization and control problems, the primary computational bottleneck is often the repeated evaluation of the forward solution for varying parameter fields \cite{shukla2024deep}. A major challenge in these settings is the lack of reliable mechanisms for predicting surrogate accuracy. Training data are typically generated from a prior distribution, whereas optimization and control procedures may require evaluations far outside this regime. \cite{jha2024residual} demonstrates that neural operators trained on a prior distribution can perform poorly when used within optimization loops. This issue is discussed in more detail in the next subsection.

\subsection{Controlling neural operator prediction accuracy}\label{ss:accuracy}
Neural operators can provide rapid surrogate evaluations, but their practical use depends on whether their prediction error can be estimated, controlled, or corrected in regimes relevant to the downstream task. The ability to predict and control accuracy is a central issue in uncertainty quantification, Bayesian inference, optimization, and control problems, where small forward-model errors may propagate through repeated evaluations. Several key works in this direction are reviewed below.

The multi-level neural network framework \cite{aldirany2024multi} iteratively refines solutions by training successive networks to minimize residual errors from previous levels. This hierarchical approach progressively reduces approximation error. By capturing high-frequency components in subsequent networks, this framework addresses limitations of standard neural network approximations, which can have low sensitivity to higher-frequency modes. The multi-stage neural network framework \cite{wang2024multi} addresses convergence plateauing in deep learning by dividing training into sequential stages, with each stage fitting the residual from the previous one. This approach progressively refines the approximation. The methods in \cite{aldirany2024multi, wang2024multi} are reported to achieve errors approaching machine precision in controlled settings.

The Galerkin neural network framework \cite{ainsworth2021galerkin} integrates the classical Galerkin method with neural networks, using the neurons in a neural network layer as basis functions to approximate variational equations. It adaptively adds new neurons to a single-layer neural network, with each neuron corresponding to a basis function. By adding more neurons, the dimension of the approximation increases in a way that produces less error than lower-dimensional subspaces.

\begin{figure}[h]
    \centering
    \includegraphics[width=0.8\linewidth]{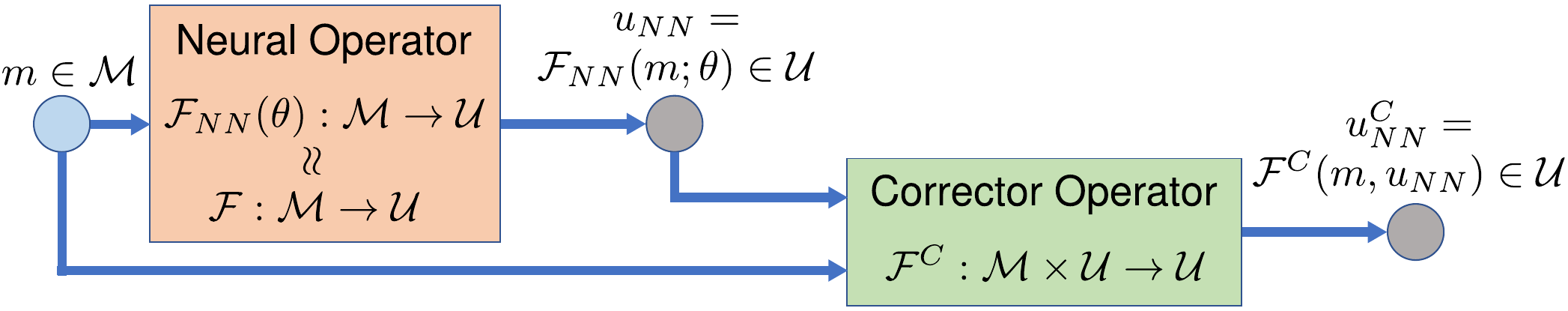}
    \caption{Residual-based error correction of neural operator predictions \cite{cao2023residual, jha2024residual}. The corrector problem is a linear boundary value problem, and if the predictor $u_{NN}$ is sufficiently close to the PDE solution $u$, the corrector gives quadratic error reduction \cite{jha2024residual}.}
    \label{fig:nnCorrector}
\end{figure}

\begin{figure}[h]
    \centering
    \vspace{-15pt} 
    \includegraphics[width=0.8\linewidth]{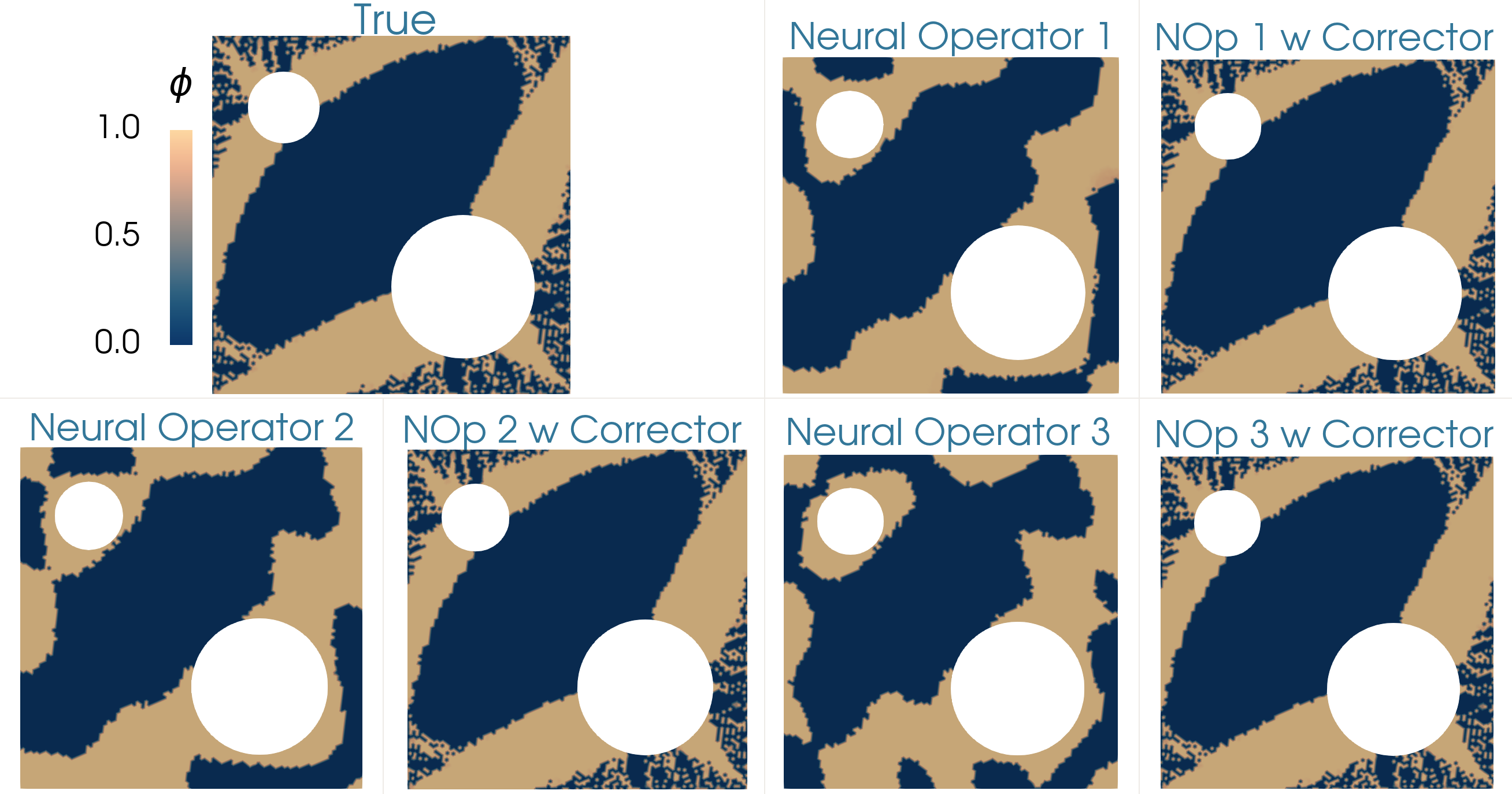}
    \vspace{-5pt} 
    \caption{Optimized diffusivity in a Poisson equation obtained using the \textit{true} model and neural operator surrogates with and without a corrector. Here, the forward problem is posed on a square domain with two circular voids. Zero Dirichlet boundary conditions are imposed on the internal void boundaries, and a constant flux is prescribed on the outer boundaries. The optimal diffusivity is obtained by minimizing the compliance, defined as the boundary integral of the product of flux and temperature over the flux boundary. For more details about the problem and results, see \cite[Section 4.3]{jha2024residual}.}
    \label{fig:nnCorrectorApplication}
\end{figure}

The residual-based error correction method \cite{cao2023residual, jha2024residual}, illustrated in \cref{fig:nnCorrector}, builds on the idea of using lower-fidelity solutions to estimate modeling errors \cite{jha2022goal}. It treats the neural operator prediction as an initial guess and solves a variational problem to correct the residual error. This approach improves accuracy in applications like Bayesian inverse problems and topology optimization, where small errors can propagate and amplify. The trade-off in the residual-based error correction method is the introduction of a linear variational problem on the modeling error field, which must be solved and added to the neural operator prediction to obtain the improved prediction. For challenging nonlinear problems, the approach leads to significant speedups. In Bayesian settings, it enhances posterior estimates without increasing computational cost \cite{cao2023residual}. For the example optimization problem of seeking a diffusivity field in the Poisson equation that minimizes the compliance, neural operators, when used as surrogates for the forward problem, produced minimizers with significant errors (about $80\%$). Here, the error is defined as the norm of the difference between the minimizer obtained using the true forward model and that obtained using the surrogate. Neural operator surrogates with the residual-based error correction produced minimizers with errors below $6\%$; see \cref{fig:nnCorrectorApplication}, where the optimized diffusivities obtained using the \textit{true} model and the surrogates are compared. A complementary direction is PhysicsCorrect \cite{Huang_Perdikaris_2026}, a training-free correction framework that stabilizes neural PDE simulations by enforcing physics-based consistency at inference time. It is conceptually aligned with residual-based correction strategies such as \cite{jha2024residual}, but avoids retraining and introduces only modest additional overhead.

Recent work also makes it clear that controlling prediction accuracy is not only a matter of training larger models. \textit{A posteriori} and residual-based indicators must be made reliable in norms that are meaningful for the PDE under consideration. In this direction, rigorous error analyses for DeepONet and FNO-type models \cite{lanthaler2022deeponeterror, kovachki2021fnoerror} provide an important theoretical baseline, while newer work on neural functional a posteriori error estimates \cite{fanaskov2024functionalaposteriori} aims to construct computable error majorants that can be used during or after training. These developments are significant because small empirical losses or averaged test errors do not, by themselves, provide reliable guarantees about the error of a specific prediction relevant to an optimization, inference, or control query.

Adaptive strategies are also playing an increasingly important role. The framework in \cite{gao2024adaptive} locally refines a pretrained surrogate in regions most relevant to posterior exploration. This is conceptually appealing because high accuracy is rarely required uniformly over the entire input space. Instead, one often seeks high fidelity only on a dynamically discovered subset of the parameter regime. Multi-fidelity and multi-level formulations provide another route to balancing cost and accuracy. Multi-fidelity deep neural operators \cite{lu2022multifidelitydno}, multi-fidelity DeepONets \cite{howard2023multifidelitydeeponet}, and application-oriented multi-fidelity operator fusion strategies \cite{xu2024multifidelitydeeponet} exploit the fact that coarse simulations, reduced models, or monitoring data may be much cheaper to obtain than uniformly high-fidelity labels.

Out-of-distribution behavior remains especially challenging. PDEBench- and PDEArena-style evaluations have shown that generalization across coefficients, resolutions, time horizons, and physical regimes is uneven across architectures \cite{takamoto2022pdebench, gupta2023pdearena}. Theoretical work on out-of-distribution risk bounds \cite{benitez2023oodrisk}, operator complexity \cite{lanthaler2023pcanetcomplexity, lanthaler2023parametriccomplexity}, and data complexity \cite{kovachki2024datacomplexity} reinforces the same message from another direction: one should not expect uniformly reliable accuracy over broad classes of operators without either exploiting stronger problem structure or incurring additional cost in data, model complexity, or online correction. Related results on DeepONet size lower bounds and neural scaling laws \cite{mukherjee2024sizelowerbounds, liu2024neuralscalingdeeponet} further indicate that practical accuracy limits are tied not only to optimization but also to intrinsic approximation and sample-complexity barriers.

Therefore, the current picture is that improvements in accuracy are possible through residual correction, adaptive refinement, multi-level training, multi-fidelity learning, and stronger theory, but each mechanism introduces a cost-accuracy trade-off. Extra linear solves, extra training stages, local fine-tuning, larger pretrained backbones, and more structured loss functions can all improve fidelity, yet they reduce some of the computational speed advantage that originally motivated neural operators. For many downstream applications, this could still be a favorable trade-off, but it highlights why predictability of accuracy, rather than average-case speed alone, remains the central challenge. 

The works reviewed above highlight several important limitations. First, methods that substantially improve the accuracy of neural operator predictions generally introduce additional computation, for example, by adding extra training stages, adaptive updates, or the solution of correction problems. Second, the relevant notion of accuracy is usually application-dependent; an error level acceptable for forward prediction may be unusable for optimization, posterior inference, or control. Third, distributional shifts remain unavoidable in many realistic workflows, and current error indicators are still far from being fully calibrated under such shifts. Addressing these issues will likely require tighter integration between operator learning, adaptive numerical analysis, and application-specific notions of goal-oriented error.

\subsection{Foundation models for PDEs}\label{ss:foundational}

A prominent emerging trend in scientific machine learning is the construction of pretrained models that are not tied to a single PDE family, geometry, or discretization, but instead aim to learn transferable representations across broad classes of operator-learning tasks. In the context of PDEs, this direction includes multi-physics pretraining \cite{mccabe2023mpp}, unified PDE solvers \cite{shen2024ups}, Universal Physics Transformers \cite{alkin2024upt}, large-scale pretrained operator transformers such as DPOT \cite{hao2024dpot}, and PDEformer-style models that condition directly on a representation of the equation itself \cite{ye2024pdeformer, ye2025pdeformer2}. Related work such as OmniArch \cite{chen2025omniarch} further suggests that large-scale operator pretraining can improve sample efficiency, few-shot transfer, and robustness across related tasks. Compared to task-specific neural operators, these approaches seek more transferable latent representations of PDE dynamics, boundary interactions, and geometric variation.

At present, most PDE foundation models are predominantly data-driven. Their training typically relies on large collections of simulation trajectories drawn from multiple PDE families, resolutions, and parameter ranges \cite{mccabe2023mpp, hao2024dpot, ohana2024thewell}. In contrast, physics-informed variants often enter through residual regularization, alignment objectives, or fine-tuning rather than replacing large-scale simulation data altogether \cite{chen2025omniarch, zhong2025pigano}. This distinction is important. Data-driven pretraining is currently the main mechanism for obtaining broad coverage across multiple equations and regimes, while physics-informed components are more often used to improve data efficiency, enforce consistency, or adapt a pretrained model to a smaller downstream dataset. In that sense, the present landscape is less about a single ``foundation architecture'' and more about a family of pretrained operator-learning systems that differ in how they encode PDE structure, geometry, and physical constraints.

The potential applications of such models are substantial. For rapid surrogate modeling across parametric families, a pretrained operator can amortize the cost of repeated PDE solves over many-query settings. This is directly relevant to inverse problems, design optimization, control, uncertainty propagation, and real-time digital twins, where the cost bottleneck is rarely a single forward solve but rather a large sequence of related solves \cite{wagg2020digital, juarez2021digital, shukla2024deep}. A sufficiently transferable pretrained operator could also enable cross-domain transfer, for example, by moving from one geometry family to another, from one boundary-condition regime to another, or from one PDE class to a related coupled system with comparatively little task-specific retraining \cite{shen2024ups, alkin2024upt}. Recent scaling efforts such as AB-UPT \cite{alkin2025abupt} and emerging multi-physics pretraining frameworks \cite{masliaev2025universalmultiphysics} indicate that this direction is especially attractive for industrial simulation settings where geometry diversity and repeated evaluations dominate runtime.

At the same time, foundation models for PDEs raise difficult technical questions. The first is the lack of reliable error bounds. Even when zero-shot or few-shot performance is strong on held-out benchmarks, there is rarely a computable guarantee on the error for a new geometry, a new boundary condition, or a new coefficient regime. The second challenge is generalization outside the training regime. Broader pretraining improves average transfer performance, but it does not remove the need to reason carefully about out-of-distribution inputs, unseen physical couplings, or extrapolative parameter settings \cite{cao2023residual,jha2024residual}. The third challenge is scale: higher-dimensional PDEs, strongly coupled multi-physics systems, and industrial geometries still place severe demands on both model capacity and training data. Benchmark ecosystems such as PDEBench, PDEArena, and The Well \cite{takamoto2022pdebench, gupta2023pdearena, ohana2024thewell} help standardize evaluation, but they also make clear that data generation remains expensive and potentially biased toward the numerical solvers, regimes, and discretizations used to create the datasets.

There are also challenges of interpretability and physics consistency. A pretrained model that performs well across many systems may still obscure which physical mechanisms it has learned. It may also be unclear which conservation laws are embedded in the architecture and when a prediction should trigger solver-based verification. For this reason, foundation models for PDEs are unlikely to replace classical numerical methods outright. A more realistic direction is tight coupling with numerical solvers through warm starts, residual correction, adaptive verification, and solver-in-the-loop refinement. In that workflow, pretrained operator models provide fast approximate maps over broad classes of problems, while numerical analysis provides the local certification and correction required for high-consequence decisions. From the perspective of scientific computing, this hybrid view appears more plausible than a purely black-box replacement of established PDE solvers.

\subsection{Final thoughts}\label{ss:finalThoughts}
Neural operators have established themselves as powerful surrogates for parametric PDEs, offering a different computational paradigm in which mappings between function spaces are learned once and evaluated rapidly. This shift has clear advantages in many-query settings such as uncertainty quantification, Bayesian inference, optimization, and control, where the dominant cost lies in repeated forward solves rather than a single high-fidelity simulation. 

At the same time, the core limitation is now well understood: accuracy is not inherently predictable in the same way as in classical numerical methods. Neural operators do not provide built-in guarantees tied to discretization, mesh refinement, or variational structure. This gap becomes critical in downstream tasks where small errors can lead to qualitatively incorrect decisions, particularly in optimization, inverse problems, and control.

Recent advances suggest that this limitation is not insurmountable, but it requires a shift in perspective. Accuracy must be actively controlled rather than passively expected. Residual-based correction, adaptive refinement, multi-level training, and multi-fidelity formulations all point toward a paradigm in which neural operators provide fast approximate predictions, while additional mechanisms ensure reliability in task-relevant regimes. The emerging theme is that neural operators are most effective when embedded within a broader computational workflow rather than used as standalone replacements for PDE solvers.

Looking forward, three directions appear particularly important. First, developing reliable, computable, and meaningful error indicators for specific applications remains a central challenge. Second, scalable strategies for handling distributional shifts, especially in high-dimensional, multi-physics settings, are essential for real-world deployment. Third, integrating pretrained operator models with physics-based correction and solver-in-the-loop verification offers a promising path to combining generality with reliability. These observations suggest that progress in neural operators will depend less on isolated architectural advances and more on principled integration with numerical analysis and application-driven error control.

In this sense, the future of neural operators is likely to involve data-driven operator learning augmented by physics-based correction, adaptive numerical analysis, problem-specific validation, and goal-oriented error certification. Advancing this integration will be key to realizing the full potential of neural operators in scientific computing.

\section*{Acknowledgements}

This work was supported in part by the National Science Foundation under the Engineering Research Initiation (ERI) award NSF \#2502279. The author also acknowledges support from the South Dakota Board of Regents Competitive Research Grant (SDBOR CRG) program. The author dedicates this work to the late Professor J. Tinsley Oden, to whom he is deeply grateful for his mentorship and support during his four years as a postdoctoral fellow and research associate at the University of Texas at Austin.

\addcontentsline{toc}{section}{References}
\biboptions{sort,numbers,comma,compress}                 
\bibliographystyle{elsarticle-harv}
\bibliography{main}

\appendix

\section{Neural operator predictions for elasticity problems}\label{s:appNNPredictions}
This appendix collects additional neural operator prediction results for the linear elasticity and hyperelasticity problems. The three neural operator architectures, DeepONet, PCANet, and FNO, are described in \cref{s:nn}. Their implementation details are outlined in \cref{sss:DeepONet}, \cref{sss:PCANet}, and \cref{sss:FNO}, respectively, and the neural network parameters are listed in \cref{tab:parameters}. 

\begin{figure}[!h]
    \centering
    \includegraphics[width=0.75\linewidth]{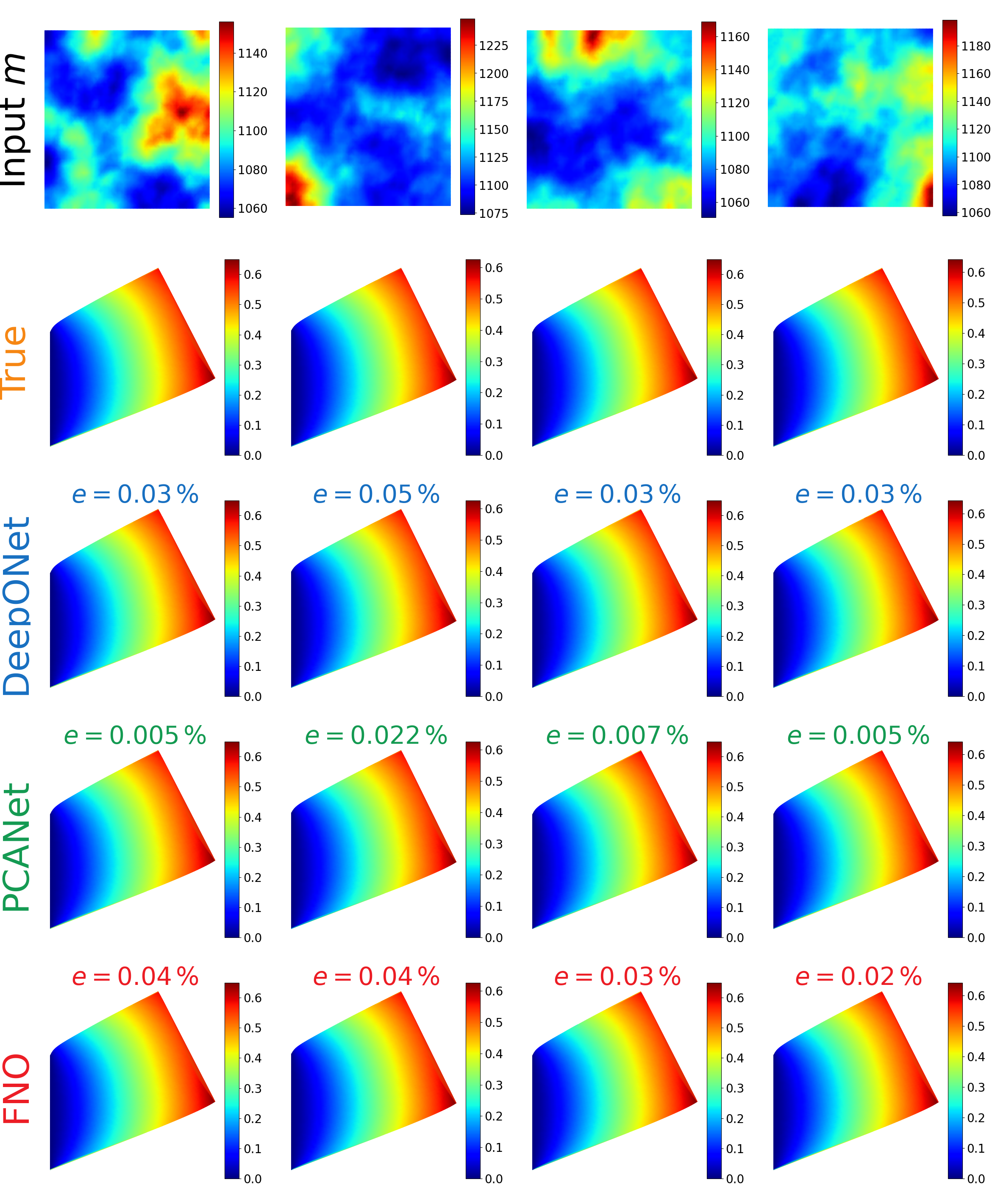}
    \caption{Comparison of DeepONet, PCANet, and FNO predictions with the finite element solution for four samples of the Young's modulus field in the linear elasticity problem. Here, $e$ denotes the relative percentage $\ell^2$ error between the true finite element solution and the surrogate solution. The displacement field $u$ is visualized by plotting the deformed configuration of the domain with coordinates $z = x + u(x)$ for $x\in D_{U}$. The color map indicates the displacement magnitude $|u(x)|$.}
    \label{fig:elasticityCompareNOps}
\end{figure}

\begin{figure}[!h]
    \centering
    \includegraphics[width=0.75\linewidth]{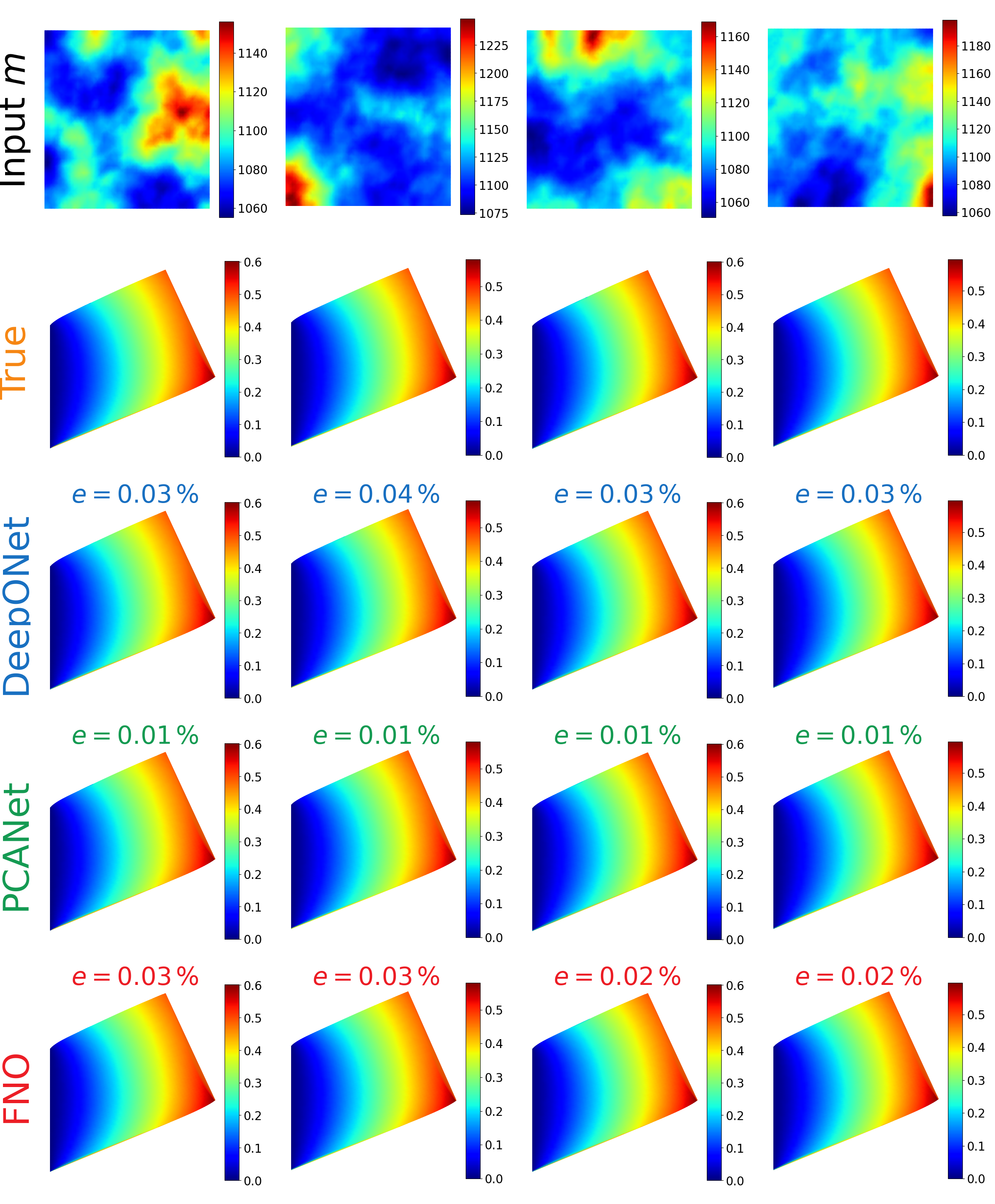}
    \caption{Comparison of DeepONet, PCANet, and FNO predictions with the finite element solution for four samples of the Young's modulus field in the hyperelasticity problem. The notation and visualization follow \cref{fig:elasticityCompareNOps}.}
    \label{fig:hyperelasticityCompareNOps}
\end{figure}

\cref{fig:elasticityCompareNOps,fig:hyperelasticityCompareNOps} compare the finite element solutions with predictions obtained using DeepONet, PCANet, and FNO for four samples of Young's modulus in the linear elasticity and hyperelasticity problems, respectively. The relative percentage $\ell^2$ error is also shown in each figure. For the elasticity problems, DeepONet and FNO yield errors close to $0.04\%$, while PCANet achieves errors that are approximately one order of magnitude smaller. This high accuracy also enables these surrogates to perform well in Bayesian inference problems, as shown in \labelcref{s:appBayesianInference}. However, the next subsection shows decreased predictive performance for OOD samples.

\subsection{Neural operator predictions for out-of-distribution samples}\label{ss:appNopPerformance}

\begin{figure}[!h]
    \centering
    \includegraphics[width=0.75\linewidth]{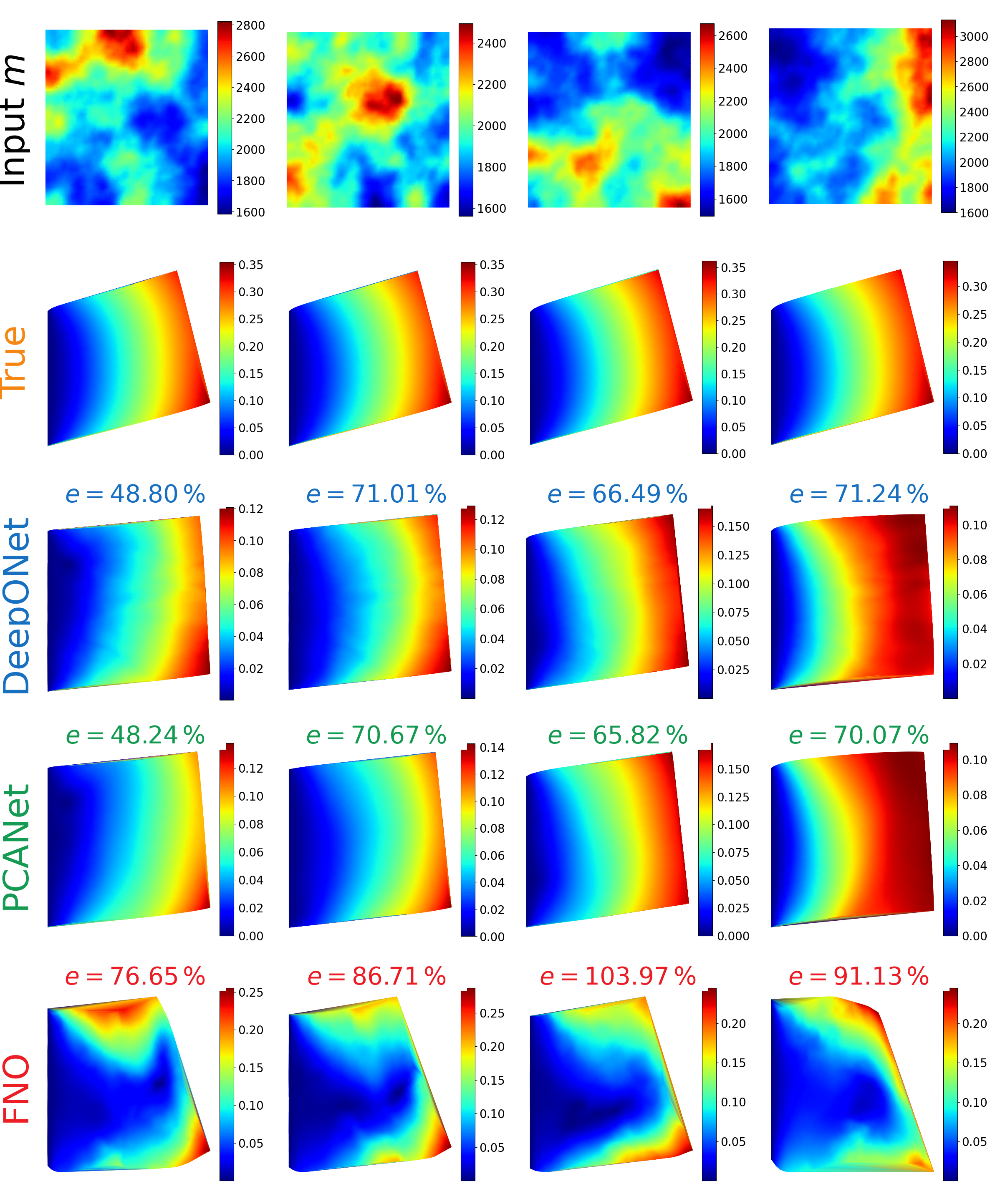}
    \caption{Representative out-of-distribution input samples and corresponding forward and surrogate solutions for the linear elasticity problem. The OOD samples correspond to Case 2 in \cref{ss:oodPerformance}. The notation and visualization follow \cref{fig:elasticityCompareNOps}.}
    \label{fig:elasticityOutDistNopSamples}
\end{figure}

\begin{figure}[!h]
    \centering
    \includegraphics[width=0.75\linewidth]{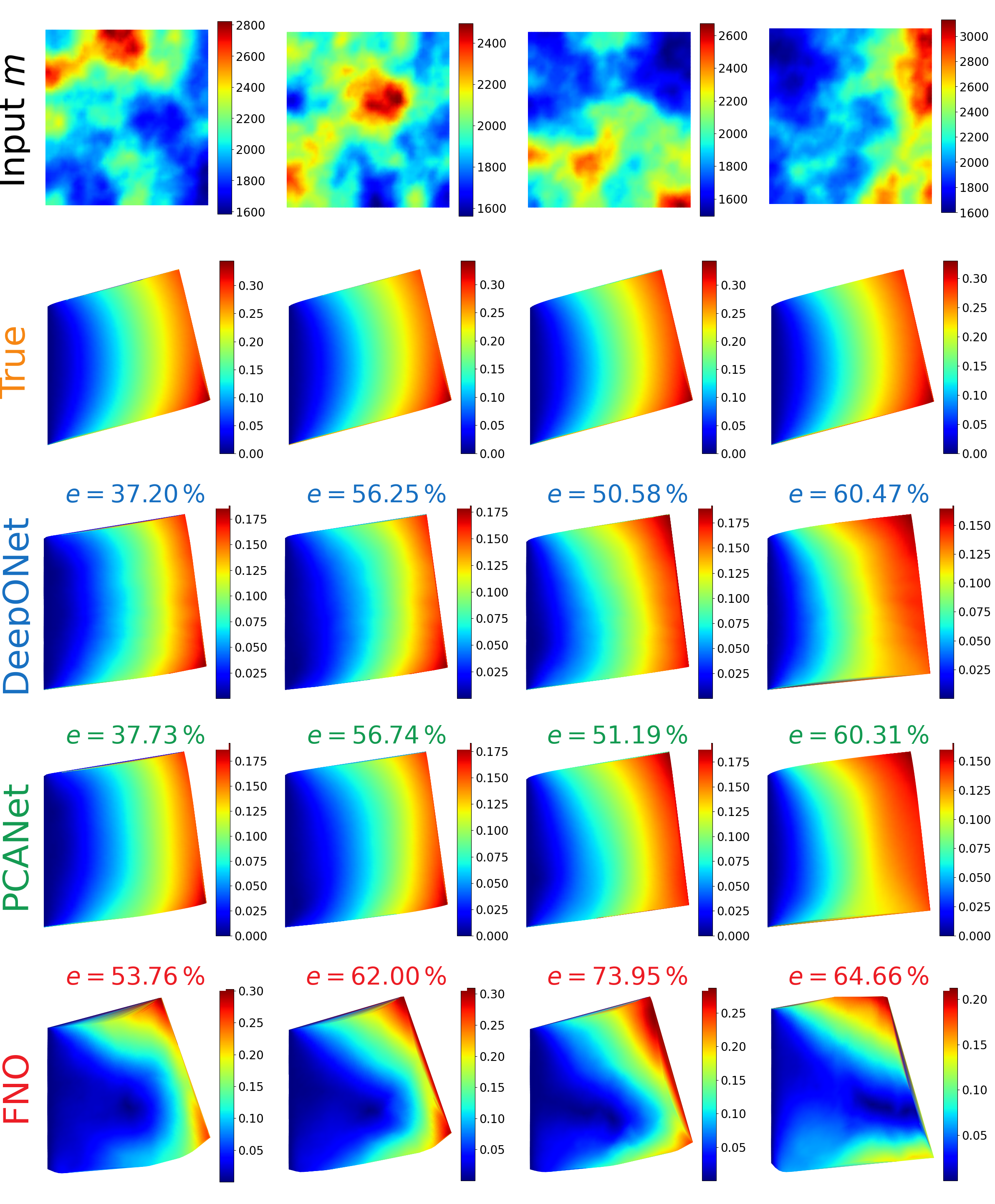}
    \caption{Representative out-of-distribution input samples and corresponding forward and surrogate solutions for the hyperelasticity problem. The OOD samples correspond to Case 2 in \cref{ss:oodPerformance}. The notation and visualization follow \cref{fig:elasticityCompareNOps}.}
    \label{fig:hyperelasticityOutDistNopSamples}
\end{figure}

In \cref{fig:elasticityOutDistNopSamples,fig:hyperelasticityOutDistNopSamples}, representative out-of-distribution input samples and the corresponding forward and surrogate solutions for the linear elasticity and hyperelasticity problems are shown, respectively. The OOD samples correspond to Case 2 in \cref{ss:oodPerformance}. The figures show that the surrogate solutions deviate significantly from the true finite element solutions, which is consistent with the large errors reported in \cref{tab:nopPerformance}.

\section{Bayesian inference of Young's modulus in the linear elasticity and hyperelasticity problems}\label{s:appBayesianInference}

This section presents the use of neural operators as surrogates in the Bayesian inference of the Young's modulus field $m$ in two elasticity problems. As in the Poisson case, positivity of Young's modulus is enforced by posing the inference problem in terms of the function $w \in W$, with $m = \alpha_m \exp(w) + \beta_m$.

\subsection{Setup of the forward problem, prior measure, and synthetic data}
The setup of the forward problem is the same as in \cref{sss:elasticitySetup} for linear elasticity and \cref{sss:hyperelasticitySetup} for hyperelasticity. The Gaussian prior measure $\mu^{\rmo}$ on $W$ is the same as $\mu_{\mscZ}$ used to generate the training data for the neural operators. The values of $\alpha_m$ and $\beta_m$, along with the covariance operator parameters, are given in \cref{sss:elasticitySetup} and \cref{sss:hyperelasticitySetup} for the two elasticity problems.  

\begin{figure}[!h]
    \centering
    \includegraphics[width=0.98\linewidth]{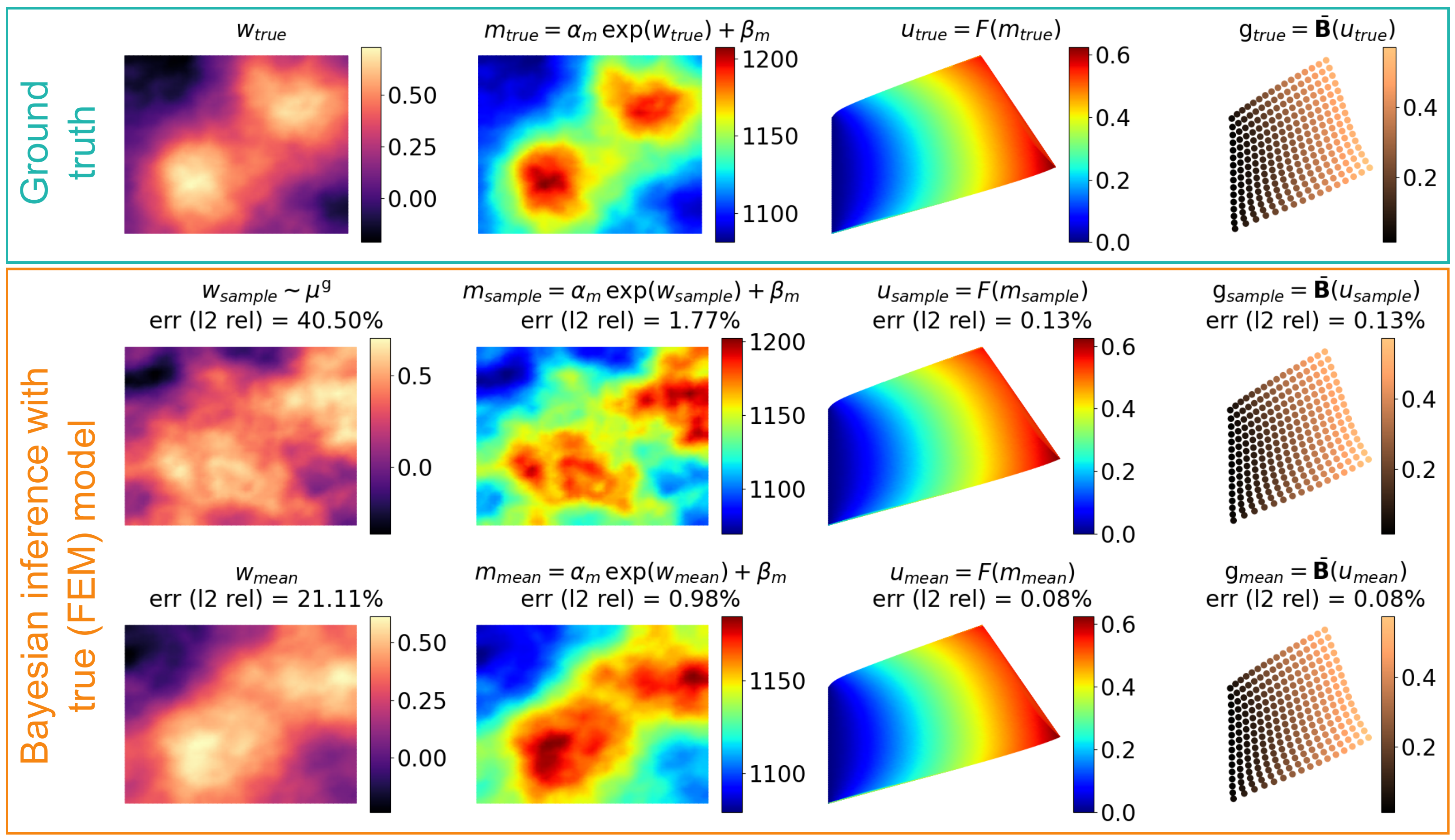}
    \caption{Bayesian inference of Young's modulus in the linear elasticity problem using the \textit{true} forward model (numerical approximation of the PDE). The top panel displays the synthetic field $w_{true}$, the corresponding Young's modulus field $m_{true}=\alpha_m\exp(w_{true})+\beta_m$, the displacement field $u_{true}=F(m_{true})$ (solution of the linear elasticity problem), and the observations $\rmg_{true}=\bar{\rb{B}}(u_{true})$. The displacement field $u_{true}$ is visualized by plotting the deformed configuration of the domain, with the color map indicating the magnitude $|u_{true}(x)|$. The observations $\rmg_{true}$ are obtained by interpolating $u_{true}$ on a $16\times16$ grid over $D_{U}=(0,1)^2$ and are visualized at the corresponding deformed observation locations $z_i=x_i+\rmg_{true,i}$, with each point colored by the magnitude $|\rmg_{true,i}|$ of the observed displacement vector. Thus, $\rmg_{true}\in\Rpow{d_{\rmg}}$ with $d_{\rmg}=2\times 16^2$. The lower panel displays a posterior sample $w_{sample}$ drawn from the posterior measure $\mu^{\rmg}$ (one realization from the MCMC chain) and the posterior mean $w_{mean}$ defined in \eqref{eq:posteriorMean}. The remaining columns display the corresponding Young's modulus field $m$, the forward solution $u=F(m)$ used in the MCMC simulation, and the predicted observations $\rmg=\bar{\rb{B}}(u)$ associated with the corresponding $w$.}
    \label{fig:elasticityBayesianInferenceTrue}
\end{figure}

The data are generated synthetically following the same procedure as in the Poisson problem; see \cref{sss:setupInferencePoisson}. In the elasticity problems, the data at each observation point consist of two displacement components, and therefore $\rmg \in \Rpow{d_{\rmg}}$ with $d_{\rmg} = 2 \times 16^2$. The top panels of \cref{fig:elasticityBayesianInferenceTrue} and \cref{fig:hyperElasticityBayesianInferenceTrue} display the synthetic data for the linear elasticity and hyperelasticity problems, respectively. The implementations can be found in the notebooks \lstinline{LinearElasticity: Generate_GroundTruth.ipynb}\footnote{\url{https://github.com/CEADpx/neural_operators/blob/survey26_v2/survey_work/applications/bayesian_inverse_problem_linear_elasticity/Generate_GroundTruth.ipynb}} and \lstinline{Hyperelasticity: Generate_GroundTruth.ipynb}\footnote{\url{https://github.com/CEADpx/neural_operators/blob/survey26_v2/survey_work/applications/bayesian_inverse_problem_hyperelasticity/Generate_GroundTruth.ipynb}}.

\subsection{Inference results for the linear elasticity problem}
The MCMC parameters are $k_{mcmc}$ = 10,500, $k_{burn} = 500$, $\beta_{pCN} = 0.15$, and $\sigma_{\rmg} = 0.0411$. Here, $\sigma_{\rmg}$ is chosen as $1\%$ of the mean of $\rmg$. The notebook \lstinline{BayesianInversion.ipynb}\footnote{\url{https://github.com/CEADpx/neural_operators/blob/survey26_v2/survey_work/applications/bayesian_inverse_problem_linear_elasticity/BayesianInversion.ipynb}} sets up the problem, loads the trained neural operators, and runs the MCMC simulations.

\begin{figure}[!h]
    \centering
    \includegraphics[width=0.98\linewidth]{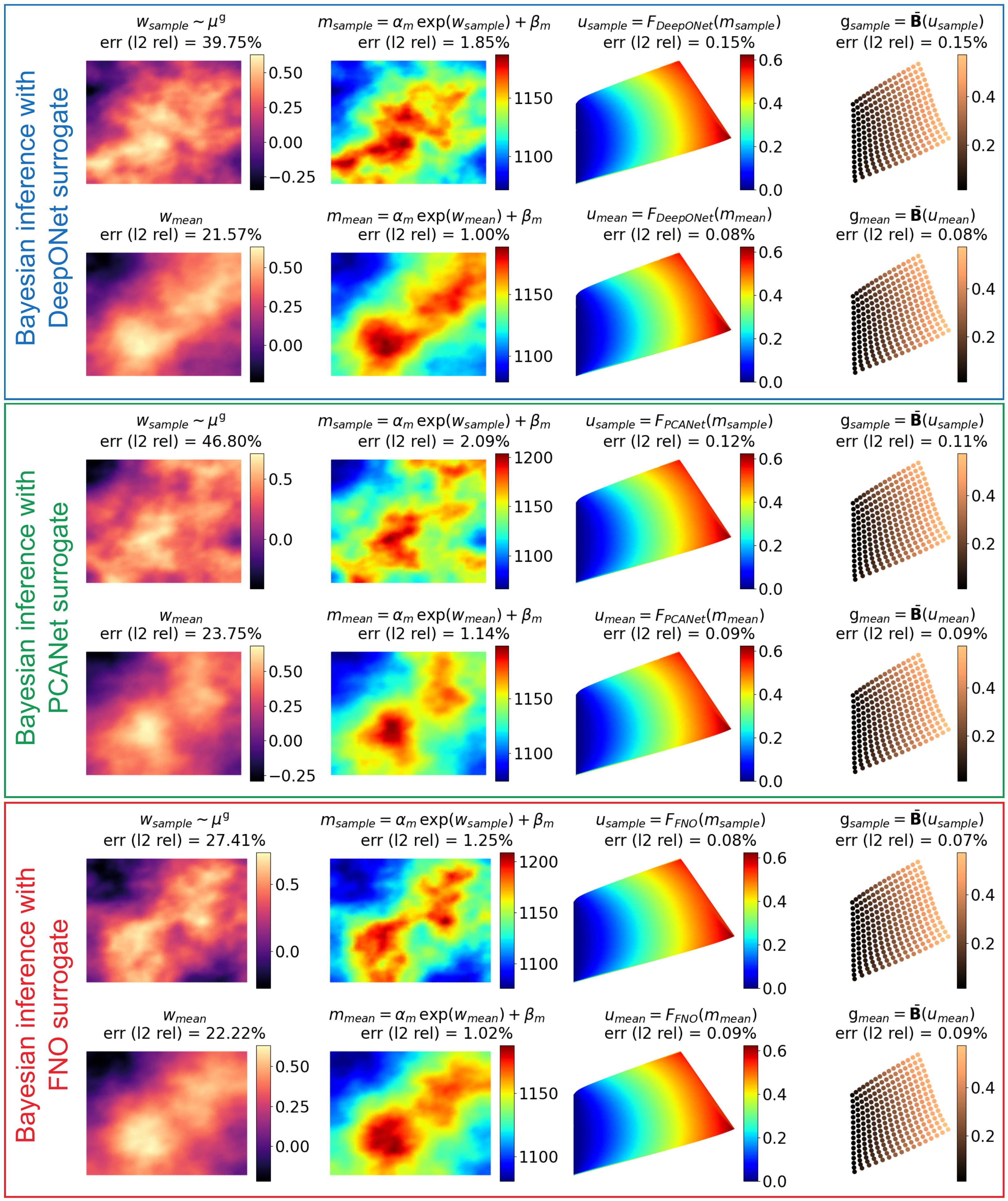}
    \caption{Comparison of Bayesian inference results for the linear elasticity problem using DeepONet, PCANet, and FNO surrogates. Each panel displays a posterior sample $w_{sample}$ drawn from the posterior measure $\mu^{\rmg}$ and the posterior mean $w_{mean}$. The corresponding Young's modulus field $m$, the forward solution $u = F(m)$, and predicted observations $\rmg = \bar{\rb{B}}(u)$ are also shown, where $F_{NOp}$, $NOp \in \{DeepONet, PCANet, FNO\}$, is the neural operator approximation of the forward operator used in the MCMC simulation and $\bar{\rb{B}}$ is the state-to-observable map. These results should be compared with the inference results obtained using the \textit{true} forward model in \cref{fig:elasticityBayesianInferenceTrue}. The visualizations of the displacement $u$ and observable $\rmg$ are described in \cref{fig:elasticityBayesianInferenceTrue}.}
    \label{fig:elasticityBayesianInferenceNOps}
\end{figure}

\begin{figure}[!h]
\centering
\subfloat[\centering Acceptance rate\label{fig:elasticityMCMCStatSub1}]{
  \begin{minipage}{.47\textwidth}
  \centering
  \includegraphics[width=\linewidth]{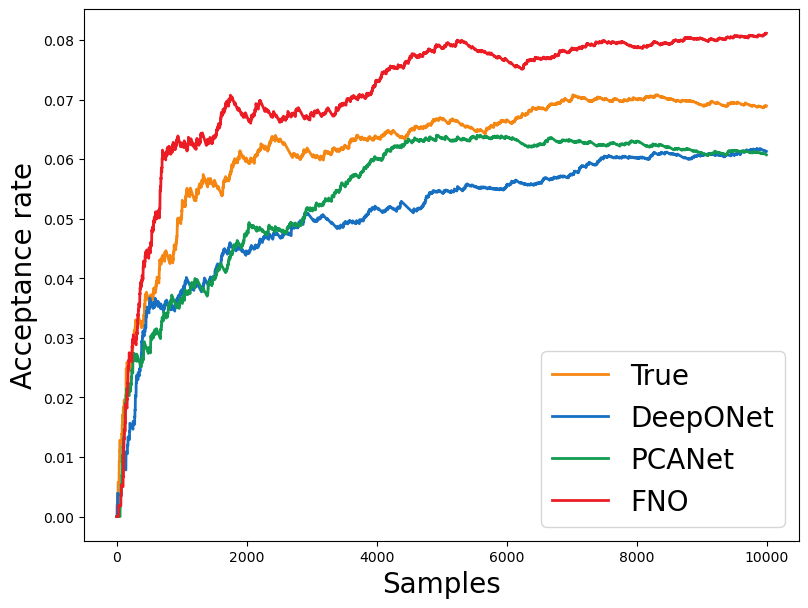}
  \end{minipage}
}%
\subfloat[\centering Cost\label{fig:elasticityMCMCStatSub2}]{
  \begin{minipage}{.47\textwidth}
  \centering
  \includegraphics[width=\linewidth]{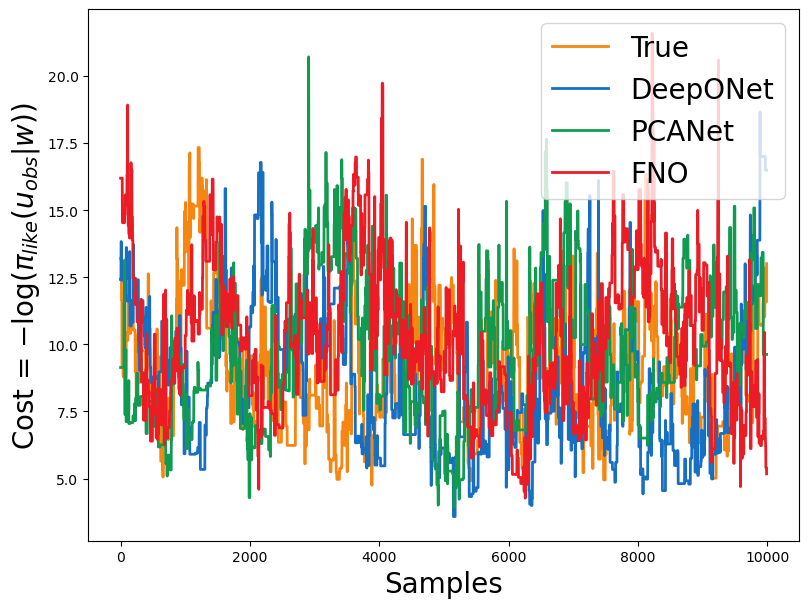}
  \end{minipage}
}
\caption{Acceptance rate and cost during MCMC simulation for the linear elasticity problem.}
\label{fig:elasticityMCMCStat}
\end{figure}

The results obtained using the \textit{true} forward model, including the posterior mean and a representative posterior sample, are shown in \cref{fig:elasticityBayesianInferenceTrue}. The results obtained using DeepONet, PCANet, and FNO surrogates are shown in \cref{fig:elasticityBayesianInferenceNOps}. The acceptance rate and cost histories for the four MCMC simulations corresponding to the \textit{true} and surrogate forward models are shown in \cref{fig:elasticityMCMCStat}.

In this case, the posterior means obtained using the \textit{true} forward model and the surrogate models are very similar. The error in the posterior mean of $m$ for both the \textit{true} model and the surrogates is less than one percent. Notably, the synthetic field $w_{true}$ exhibits higher values along the diagonal, and the posterior mean successfully captures this variation.

\subsection{Inference results for the hyperelasticity problem}
\begin{figure}[!h]
    \centering
    \includegraphics[width=0.98\linewidth]{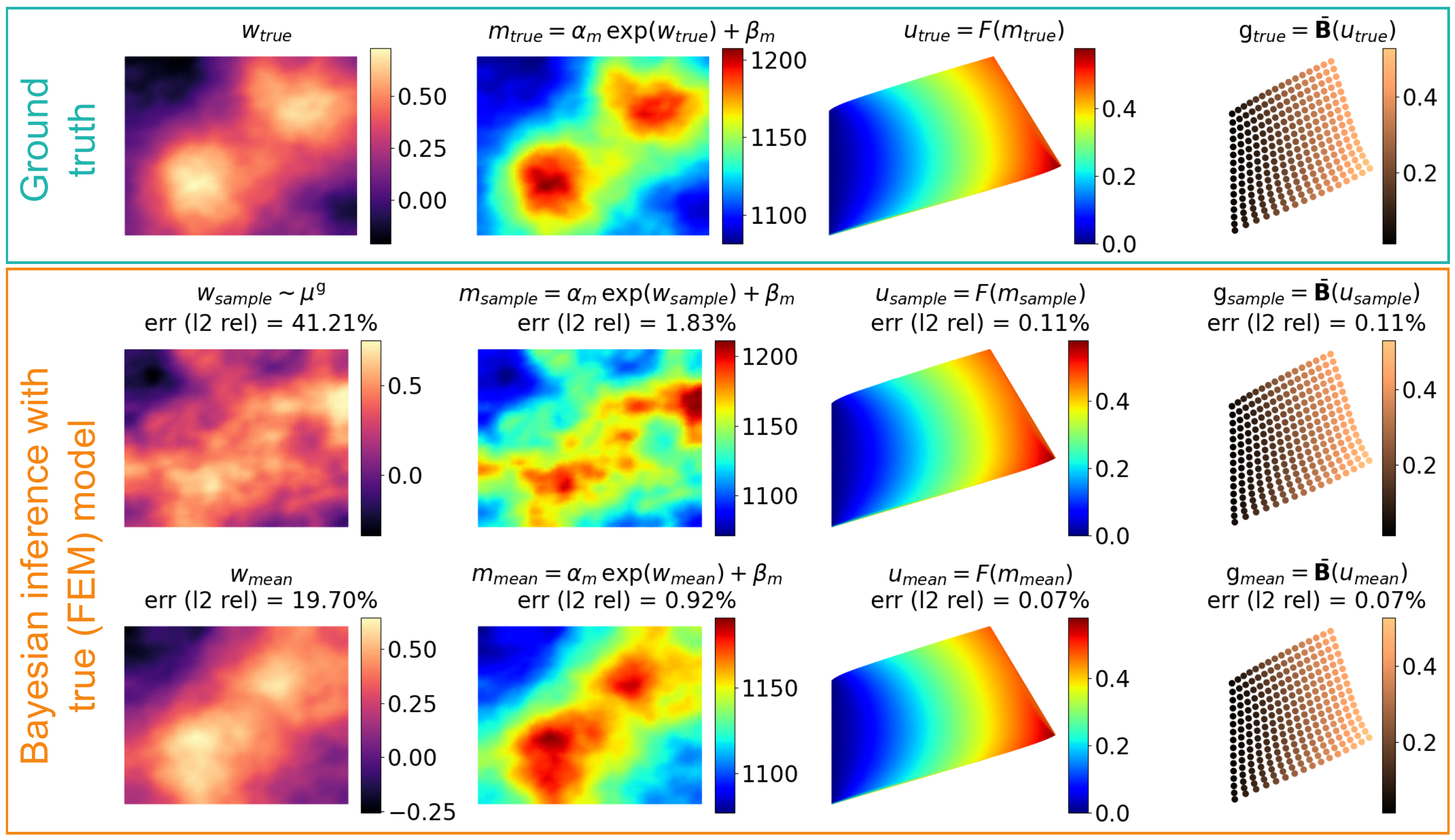}
    \caption{Bayesian inference of Young's modulus in the hyperelasticity problem using the \textit{true} forward model (numerical approximation of the PDE). The top panel displays the synthetic field $w_{true}$, the corresponding Young's modulus field $m_{true}=\alpha_m\exp(w_{true})+\beta_m$, the displacement field $u_{true}=F(m_{true})$ (solution of the hyperelasticity problem), and the observations $\rmg_{true}=\bar{\rb{B}}(u_{true})$. The observations are obtained by interpolating the displacement field $u_{true}$ on a $16\times16$ grid over $D_{U}=(0,1)^2$, so that $\rmg_{true}\in\Rpow{d_{\rmg}}$ with $d_{\rmg}=2\times 16^2$. The lower panel displays a posterior sample $w_{sample}$ drawn from the posterior measure $\mu^{\rmg}$ (one realization from the MCMC chain) and the posterior mean $w_{mean}$ defined in \eqref{eq:posteriorMean}. The remaining columns display the corresponding Young's modulus field $m$, the forward solution $u=F(m)$ used in the MCMC simulation, and the predicted observations $\rmg=\bar{\rb{B}}(u)$ associated with the corresponding $w$. The visualizations of the displacement $u$ and observable $\rmg$ are described in \cref{fig:elasticityBayesianInferenceTrue}.}
    \label{fig:hyperElasticityBayesianInferenceTrue}
\end{figure}

\begin{figure}[!h]
    \centering
    \includegraphics[width=0.98\linewidth]{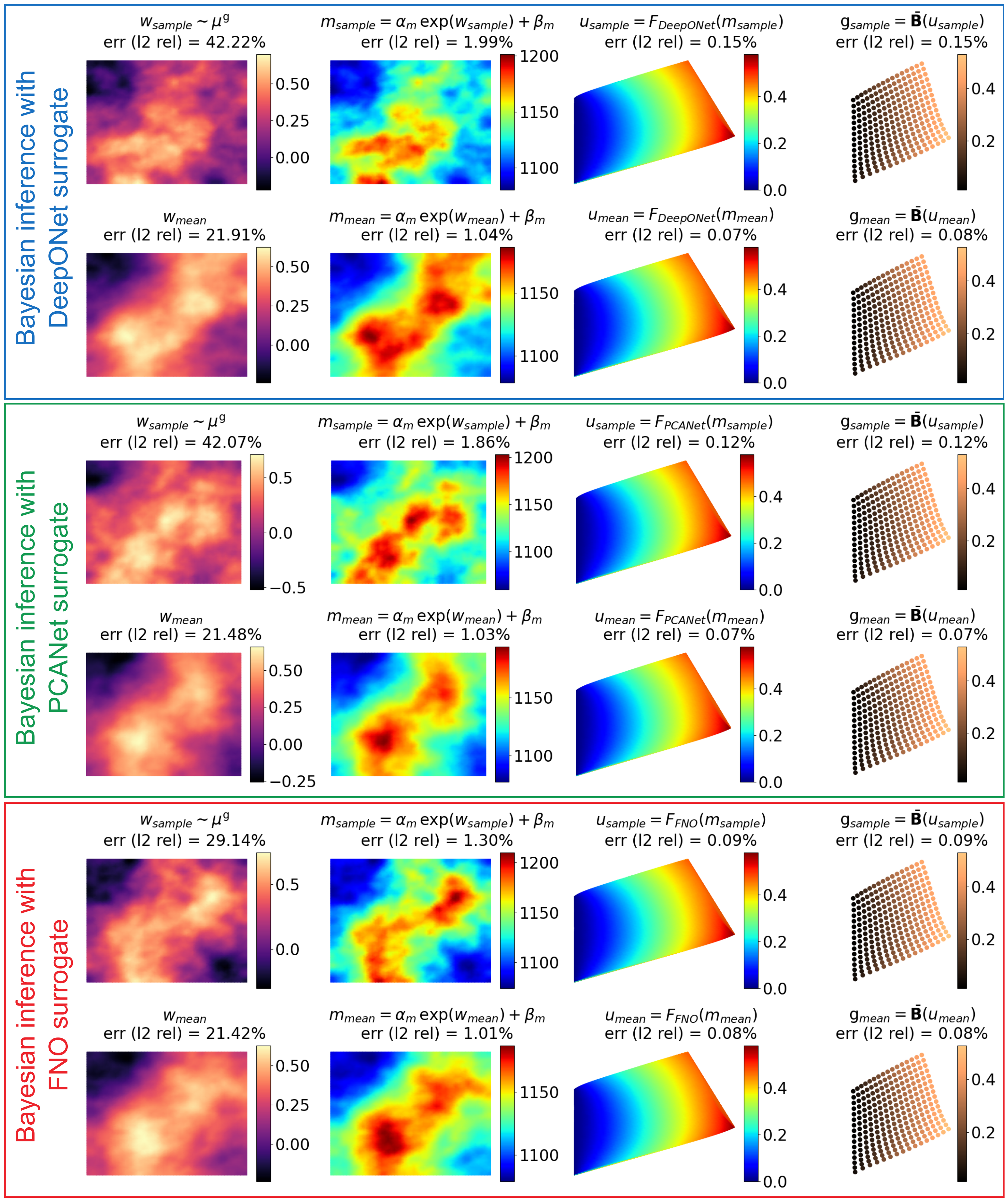}
    \caption{Comparison of Bayesian inference results for the hyperelasticity problem using DeepONet, PCANet, and FNO surrogates. Each panel displays a posterior sample $w_{sample}$ drawn from the posterior measure $\mu^{\rmg}$ and the posterior mean $w_{mean}$. The corresponding Young's modulus field $m$, the forward solution $u = F(m)$, and predicted observations $\rmg = \bar{\rb{B}}(u)$ are also shown, where $F_{NOp}$, $NOp \in \{DeepONet, PCANet, FNO\}$, is the neural operator approximation of the forward operator used in the MCMC simulation and $\bar{\rb{B}}$ is the state-to-observable map. These results should be compared with the inference results obtained using the \textit{true} forward model in \cref{fig:hyperElasticityBayesianInferenceTrue}. The visualizations of the displacement $u$ and observable $\rmg$ are described in \cref{fig:elasticityBayesianInferenceTrue}.}
    \label{fig:hyperElasticityBayesianInferenceNOps}
\end{figure}

\begin{figure}[!h]
\centering
\subfloat[\centering Acceptance rate\label{fig:hyperElasticityMCMCStatSub1}]{
  \begin{minipage}{.47\textwidth}
  \centering
  \includegraphics[width=\linewidth]{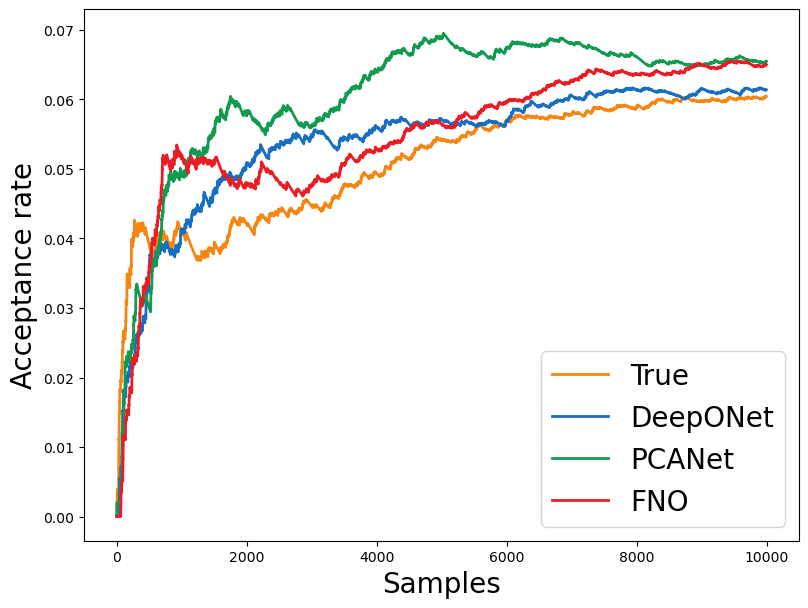}
  \end{minipage}
}%
\subfloat[\centering Cost\label{fig:hyperElasticityMCMCStatSub2}]{
  \begin{minipage}{.47\textwidth}
  \centering
  \includegraphics[width=\linewidth]{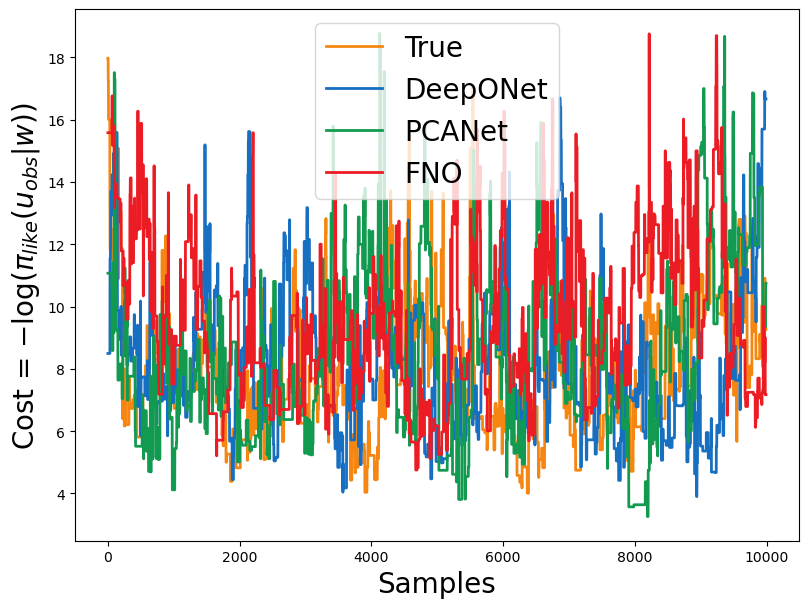}
  \end{minipage}
}
\caption{Acceptance rate and cost during MCMC simulation for the hyperelasticity problem.}
\label{fig:hyperElasticityMCMCStat}
\end{figure}

The MCMC parameters are the same as those used in the linear elasticity case. The results obtained using the \textit{true} forward model, including the posterior mean and a representative posterior sample, are shown in \cref{fig:hyperElasticityBayesianInferenceTrue}. The results obtained using DeepONet, PCANet, and FNO surrogates are shown in \cref{fig:hyperElasticityBayesianInferenceNOps}. The acceptance rate and cost histories for the four MCMC simulations corresponding to the \textit{true} and surrogate forward models are shown in \cref{fig:hyperElasticityMCMCStat}. The results are similar to the linear elasticity case, with the posterior means from the surrogate models closely matching the posterior mean obtained using the \textit{true} forward model.

\end{document}